\documentclass[aps,superscriptaddress,twocolumn,reprint]{revtex4-2}

\usepackage{times}
\usepackage{graphicx}
\usepackage{amsmath}
\usepackage{amsfonts}
\usepackage{amsmath}
\usepackage{amssymb}
\usepackage{amsthm}
\usepackage{bm}
\usepackage{amsopn}
\usepackage{xspace}
\usepackage{array}
\usepackage{placeins}
\usepackage{geometry}
\usepackage{comment}
\usepackage{subcaption}
\usepackage{cases}
\usepackage[outdir=.]{epstopdf}
\usepackage{tikz}
\usepackage{3dplot} 
\usepackage{xcolor}
\usepackage{makecell}

\newgeometry{right=1.5cm,left=1.5cm,top=1.8cm,bottom=1.8cm}

\newdimen\XCoord
\newdimen\YCoord
\newcommand*{\ExtractCoordinate}[1]{\path (#1); \pgfgetlastxy{\XCoord}{\YCoord};}%
\usetikzlibrary{arrows.meta,positioning,calc,decorations.markings}

\DeclareMathOperator\erf{erf}
\begin{document}
\author{C. Livi}
\affiliation{Fluids and Flows group and J.M. Burgers Centre for Fluid Dynamics, Department of Applied Physics, Eindhoven University of Technology, P.O. Box 513, 5600MB, Eindhoven, The Netherlands}
\author{G. Di Staso}
\affiliation{Fluids and Flows group and J.M. Burgers Centre for Fluid Dynamics, Department of Applied Physics, Eindhoven University of Technology, P.O. Box 513, 5600MB, Eindhoven, The Netherlands}
\affiliation{FLOW Matters Consultancy B.V., Groene Loper 5, 5612AE, Eindhoven, The Netherlands}
\author{{H. J. H.} Clercx}
\affiliation{Fluids and Flows group and J.M. Burgers Centre for Fluid Dynamics, Department of Applied Physics, Eindhoven University of Technology, P.O. Box 513, 5600MB, Eindhoven, The Netherlands}
\author{F. Toschi}
\affiliation{Fluids and Flows group and J.M. Burgers Centre for Fluid Dynamics, Department of Applied Physics, Eindhoven University of Technology, P.O. Box 513, 5600MB, Eindhoven, The Netherlands}
\begin{abstract}
The capability to simulate a two-way coupled interaction between a rarefied gas and an arbitrary-shaped colloidal particle is important for many practical applications, such as aerospace engineering, lung drug deliver and semiconductor manufacturing. By means of numerical simulations based on the Direct Simulation Monte Carlo (DSMC) method, we investigate the influence of the orientation of the particle and rarefaction on the drag and lift coefficients, in the case of prolate and oblate ellipsoidal particles immersed in a uniform ambient flow. This is done by modeling the solid particles using a cut-cell algorithm embedded within our DSMC solver. In this approach, the surface of the particle is described by its analytical expression and the microscopic gas-solid interactions are computed exactly using a ray-tracing technique. The measured drag and lift coefficients are used to extend the correlations available in the continuum regime to the rarefied regime, focusing on the transitional  and free-molecular regimes. The functional forms for the correlations for the ellipsoidal particles are chosen as a generalization from the spherical case. We show that the fits over the data from numerical simulations can be extended to regimes outside the simulated range of $Kn$ by testing the obtained predictive model on values of $Kn$ that where not included in the fitting process, allowing to achieve an higher precision when compared with existing predictive models from literature. Finally, we underline the importance of this work in providing new correlations for non-spherical particles that can be used for point-particle Euler-Lagrangian simulations to address the problem of contamination from finite-size particles in high-tech mechanical systems.
\end{abstract}
\title{On the drag and lift coefficients of ellipsoidal particles under rarefied flow conditions}
\maketitle

\section{Introduction}
Multiphase flows including particulate suspensions in conditions where the flow around the particles is rarefied are important in many different natural, medical and industrial applications. Examples can be found in the formation of cloud droplets and in ozone depletion in the stratosphere \cite{wang}, contamination from particle debris in high tech mechanical systems \cite{lito} and lung drug delivery \cite{lung}. In all these cases the typical size of the particles is small when compared to the mean free path of the surrounding gas molecules and non-equilibrium effects are important in gas-surface interactions.\\ Numerous studies have been proposed to address the problem of shape effects on the transport of particles in the continuum regime. From the pioneering theoretical work of Oberbeck \cite{oberbeck} and Jeffery \cite{jeffery}, who firstly investigated the motion of an ellipsoid immersed in a fluid in the Stokes limit, an increasingly growing effort has been dedicated to understand shape and orientation effects on the drag, lift and torque experienced by particles in different flow conditions \cite{haider, ganser, holzer,zastawny,richter,ouchene,livi}.\\
In this context, and in the continuum regime, Sanjeevi \textit{et al.} \cite{sanjeevi1,sanjeevi2} performed accurate simulations using the Lattice-Boltzmann method, extending available correlations for the drag, lift and torque coefficients for particles with different shapes, including ellipsoidal particles, from the Stokes limit to high Reynolds number cases. They show that the sine-squared scaling of the drag force, with respect to the orientation of the particles, firstly proposed by Happel and Brenner \cite{happel} for the Stokes limit, can be extended to a large range of Reynolds numbers. The reason for the sine-squared drag law to hold at high Reynolds resides in pressure effects around the particles \cite{sanjeevi1}, rather than in the linearity theory of Stokes flows. More specifically, the interplay between the two components of the drag, namely the upstream (i.e. in front of the particle) and the wake (i.e. in the rear of the particle) drag, adequately compensates non-linear effects from the velocity field, making the total drag section appear to scale in a sine-squared manner.\\ 
\begin{figure}[h!]
\begin{center}
\tdplotsetmaincoords{60}{110}

\pgfmathsetmacro{\avec}{0.3}
\pgfmathsetmacro{\bvec}{0.15}
\pgfmathsetmacro{\rvec}{0.3}

\pgfmathsetmacro{\thetavec}{-160}
\pgfmathsetmacro{\phivec}{80}

\begin{tikzpicture}[>=stealth,scale=5,tdplot_main_coords]

\shadedraw[tdplot_screen_coords,ball color = white,rotate=-10] (0,0) circle (0.3cm and  0.15cm);

\coordinate (O)   at (0,0,0);

\tdplotsetcoord{P}{\rvec}{90}{-20}




\def\ufx{0.35+0.2}
\def\ufy{0.38+0.2}
\def\ufz{0.15+0.12}

\def\usx{0}
\def\usy{0}
\def\usz{0}

\coordinate (Us) at (\usx,\usy,\usz);
\coordinate (Uf)  at (\ufx,\ufy,\ufz);
\draw[thick,->, line width = 0.8mm,blue] (Us) -- (Uf) node[anchor=north]{\textcolor{black}{\Large{$\mathbf{U}$}}};

\draw[thick,->, line width = 0.6mm,black] (Us) -- (\usx+0.7,\usy,\usz) node[anchor=north east]{\Large{$z'$}};
\draw[thick,->, line width = 0.6mm,black] (Us) -- (\usx,\usy+0.6,\usz) node[anchor=north west]{\Large{$x'$}};
\draw[thick,->, line width = 0.6mm,black] (Us) -- (\usx,\usy,\usz+0.5) node[anchor=south]{\Large{$y'$}};

\draw[thick,->, line width = 0.8mm,blue] (Us) -- (\ufx,\usy,\usz) node[anchor=north west]{\textcolor{black}{\Large{$\mathbf{U}_{z'}$}}};
\draw[thick,->, line width = 0.8mm,blue] (Us) -- (\usx,\ufy,\ufz) node[anchor=south]{\textcolor{black}{\Large{$\mathbf{U}_{x'y'}$}}};


\tdplotdrawarc[thick,line width=0.35mm,blue]{(Us)}{0.12}{0}{75}{anchor=north}{\Large{\textcolor{black}{$\theta$}}}
\tdplotdrawarc[thick,line width=0.35mm,blue]{(Us)}{0.45}{90}{132}{anchor=west}{\Large{\textcolor{black}{$\Phi$}}}

\tdplotsetthetaplanecoords{\phivec}



\end{tikzpicture}
\end{center}
\caption{\small{Sketch of an ellipsoidal particle immersed in a uniform Stokes-flow with velocity $\mathbf{U}$ for an arbitrary orientation. It is always possible to describe the ambient flow in the body-centered reference frame $(x'y'z')$ aligned with the semi-axes of the particle. In this reference, the ambient velocity $\mathbf{U}$ can be decomposed in its components $\mathbf{U}_{x'y'}$, laying on the $x'y'$-plane forming an angle $\Phi$ with respect to the $x'$ axis, and $\mathbf{U}_{z'}$, laying on the $z'$-axis. Since $\mathbf{U}_{z'}$ is independent on the relative orientation between the particle and the flow by construction (as it is always orthogonal to the principal axis of the particle), orientation effects are described by the single angular variable $\Phi$, or angle of attack, without loss of generality.}}
\label{fig:general_geom}
\end{figure}
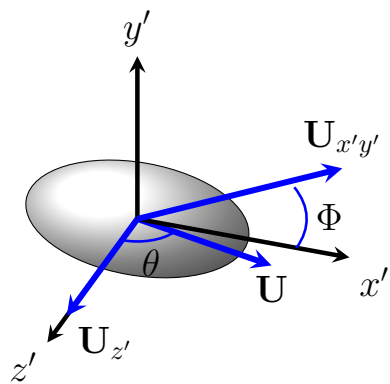
While the understanding of particle-flow interactions in the continuum regime is consistently increasing through the years, our knowledge of the impact of rarefaction conditions, related to non-equilibrium effects, on the dynamics of the particles is still limited due to the difficulties in addressing such effects both numerically and experimentally. From the numerical point of view, limitations arise as the conventional Navier-Stokes solvers fail due to the breakdown of the continuum assumption, while from the experimental point of view it is difficult to create ideal conditions to investigate the dynamics of nano-sized particles in rarefied flow conditions.\\
Typically, in numerical studies related to the transport of nano-sized particles in micro-mechanical devices such as micro-channels \cite{xuzheng}, hard-drives \cite{shen,zhang}, aerodynamic lenses \cite{abduali} or suspension plasma sprayers  \cite{kaizhang}, particles are simulated through Euler-Lagrangian approaches. The flow field is evaluated with an Eulerian approach, while the trajectories of the particles are computed with a Lagrangian approach where the particles are modeled as (spherical) point particles. Rarefaction effects are then included through phenomenological corrections to the drag force of the continuum limit, such as the classical Cunningham corrections \cite{cunningham,liu}.\\
While, in general, the Euler-Lagrangian description provides a reasonable approximation for the dynamics of micro- and nano-metric particles immersed in a gas, using a point-particle approach, any effect related to the finite size of the particles, their shape and orientation is neglected.\\
Different authors addressed, both analytically and numerically, the interaction between rarefied gas flows and a finite-size spherical particle. Epstein \cite{epstein} firstly derived a drag relation for a sphere translating in a gas at thermal equilibrium in the free-molecular regime, focusing on the case of fully diffusive reflections between the impinging gas molecules and the solid surface. The analysis proposed by Epstein is based on the assumption that the flow velocity is small compared with the thermal molecular speed. This approach was later extended by Baines \cite{baines} for the case of specular reflections. More recently, Li \textit{et al.} \cite{li1,li2} proposed an alternative formulation for the mobility of small particles in the free-molecular regime based on kinetic theory. Their approach takes into account electric mobility in the collisional cross-section, allowing to include van der Waals and other interactions on the drag force. They show that their results are consistent with the formulation from Epstein in the limit of rigid-body hard-sphere intermolecular interactions.\\
Phillips \cite{phillips}  provides an analytical expression for the drag force on a sphere through an approximate solution of the Boltzmann equation. The results from Phillips match with the experimental observations from Millikan \cite{millikan1,millikan2} for a large range of the Knudsen numbers, covering the slip and transition regimes ($0.0865\leq Kn \leq 3.36$). Gallis \textit{et al.} \cite{gallis,gallis2} proposed an approach based on the use of Green's function to calculate drag and heat flux experienced by a sphere in the free-molecular regime for monatomic and diatomic gases. The aforementioned approaches are, however, limited to spherical particles and have not been extended, so far, to particles with more complex shapes.\\
Some works are available in the literature that tackle the problem of gas-solid interactions in the case of non-spherical particles from a theoretical perspective: Halbritter \cite{halbritter} derived a theoretical formulation for the torque exerted by a rarefied gas on an ellipsoidal particle at thermal equilibrium. Dahneke \cite{dahneke} extended the analytical formulation from Epstein, valid in the free-molecular regime, to particles with different shapes, including cylinders, prolate and oblate ellipsoids. Martinetz \textit{et al.} \cite{martinetz} derived a theoretical formulation based on the Boltzmann equation for the roto-translational dynamics of particles with different shapes immersed in a background free-molecular gas. While the aforementioned theoretical advancements are of great importance in understanding the underlying physics of rarefied gas dynamics, they are limited to free-molecular flows.\\
Some phenomenological models have been proposed \cite{dahneke3} to investigate the drag corrections experienced by non-spherical particles in the transition and slip flow regimes, such as the Equivalent Sphere Approximation (ESA) and the Adjusted Sphere Approximation (ASA). The former consists in the direct application of the Cunningham corrections \cite{cunningham} on the sphere with equivalent volume of the investigated particle, so that any information on the orientation is lost and this model offers a good accuracy only for slightly non-spherical bodies. The latter is a more sophisticated model where the rarefaction corrections are modeled by equating the Cunningham corrections of a spherical particle having an effective radius (to be determined) with the ratio of continuum drag force to free-molecule drag force of the investigated particle. The main feature of the ASA model is that it allows to keep into consideration orientation effects, but it requires the knowledge of the drag force on the body in the continuum and free-molecular regimes, and this information is only available for a limited number of shapes.\\
To address rarefied gas flow problems in a broad spectrum of $Kn$, ranging from slip to free-molecular flows, the Direct-Simulation Monte Carlo (DSMC) method  \cite{bird,stefanov} has proven to be a stable and accurate numerical approach to model a wide range of applications, from classical rarefied gas dynamics \cite{karnidakis} to, more recently, micro-fluidic devices \cite{reese}. In standard DSMC simulations, the flow domain is discretized into a finite-size grid where the simulated gas molecules can move ballistically. Gas molecules within the same DSMC cell undergo stochastic binary collisions depending on their relative velocity and the flow macroscopic quantities such as the gas density, temperature and pressure.\\   
In the framework of the DSMC method, finite-size solid particles can be modeled as finite volumes enclosed by moving boundaries suspended in the fluid domain. Fluid-solid interactions are addressed via appropriate molecule-surface collision laws applied at the surface of the particle. Momentum exchange between the gas molecules and the solid particle is used to compute the force and the torque exerted on the solid particle. \\
Two main approaches are usually applied to model the surface of the particle in the DSMC domain:  in one case the surface of the particle is approximated  by a non-Cartesian body-fitted mesh, and every face on the meshed surface coincides with a DSMC grid cell face on the gas domain. In the second approach, the surface of the particle is represented with its analytical expression and it is free to move on the Cartesian DSMC grid, somewhat similar to the immersed boundary method proposed by Peskin \cite{Peskin}. \\
The latter approach has been called the cut-cell method \cite{cut1,cut2}, as the super-imposition of the solid particle volume on the DSMC cartesian grid imposes that some of the DSMC grid cells (i.e. the boundary cells at the gas-solid interface) are cut by the surface of the particle, requiring to dynamically compute and update the volume of such cells. The cut-cell method provides two main important advantages with respect to the body-fitted mesh approach: firstly, the surface of the particle can be described analytically and the collision points between gas molecules and the solid surface computed exactly using, for example, a ray-tracing technique. Secondly, when the motion of the particle is taken into account, it overcomes the complicated problem of adaptive re-meshing of the whole simulation grid at every time step, as only the cut-cells volumes have to be recomputed. \\
Examples of recent successful applications of the cut-cell method to investigate interactions between a rarefied gas flow and different solid particles can be found in the literature: Jin \textit{et al.} \cite{jin} propose an efficient approach to recalculate the cut-cell volume based on the computation of the intersected edges at which the solid surface intersects the DSMC boundary cells. The surface of the particle is then approximated, on each boundary cell, through a polyhedron passing from the intersected edge, and the solid fraction of the cell is computed accordingly to this local approximation. They successfully apply this approach to spherical particles as well as to particles with more complex surfaces. Shrestha \textit{et al.} \cite{shrestha} apply the cut-cell algorithm to the study of different problems, such as the Brownian diffusion of a spherical particle in the free-molecular regime and the transport of an arbitrary-shape particle driven by the thermophoretic force. Baier \textit{et al.} \cite{baier} investigated the thermophoretic force exerted on spherical Janus particles at different orientations with respect to the thermal gradient. Chinnappan \textit{et al.} \cite{chinnappan} investigated the transport dynamics of ellipsoidal particles in the free-molecular gas flow regime. The aforementioned studies are, however, mostly limited to gas-solid interactions in the free-molecular regimes, and up to our knowledge an extensive investigation of the drag and lift correlations between particles and gas in the transitional regime is still missing in the literature. This regime is of particular interest for many of the high-tech applications, as most of the gas flows are not in the free-molecular regime, although still at very low pressures and densities, and the Knudsen number based on the particle size of the contaminant particles often exceeds unity.\\
In this work we aim to cover this gap by addressing the impact of a finite Knudsen number ($1\leq Kn \leq 10$) in the interactions between a gas flow and ellipsoidal particles. We do so by proposing a cut-cell algorithm which is able to describe a spheroidal particle exactly, at any aspect ratio. Through the use of a standard ray-sphere intersection approach, the collision points between the gas molecules and the ellipsoidal solid particle are obtained at the exact position on the surface of the particle and the volume of the boundary cells (cut-cells) is computed through a Monte-Carlo approach. We address orientation effects of an impinging uniform gas flow on different ellipsoidal particles. This class of problems can be studied, without loss of generality, by changing the angle of attack at which the gas flow impinges on the simulated particles, as sketched in Fig. \ref{fig:general_geom}. We repeat this analysis for different Knudsen numbers ranging from the transitional to the free-molecular regimes, with the final objective to provide a heuristic model able to predict rarefaction and orientation effects on the hydrodynamic forces acting on the ellipsoidal solid particles.\\
In the first part of this paper we present an analysis of the cut-cell algorithm performances through the computation of the drag force experienced by a spherical particle immersed in a uniform ambient flow for different Knudsen numbers, showing a good agreement with similar approaches available in the literature. We then investigate the accuracy scaling of the mean value and of the standard deviation of the drag force experienced by a solid spherical particle in different conditions, showing the impact of spatial and kinetic resolutions on the accuracy of the simulations. While in related publications \cite{jin,chonling} great care is dedicated towards the calculation of the accuracy at which the cut-cells volume is recovered, the convergence analysis presented in this work is, up to our knowledge, still not available in the literature.\\
We then address the impact of shape, orientation and rarefaction on the drag and lift coefficients for different ellipsoidal particles. We firstly present a suitable definition of the Knudsen number for ellipsoidal particles based on the sphere with equivalent volume. The presence of multiple characteristic lengths (embodied by the major and minor axes) must be taken into account in the definition of the dimensionless numbers that include a typical dimension of the particle, such as the Reynolds and Knudsen numbers. We show that using a definition of Knudsen based on the radius of the sphere with equivalent volume, the rarefaction effects on ellipsoidal particles can be correctly described without the need of a shape-related parameter in the Knudsen number, such as the ellipsoid major or minor radius used by Dahneke \cite{dahneke}. We finally derive a predictive model that includes rarefaction effects for the drag and lift coefficients, focusing on the transition and free-molecular regimes. The predictive model is able to capture rarefaction and orientation effects in the transition and free-molecular regimes, and we show that it can be successfully applied to predict rarefaction effects to Knudsen numbers that were not included in the fitting process. Moreover, the performances of the model proposed in this work outperform the ESA and the ASA models proposed by Dahneke \cite{dahneke3} in the prediction of the drag coefficients for the particles under investigation.\\
Our results can be used to improve existing Euler-Lagrangian simulations of particle transport in rarefied conditions, as it would allow to model ellipsoidal particles and to include orientation effects in the dynamics of the simulated particles.\\
The paper is structured as follows: in Section \ref{sec:num_meth} we present a detailed analysis and validation of the proposed numerical scheme, showing its capability to recover the drag force exerted by a uniform ambient flow on a spherical particle as well as the accuracy scaling with respect to the spatial and kinetic resolutions of the simulations. In Section \ref{sec:kn} we introduce and discuss the definition for the Knudsen number for ellipsoidal particles based on the equivalent sphere. In Section \ref{sec:drag} we propose the predictive model for the drag and lift coefficients of a prolate and oblate ellipsoidal particle. We summarize and discuss our results in Section \ref{sec:conclusions}.

\section{Numerical Method}
\label{sec:num_meth}
\subsection{The Direct Simulation Monte Carlo (DSMC) in a nutshell}

We approach the solution of the Boltzmann equation using the DSMC method featuring the No-Time Counter (NTC) collision scheme, as firstly proposed by Bird \cite{bird}. In this approach, the real gas molecules are approximated by a finite set of $N$ model particles denoted by their positions, $\mathbf{x}_i$ and velocities, $\mathbf{c}_i$, that move and collide in a physical space domain. Binary intermolecular collisions and interactions with solid boundaries are then modeled through a stochastic approach.\\
 The time evolution for the computational molecules is split in two separate parts: a streaming step and a collision step. During the streaming step, the position of the molecules is updated ballistically and, during the update, the boundary conditions are taken into account.  To address the intermolecular collisions, the physical simulation domain is divided into a computational grid and the DSMC molecules within the same grid cell undergo stochastic binary collisions with a probability given by:
\begin{align}
P = F_N \sigma_T c_r \Delta t / V_c,
\label{eq:prob}
\end{align}
where $F_N$ is the kinetic resolution, representing the number of real molecules represented by a single computational molecule, $\sigma_T$ is the collision cross-section, $c_r$ is the relative velocity between the colliding molecules, $\Delta t$ is the computational time step and $V_c$ is the volume of the DSMC grid cell. The total number of intermolecular collisions that are imposed in every DSMC cell can be defined \cite{garcia} as:\\
\begin{align}
M_{coll} =\frac{N_c(N_c-1) F_N \sigma_T c_r^{max} \Delta t}{2V_c},
\label{eq:tot_col}
\end{align}
where $N_c$ is the total number of DSMC molecules contained in the DSMC cell and $c_r^{max}$ the maximum relative velocity in between molecules pairs within inside the cells. The collision pair are then selected through an acceptance-rejection algorithm based on their relative velocity.\\
In this work we describe the intermolecular interactions using the Variable Hard-Sphere (VHS) model, in which, following Bird \cite{bird},  $\sigma_T$ is given by: 
\begin{align}
\sigma_{T,VHS} = \pi d^2\frac{\qquad \left(\frac{2k_B T}{mc_r^2}\right)^{\omega-0.5}}{\Gamma(2.5 - \omega)},
\end{align}
where $d$ is the molecular diameter at the reference temperature $T$, $k_B$ is the Boltzmann constant, $m$ is the molecular mass, $\Gamma$ is the gamma-function and $\omega$ is the viscosity coefficient used to recover the correct scaling of the viscosity, with respect to the temperature, in the VHS model. For argon gas the reference quantities are \cite{bird} $T= 273.15$K, $d = 4.17\cdot 10^{-10}$m, $m =  6.63\cdot 10^{-26}$kg and $\omega = 0.81$.\\
To ensure the accuracy of the DSMC simulations, it is necessary that the spatial and temporal resolutions are small enough with respect to the characteristic kinetic scales. These conditions are typically obtained by enforcing the following rule-of-thumb relations \cite{bird,garcia} on the cell size $L_c$ and the simulation time step $\Delta t$:
\begin{align}
L_c \leq 0.3 \lambda,
\label{eq:cellsize}
\end{align}
\begin{align}
\Delta t \leq 0.2 \frac{L_c}{\bar{c}_m + U_0},
\label{eq:dt}
\end{align}
where $\lambda$ is the mean free path of the simulated gas molecules and $\bar{c}_m = \sqrt{8k_BT/\pi m}$ is the mean thermal velocity for a gas molecule with temperature $T$ and mass $m$ and $U_0$ is the ambient flow velocity.\\
In the rest of this work we will define $\lambda$ according to Phillips \cite{phillips} as:
\begin{align}
\lambda = \frac{2\mu}{\bar{c}_m \rho},
\label{eq:lambda_mfp}
\end{align}
where $\mu$ is the dynamic viscosity of the gas. In the VHS model $\mu$ is defined as:
\begin{align}
\mu_{VHS} = \mu(T/T_{ref})^\omega,
\label{eq:mu}
\end{align}
where for argon gas $\mu =2.12\cdot 10^{-5}\mbox{ kg m}^{-1} \mbox{s}^{-1}$ at the reference temperature $T_{ref} = 273.15\mbox{K}$ and $\omega$ as before.\\
\subsection{Fluid-solid interactions and the cut-cell method}
\label{subsec:validation}
\begin{figure}[h!]
\begin{center}
\tdplotsetmaincoords{60}{110}

\pgfmathsetmacro{\rvec}{.7}
\pgfmathsetmacro{\thetavec}{-160}
\pgfmathsetmacro{\phivec}{80}

\begin{tikzpicture}[scale=5,tdplot_main_coords]

\shadedraw[tdplot_screen_coords,ball color = white] (0,0) circle (\rvec);

\coordinate (O)   at (0,0,0);

\tdplotsetcoord{P}{\rvec}{90}{-20}
\tdplotsetcoord{B}{1.5}{90}{-10}
\tdplotsetcoord{C}{1.1}{90}{-90}

\tdplotsetcoord{N}{1.3}{90}{-20}
\tdplotsetcoord{T}{0.8}{90}{-90}

\tdplotsetcoord{F}{0.5}{90}{150}

\shadedraw[tdplot_screen_coords,ball color = red] (Bxy) circle (0.06)  node[anchor=north]{};


\draw[-stealth,thick, line width = 0.8mm] (0,0,0) -- (0.5,0,0) node[anchor=north east]{\Large{$y'$}};
\draw[-stealth,thick, line width = 0.8mm] (0,0,0) -- (0,0.5,0) node[anchor=north west]{\Large{$x'$}};
\draw[-stealth,thick, line width = 0.8mm] (0,0,0) -- (0,0,0.5) node[anchor=south]{\Large{$z'$}};

\fill[tdplot_screen_coords,blue,thick] (Pxy) circle (0.04);

 \ExtractCoordinate{Pxy}

\draw[-stealth,color=black, line width = 0.6mm,shorten >= 2mm] (Bxy) -- (Pxy) node[midway,above] {};

\draw[-stealth,color=black, line width = 0.6mm] (Pxy) -- (Cxy) node[midway,above] {};
\draw[tdplot_screen_coords,red,dashed,thick] (Cxy) circle (0.06);

\draw[-stealth,color=black, line width = 0.5mm,dashed,purple] (Pxy) -- (Nxy) node[anchor=east]{\textcolor{purple}{\Large{$\hat{n}$}}};
\draw[-stealth,color=black, line width = 0.5mm,dashed,purple] (Pxy) -- (Txy) node[anchor=south]{\textcolor{purple}{\Large{$\hat{t}$}}};

\draw[-stealth,color=blue, line width = 0.8mm] (O) -- (Fxy) node[midway,above] {};

\node (x) at (Pxy) {\textcolor{red}{\Large{$\mathbf{\times}$}}};
\node (a) at (1.6,-0.1,0) {\Large{$\mathbf{p}_i$}};
\node (a) at (0,-1,0.15) {\Large{$\mathbf{p}'_i$}};
\node (c) at (Fxy |- 0,-0.1,0.3) {\Large{$\mathbf{F}_i = \frac{\mathbf{p}_i - \mathbf{p}'_i}{\Delta t}$}};


\coordinate (Ti)   at (0,0,0.15);
\coordinate (Tm)   at (0.12,-0.12,0);

\coordinate (Tf)   at (0,0,-0.215);
\draw [red, line width = 0.8mm] (Ti.west) to [out=180,in=90] (Tm.center);
\draw [-stealth,red, line width = 0.8mm] (Tm.center) to [out=-90,in=180] (Tf.west);
\node (c) at (0.6,0.5,0) {\makecell{\Large{$\mathbf{T}_i =$}$ \left(\mathbf{x}_i - \mathbf{X} \right) \times$ \\ \Large{$\ \  \frac{ \mathbf{p}_i - \mathbf{p}'_i}{\Delta t}$}}};




\tdplotsetthetaplanecoords{\phivec}



\end{tikzpicture}
\end{center}
\caption{\small{Sketch of the gas-solid interaction scheme. An impinging gas molecule $i$ (red sphere) with initial momentum $\mathbf{p}_i$  hits the surface of the solid particle (gray sphere) and undergoes a diffusive reflection with post-collisional momentum $\mathbf{p}'_i$. The exact collision point (red cross) on the surface of the particle is obtained through a ray-sphere intersection algorithm which allows to detect the exact intersection between the DSMC particles trajectories and the solid surface, described by its analytical expression. The interaction procedure is the following: firstly the system reference frame is transformed into the reference centered on the particle center ($x'y'z'$), then the DSMC molecules are advanced during the streaming step and the collision points with the solid surface are evaluated. To apply the diffusive reflection, the normal and tangent unit vectors ($\hat{n},\hat{t}$), with respect to the solid surface, are obtained for each collision point and used to compute the final position and velocity of the reflected particles. Finally, the coordinates and the velocities of the DSMC particles are transformed back in the system reference frame. Each reflected molecule exerts a force on the solid particle (blue arrow) given by $\mathbf{F}_i = (\mathbf{p}_i - \mathbf{p}'_i)/\Delta t$ and a torque (red arrow) $\mathbf{T}_i =  \left(\mathbf{x}_i - \mathbf{X} \right) \times  \frac{ \mathbf{p}_i - \mathbf{p}'_i}{\Delta t}$, where $\mathbf{x}_i$ and $ \mathbf{X}$ are the collision point on the surface of the particle and its center of mass, respectively.}}
\label{fig:coll_sketch}
\end{figure}
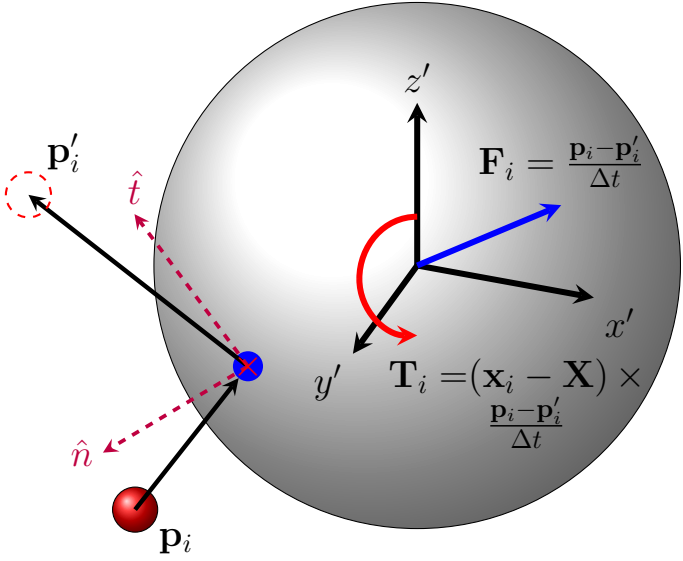

\begin{figure}
\centering
\includegraphics[width=0.45\textwidth]{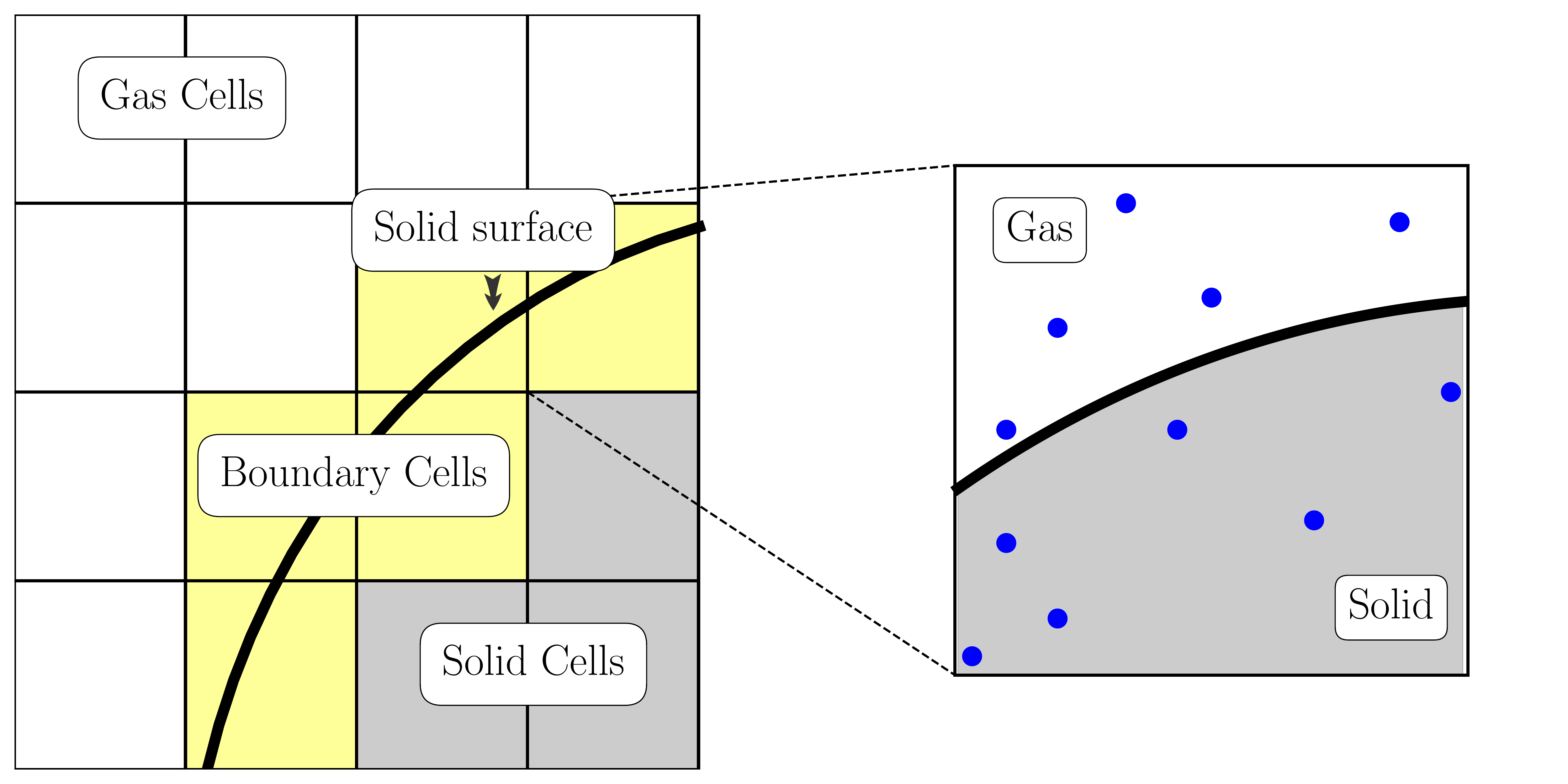}
\caption{\small{Sketch of the decomposition of the DSMC computational grid in cells occupied only by the gas (white), cells completely occupied by the solid particle (gray) and boundary cells (yellow) partially occupied by both the gas and the solid. These boundary cells are cut by the surface of the simulated particle (black curved line) and the volume fraction occupied by the gas must be calculated in order to obtain the correct intermolecular collision probability. In this work we use a Monte Carlo approach to evaluate the gas and solid volume fractions of the boundary cells: a set of random points $N_t$ (blue dots) is generated in each boundary cell and the final solid volume fraction of the cell is given by the ratio between the points belonging to the solid region and the total number of points generated in the boundary cell, as defined in Eq. (\ref{eq:montecarlo}).}}
\label{fig:cells_sketch}
\end{figure}

\begin{figure}[h!]
\centering
\includegraphics[width=0.4\textwidth]{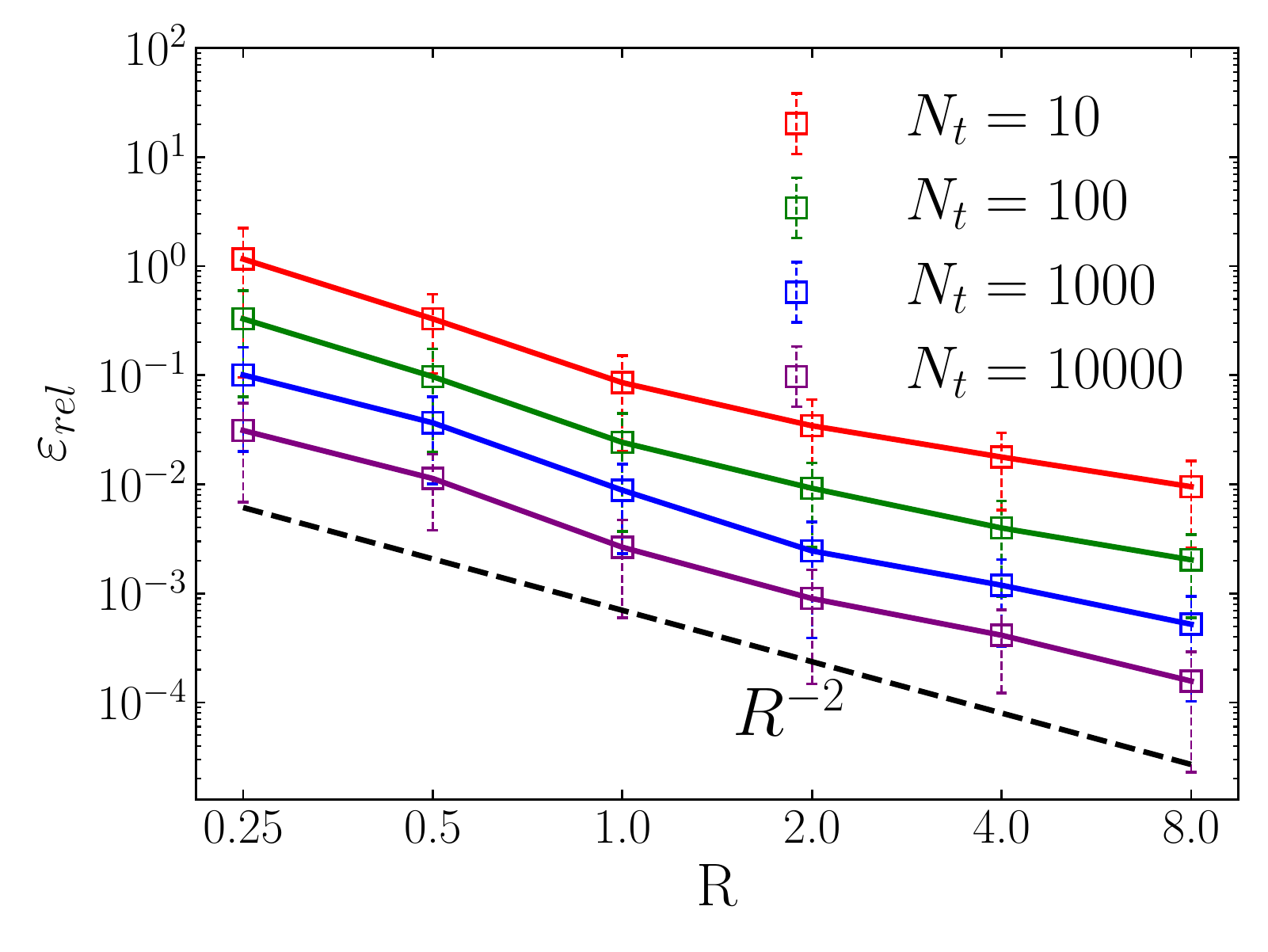}
\caption{\small{Relative error, as defined in Eq. (\ref{eq:rel_er}), in the evaluation of the boundary cell solid volume using a Monte Carlo approach as a function of the radius of the particle, $R$ (cell units), for different values of the Monte Carlo trials $N_{t}$. The solid volume fraction of the boundary cells is reproduced with an accuracy higher than $90\%$ for $N_{t}\geq 1000$ also in cases where the solid particles are very small when compared with the DSMC grid. The scaling of the error with respect to the numerical resolution of the particle is second-order. The error bars are calculated as the standard deviation calculated on a sample of $100$ independent measurements.} }
\label{fig:volume_sphere}
\end{figure}

\begin{figure}
\centering
\includegraphics[width=0.45\textwidth]{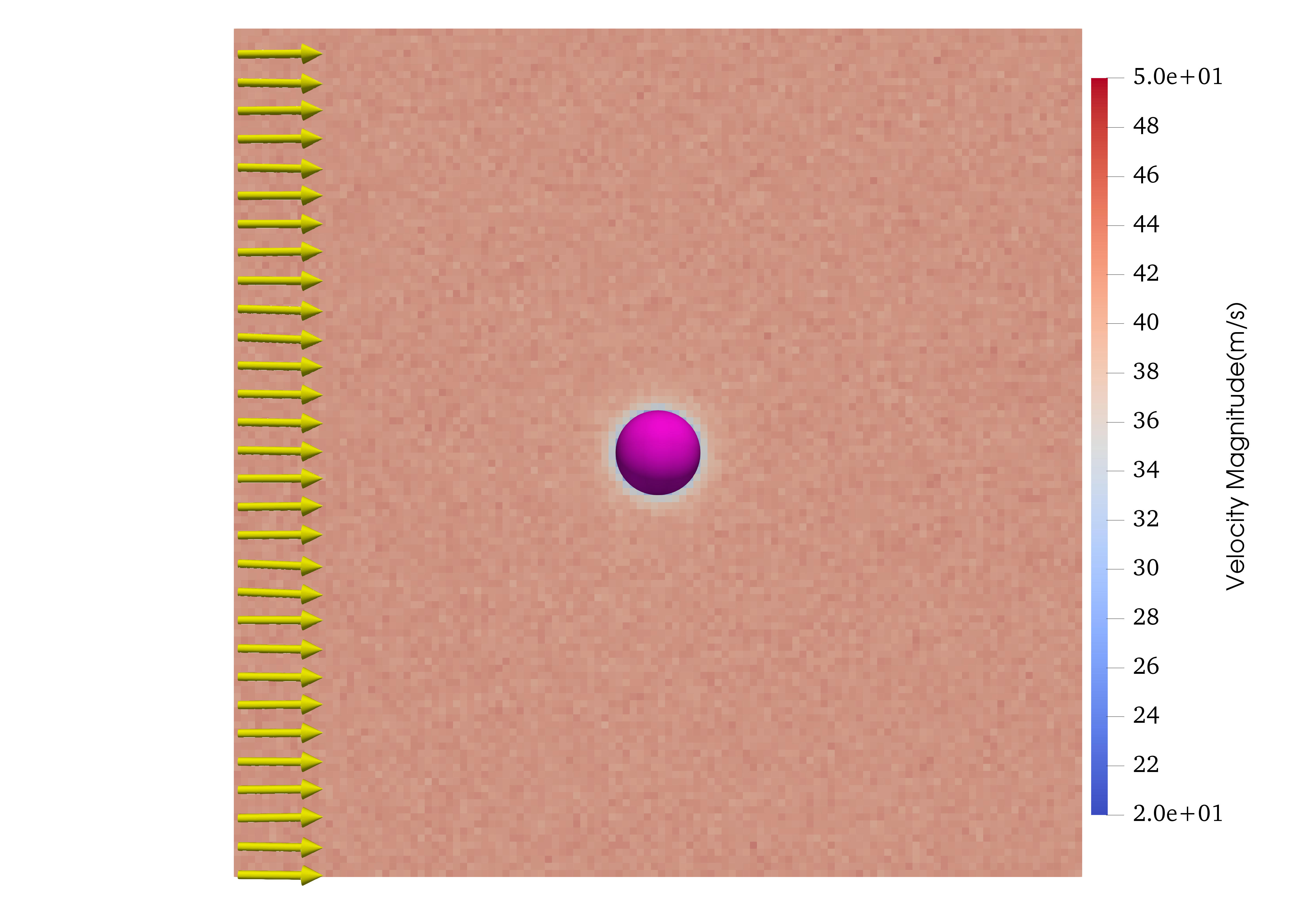}
\caption{Snapshot of the velocity field around a particle from a DSMC simulation. The plot represents a cut on the $xy$ plane, crossing the particle center, of a spherical particle (colored in magenta) with radius $R=0.25\mu \mbox{m}$ and particle-based Knudsen number of $Kn=10$, immersed in an argon gas flow with free stream velocity $\mathbf{U} = 43.8\hat{x}\mbox{ m/s}$. The stream velocity direction is indicated by the yellow arrows. The simulation domain size is set to $L=5\mu \mbox{m}$, so that $L=20\cdot R$.}
\label{fig:sketch_sphere}
\end{figure}

\begin{figure}[h!]
\centering
\includegraphics[width=0.40\textwidth]{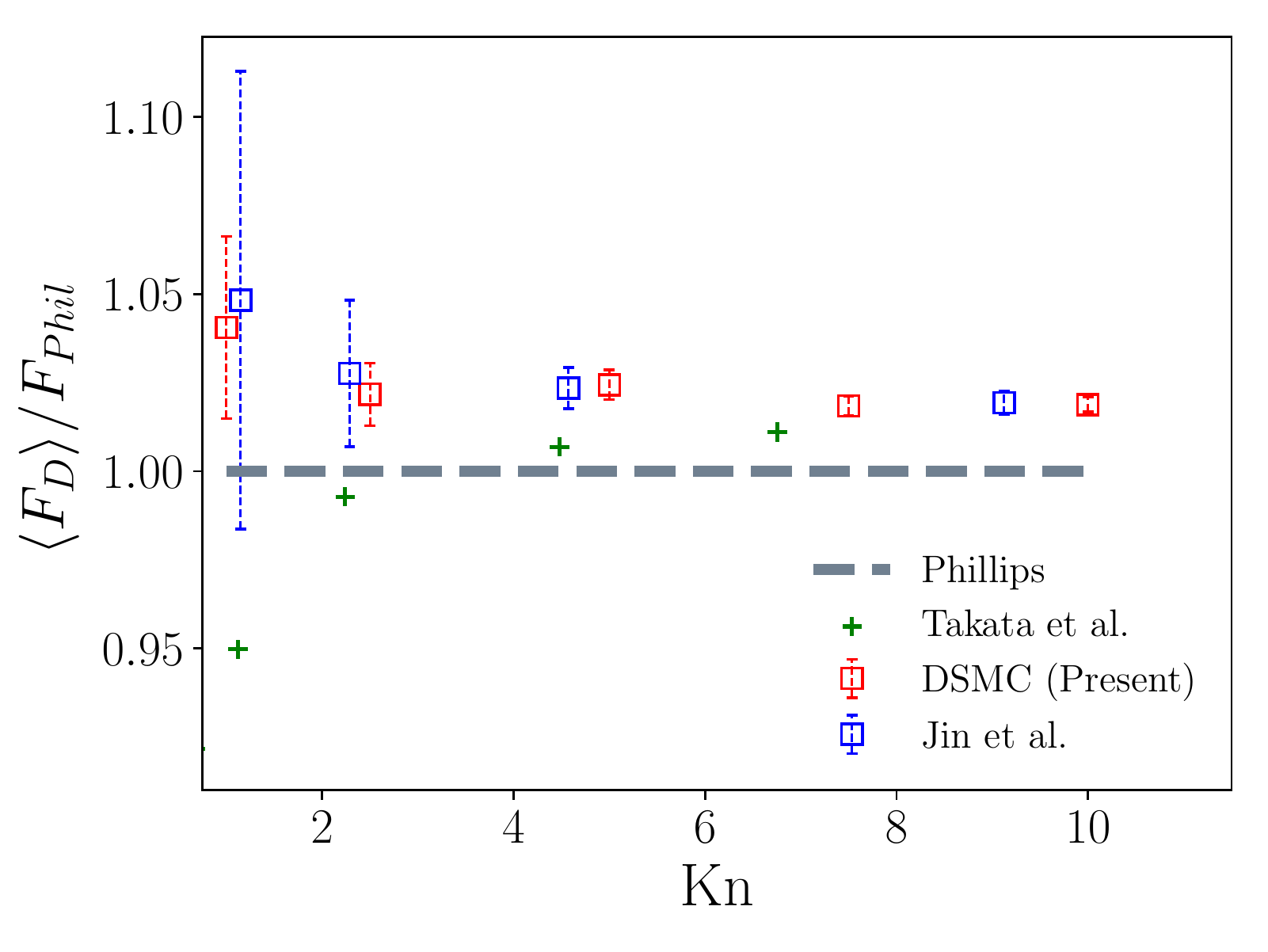}\\
\includegraphics[width=0.40\textwidth]{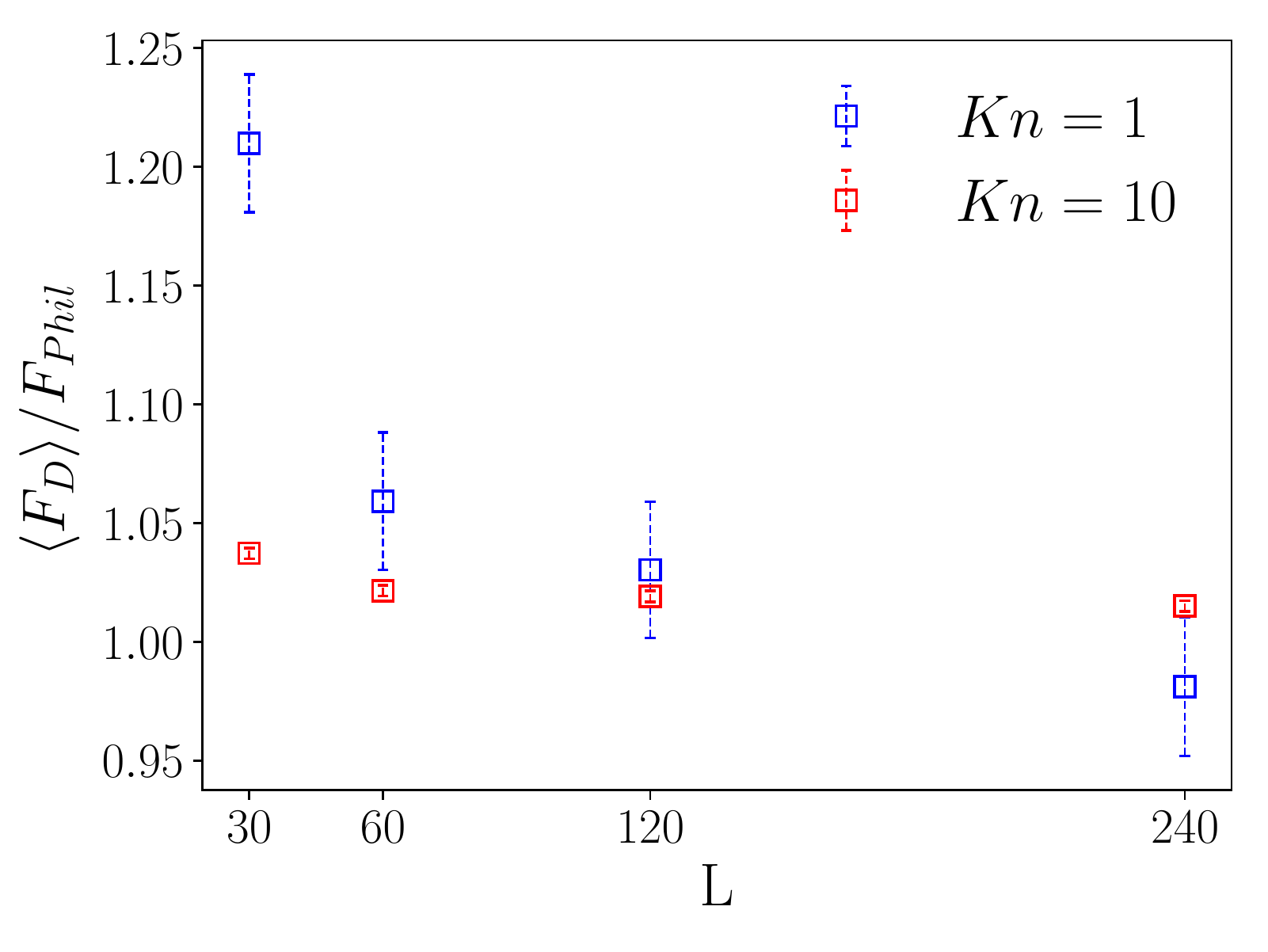}
\caption{\small{In the top plot, the average drag force, $\langle F_D\rangle$, on a spherical particle immersed in a uniform gas flow is plotted for different values of the Knudsen number. The results from our DSMC simulations (red  squares) are compared with the DSMC results from Jin \textit{et al.}\cite{jin} (blue squares), with the analytical approximation from Phillips \cite{phillips} (blue dashed line) and from Takata \textit{et al.} \cite{takata} (green crosses), both based on the direct solution of the Boltzmann equation. Our DSMC data is averaged over $10000$ time steps after an initial transient of $5000$ time steps. In the bottom plot the effects of varying the simulation box size $L$, leaving all other parameters unchanged, are reported for $Kn=10$ (red squares) and $Kn=1$ (blue squares). It is evident that for sufficiently large simulation box sizes ($L\geq 120$, corresponding to $L\geq 20\cdot R$), the detrimental effects from the finite size of the simulation grid are small, with a deviation of about $2-3\%$ for $L=120$ and of about $1-2\%$ using $L=240$, for all the $Kn$ numbers. For practical reasons we will use $L=120$ for all the simulations presented in this work. In both plots, the error bars are computed using the $95\%$ confidence interval from Eq. (\ref{eq:errorbars}). }}
\label{fig:drag_valid}
\end{figure}

In this work we present an algorithm based on the cut-cell method to describe the 2-way coupling between the gas flow and a spheroidal solid particle. The surface of the particle immersed in the gas domain is described by its analytical expression, and the momentum exchange between the gas and the solid particle is computed from the microscopic interactions between the simulated gas molecules and the solid surface. In this way we overcome the limitations of the alternative method used to evaluate the force and the torque on the particle based on the macroscopic stress tensor, which is often less accurate due to the statistical fluctuations of the higher order macroscopic fields around the particle.\\
The collision points at which the DSMC molecules impinge on the surface of the solid particle are evaluated exactly using a ray-sphere intersection algorithm \cite{rendering}, extended to include ellipsoidal particles. In very few words, the algorithm consists into applying a transformation of the space coordinates that allows to describe the ellipsoidal particle as a sphere with unit radius, whose center coincides with the origin of a new translated reference frame. The trajectories of the DSMC particles are recomputed in the transformed reference frame and the collision points are obtained analytically through the evaluation of the intersections between the new trajectories (lines) and the scaled sphere. The collision points coordinates in the original reference frame are obtained by applying the inverse transformation. A purely diffusive reflection scheme is then applied to reflect the impinging molecules, as shown in Fig. \ref{fig:coll_sketch}.\\
For each DSMC molecule $i$ hitting the surface of the solid particle at position $\mathbf{x}_i$, with initial momentum $\mathbf{p}_i$ and post-collision momentum $\mathbf{p}'_i$, the total momentum transferred from the gas to the solid particle within a single time step $\Delta t$ is:
\begin{align}
\Delta \mathbf{p} = \sum_i  \left(\mathbf{p}_i - \mathbf{p}'_i\right),
\end{align}
from which the total force $\mathcal{\mathbf{F}}$ and torque $\mathcal{\mathbf{T}}$, exerted on the rigid body, can be directly obtained:
\begin{align}
\mathcal{\mathbf{F}} =  \sum_i \left( \frac{ \mathbf{p}_i - \mathbf{p}'_i}{\Delta t}\right),
\label{eq:force}
\end{align}
\begin{align}
\mathcal{\mathbf{T}} =  \sum_i \left[ \left(\mathbf{x}_i - \mathbf{X} \right) \times  \frac{ \left(\mathbf{p}_i - \mathbf{p}'_i \right)}{\Delta t}\right],
\label{eq:torque}
\end{align}
where $\mathbf{X}$ denotes the center of mass of the solid particle. While in this work we will focus on particles that are fixed in space, Eqs. (\ref{eq:force}) and (\ref{eq:torque}) can be used to update the solid particle translational and angular velocities, position and orientation.\\
The simulation grid is divided in three regions: gas cells completely filled with gas molecules, solid cells that are completely occupied by the solid particle and boundary cells (cut cells) that are partially covered by the solid particle and partially filled with gas, as sketched in Fig. \ref{fig:cells_sketch}. In order to correctly evaluate the collision probability given by Eq. (\ref{eq:prob}) within the boundary cells, the local cell volume with gas has to be calculated. We perform this update by marking the cells close to the solid particle, so that only the marked cells are candidate for being boundary cells. The gas volume of the boundary cells is then evaluated through a Monte Carlo approach: a set of $N_t$ random points is generated in the DSMC boundary cell and the gas fraction volume, $V_{g}$, is obtained as:
\begin{align}
V_{g} = V_c - V_s = \frac{N_t - N_s}{N_t}V_c,
\label{eq:montecarlo}
\end{align}
where  $N_{s}$ represents the number of points that are generated inside the solid volume and $V_{s}$ is the volume fraction of a DSMC cell occupied by the solid volume. The relative error at which the solid volume fraction of the boundary cells is computed can be expressed as \cite{jin}:
\begin{align}
\varepsilon_{rel} = \frac{V_p - \sum_{all\ cells}V_{s}}{V_{p,\ bound.\ cells}},
\label{eq:rel_er}
\end{align}
where $V_{p} $ is the real (analytic) volume of the solid particle and $V_{p,\ bound.\ cells}$ is the real volume of the solid fraction of all boundary cells. The numerator represents the difference between the real volume of the particle and the computational volume as calculated from the Monte Carlo approach, which ultimately describes the difference in volume in the boundary cells. The denominator represents the real volume of only the boundary cells, and this can be obtained by subtracting to $V_p$ the volume of the DSMC cells completely occupied by the solid particle. In this way we can define an estimator of the accuracy at which the total volume of the boundary cells is computed.\\
The scaling of the relative error in Eq. (\ref{eq:rel_er}), with respect to different resolutions of a spherical particle with radius $R$ (in cell units) is plotted in Fig. \ref{fig:volume_sphere} for different values of the Monte Carlo trials $N_{t}$. It is shown that using a sufficiently large number of Monte Carlo trials, the volume of the boundary cells is recovered with an accuracy of at least $\sim 95\%$  also for particles with a radius that is only a fraction of the simulation grid size. Since in this work we focus on particles that are fixed in space, the gas volume fraction of the boundary cells can in principle be computed analytically. We prefer, however, to keep our approach general using the Monte Carlo approach, as it can be directly applied to different grids, particle shapes and moving objects. Moreover, using fixed particles, the volume fraction evaluation of the boundary cells needs to be performed only once and we set $N_{t} = 100000$ to ensure a very high precision of the computation. It is worth mentioning that in cases where the particle is allowed to move, the volume computation must be performed at each time step and a lower number of $N_{t}$ would allow a faster computation.\\
The cut-cell algorithm implemented and presented in this study has been incorporated in the parallel DSMC solver validated by Di Staso \cite{distaso_thesis}. The intensive computations required for the DSMC simulations presented in this work, in fact, can become feasible only by taking advantage of parallel computation. This can be easily done for a DSMC algorithm, thanks to the locality of the interactions between gas molecules, by enforcing a three-dimensional Cartesian processor grid on which the DSMC simulation domain is decomposed. The simulations presented in this work are executed on computational nodes with 2 AMD EPYC 7282 CPUs per node, and the individual run wall clock time strongly depends on the Knudsen number, ranging from $20$ hours ($Kn\geq 10$) to several days ($Kn\sim 1$) on one node.\\
To validate the algorithm, we perform simulations of a rarefied argon gas flow impinging on a spherical particle in the same conditions as Jin \textit{et al.} \cite{jin}. The simulation setup, represented in Fig. \ref{fig:sketch_sphere}, is the following: the radius of the particle is fixed at $R=0.25\mu\mbox{m}$ and the gas temperature is set to $T=300\mbox{K}$. The gas density, $\rho$, flow velocity, $U_0$, and pressure, $P$, are varied accordingly to the Knudsen number, defined as $Kn = \lambda/R$, and the (particle-based) Reynolds number, $Re=2U_0R/\nu$, is kept constant and equal to $0.022$ to match with the setup from \cite{jin}. The computational grid and time step are chosen accordingly to Eqs. (\ref{eq:cellsize})-(\ref{eq:dt}), with the additional requirement that the simulation box size $L=20\cdot R = 5\mu \mbox{m}$ to avoid as much as possible detrimental effects due to the vicinity of the particle to the boundaries of the simulation box (an example of such effects is shown in the bottom part of  Fig. \ref{fig:drag_valid}). For practical purposes, and to reduce the computational burden, $120$ DSMC cells per linear direction are used to discretize the domain for all the simulations, leading to a value of $L_c/\lambda$ ranging from $0.017 \leq   L_c/\lambda\leq 0.17 $, which is always within the rules-of-thumb limits. This discretization leads to a particle radius of $R=6$ (cells units). The number of particles-per-cell is set to $N_{c}=50$, leading to roughly $8.6\cdot 10^7$ computational particles. Free-streaming boundary conditions are imposed along the flow direction and periodic boundary conditions are applied along the transverse directions. With this configuration we reach a very high accuracy for all the investigated range of $Kn$ and, to give an example, for $Kn=10$ we have that one computational molecule represents four physical argon atoms. \\
In all the simulations presented in this paper the drag force $F_D$ is averaged over $N_{\Delta t} = 10000$ time steps after an initial transient of $5000$ time steps, which is enough to reach the steady state in all investigated cases. The error bars are calculated using the $95\%$ confidence interval given by:
\begin{align}
\varepsilon_{95} = \frac{2\sigma_{std}}{\sqrt{N_{\Delta t}}},
\label{eq:errorbars}
\end{align}
where $\sigma_{std}$ is the standard deviation on the average value of $F_D$.\\
The validation of the proposed algorithm is presented in Fig. \ref{fig:drag_valid}, where the drag force measured with our DSMC code is compared with the DSMC results from Jin \textit{et al.} \cite{jin} and with the analytical approximations from Phillips \cite{phillips} and Takata \textit{et al.} \cite {takata}, both based on the direct solution of the Boltzmann equation. The results are normalized with respect to the prediction from Phillips, given by:
\begin{equation}
\begin{split}
& F_{Phil.}(R) =  -6\pi \mu R U_0 \times \\ \times & \frac{15 - 3c_1 Kn + c_2(8+\pi \sigma)(c_1^2 +2)Kn^2}{15 +12c_1 Kn + 9 (c_1^2 +1)Kn^2 + 18c_2(c_1^2+2)Kn^3} = \\ = &  -6\pi \mu R U_0\cdot f(Kn),
\end{split}
\label{eq:phil}
\end{equation} 
where $c_1 = \frac{2-\sigma}{\sigma}$, $c_2 = \frac{1}{2-\sigma}$, $\mu$ is the gas dynamic viscosity and $\sigma$ is the momentum accommodation coefficient, with range $0 \leq \sigma \leq 1$. In our simulations $\sigma =1$ (fully-diffusive surface) and thus $c_1=c_2=1$.\\
As it can be seen from Fig. \ref{fig:drag_valid}, simulation results with our DSMC method are well aligned with the results available in the literature obtained with similar approaches (see Jin \textit{et al.} \cite{jin}). The consistent small deviation between the values obtained with DSMC solvers with respect to the approximations from Phillips \cite{phillips} and Takata \textit{et al.} \cite{takata} are related to the limitations of the different numerical approaches used to solve the Boltzmann equation (DSMC for the present work and \cite{jin}, direct solution of the Boltzmann equation using a finite-difference approach for \cite{takata} and method of moments for \cite{phillips}), as well as to some residuals of finite-size effects related to the finite simulation domain (bottom of Fig. \ref{fig:drag_valid}).\\
In the last part of this Section, we want to investigate the accuracy scaling with respect to spatial and kinetic resolution, separately. The former is related to the impact of different sizes of the particles (in cell units), and is embodied by the parameters $L_c/\lambda$ and the particle radius $R$. The latter represents the number of real particles described by a computational particle and is tuned via the number of particles-per-cells, $N_{c}$. To our knowledge an extensive convergence analysis for the cut-cell algorithm is not available in the literature, and it is important to understand the impact of the aforementioned parameters on the accuracy of the DSMC simulations.\\
In the first analysis we compare the relative error on the mean value and the standard deviation of the drag force on a spherical particle at $Kn=10$ for different values of the particle radius, $R$ (in cells units). This is done by fixing the total number of DSMC molecules (in order to maintain the kinetic resolution, i.e., the number of real particles represented by a single computational particle, the same for all the simulations) and by fixing the simulation time step, to ensure that the number of collisions per time step is unchanged between different resolutions. We do so to isolate the effects induced by varying the simulation grid size on the overall simulation accuracy.\\
In this convergence analysis, the ratio between the simulation box size and the particle radius is fixed to $L/R = 20$, so that varying the resolution of the particles is equivalent to vary the value of the DSMC spatial resolution given by $L_c/\lambda$. For this analysis, we focus on the case at $Kn=10$ due to the computational limitations (in terms of grid resolution) in investigating low values of $L_c/\lambda$ for lower $Kn$.\\
From the results of this analysis, shown in Fig. \ref{fig:scaling_r}, we observe that when intermolecular collisions are present, the relative error on the mean value of the drag force exhibits a second-order scaling with respect to the spatial resolution for small values of $R$, while it deviates from the scaling law for larger values of $R$. This deviation is related to the approaching of the DSMC resolution limits in resolving stochastic intermolecular collisions as, in this setup, the number of particles-per-cell, $N_c$, decreases for increasing $R$ as a consequence of keeping the total number of particles unchanged. Once intermolecular collisions are switched off, in fact, the error is drastically reduced and it shows a consistent third-order convergence with respect to spatial resolution. The standard deviation is constant, as we impose that the number of collisions per time step between the DSMC molecules and the solid particle is the same for all the simulations. Interestingly, the algorithm offers a remarkable good accuracy also for cases where the radius of the particle is of the order of the DSMC cell size, showing that this algorithm is effective also in cases where the curvature of the solid particle is small compared to the DSMC spatial grid resolution. \\
\begin{figure}
\centering
\includegraphics[width=0.40\textwidth]{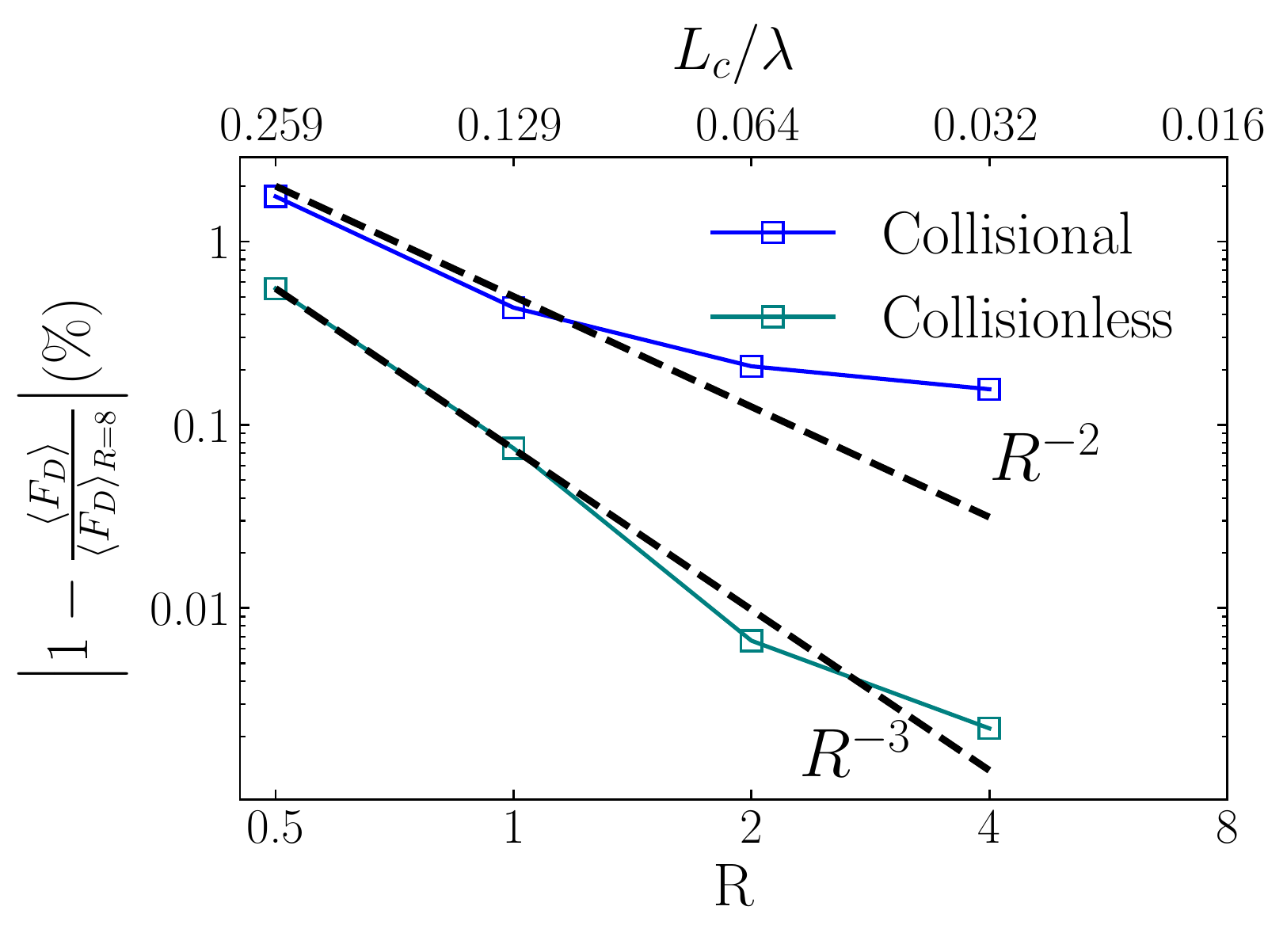}\\
\includegraphics[width=0.40\textwidth]{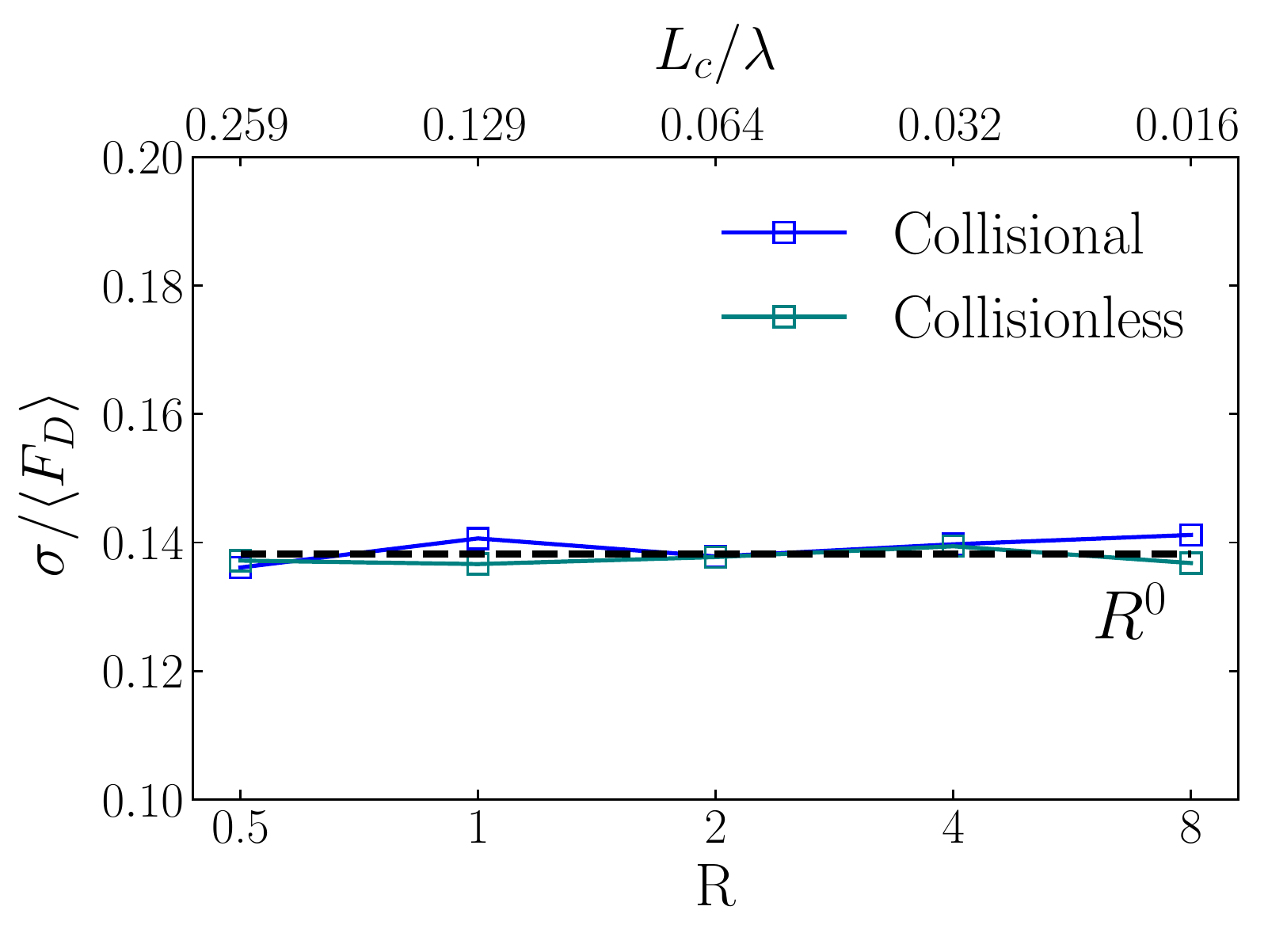}
\caption{\small{Relative error (top) of the mean value of the drag force, $\langle F_D \rangle$, experienced by a spherical particle immersed in a uniform argon gas flow at $Kn=10$, as a function of the spatial resolution of the solid particle $R$, for collisional (blue) and collisionless (green) DSMC simulations. In this analysis the ratio between the simulation box size and the radius of the particle is fixed to $L/R = 20$, the total number of DSMC particles is set to $N_{tot} = 8.6\cdot 10^7$ and the simulation time step is fixed to $dt =1.76\cdot 10^{-11}\mbox{s}$. In this way we ensure that the number of collisions per time step between the DSMC molecules and the solid particle is the same for all the simulations. The relative error is computed with respect to the value of  $\langle F_D \rangle$ measured at $R=8$. The measured convergence of  $\langle F_D \rangle$ is roughly second order for the collisional case, and we observe a deviation from the scaling curve for increasing values of $R$ as we approach the resolution limit of the DSMC method. This is confirmed by collisionless simulations, in which the resolution limitations related to intermolecular collisions are removed and the relative error shows a consistent third-order scaling. The standard deviation (bottom) on  $\langle F_D \rangle$ is constant for all the simulations, as expected.}}
\label{fig:scaling_r}
\end{figure}
We will address now the accuracy scaling with respect to the kinetic resolution. This is done by fixing the particle radius to $R=4$ (and thus $L_c/\lambda=0.032$) and by varying the number of particles-per-cell, $N_{c}$. Results are shown in Fig. \ref{fig:scaling_ppc}, where it can be observed that both the scaling of the mean value of the drag force and of its standard deviation are in agreement with the typical results of a DSMC simulation, with the former scaling linearly with $N_{c}$ and the latter scaling as $N_{c}^{-1/2}$. \\
\begin{figure}
\centering
\includegraphics[width=0.40\textwidth]{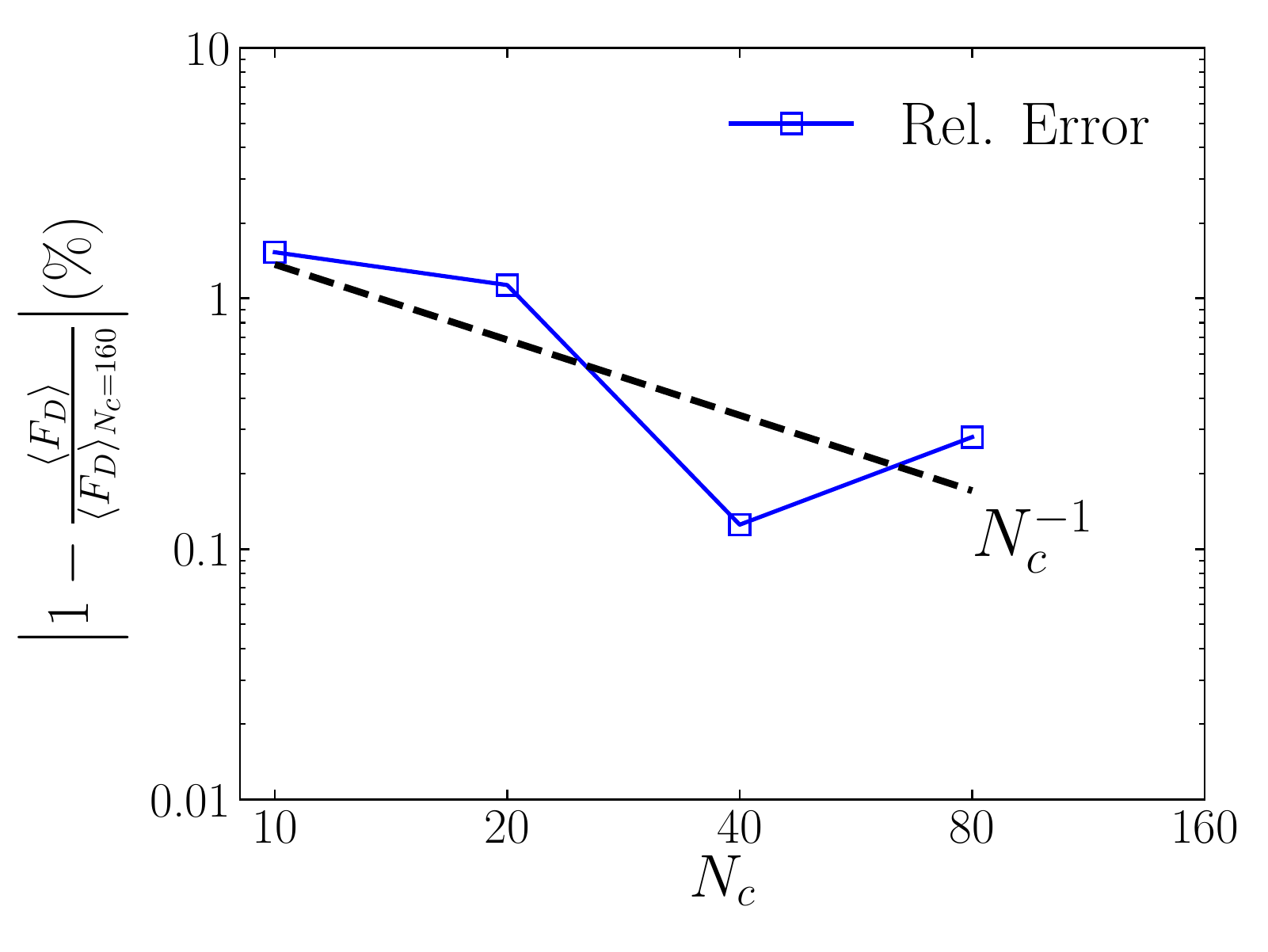}\\
\includegraphics[width=0.40\textwidth]{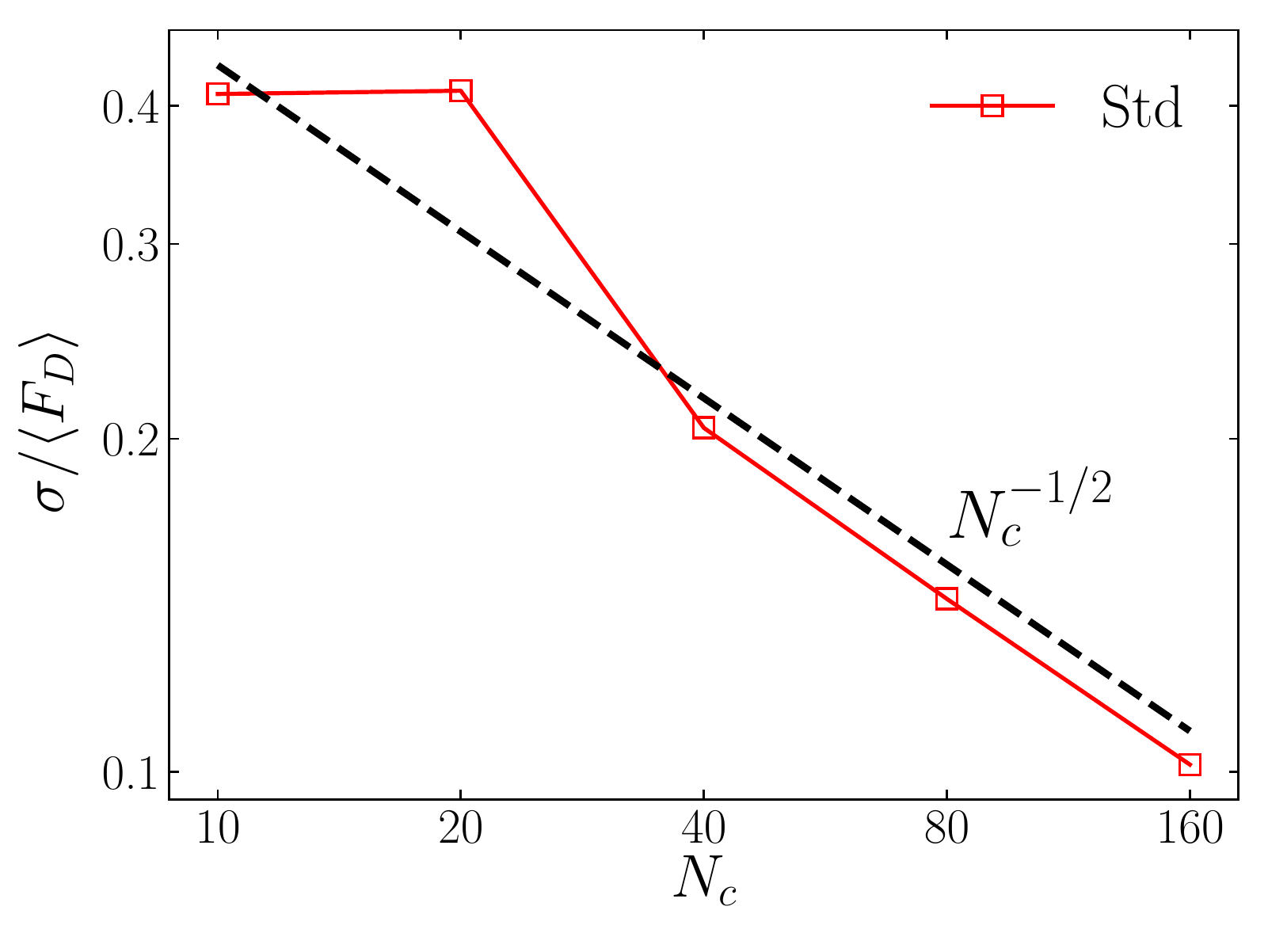}
\caption{\small{Relative error (top) of the mean value of the drag force, $\langle F_D \rangle$, experienced by a spherical particle immersed in a uniform Argon gas flow at $Kn=10$, as a function of the kinetic resolution of the DSMC simulation given by the number of particles-per-cell, $N_c$. In this analysis the ratio between the simulation box size and the radius of the particle is $L/R = 20$ and the radius of the particle is set to $R=4$. The relative error is computed with respect to the value of  $\langle F_D \rangle$ measured at $N_c=160$. The measured convergence of  $\langle F_D \rangle$ is first order with respect to $N_c$, while the convergence rate of the standard deviation (bottom) is $0.5$.}}
\label{fig:scaling_ppc}
\end{figure}
\section{Drag correlations for ellipsoidal particles at finite Knudsen number}
\label{sec:kn}
\begin{figure}
\centering
\includegraphics[width=0.40\textwidth]{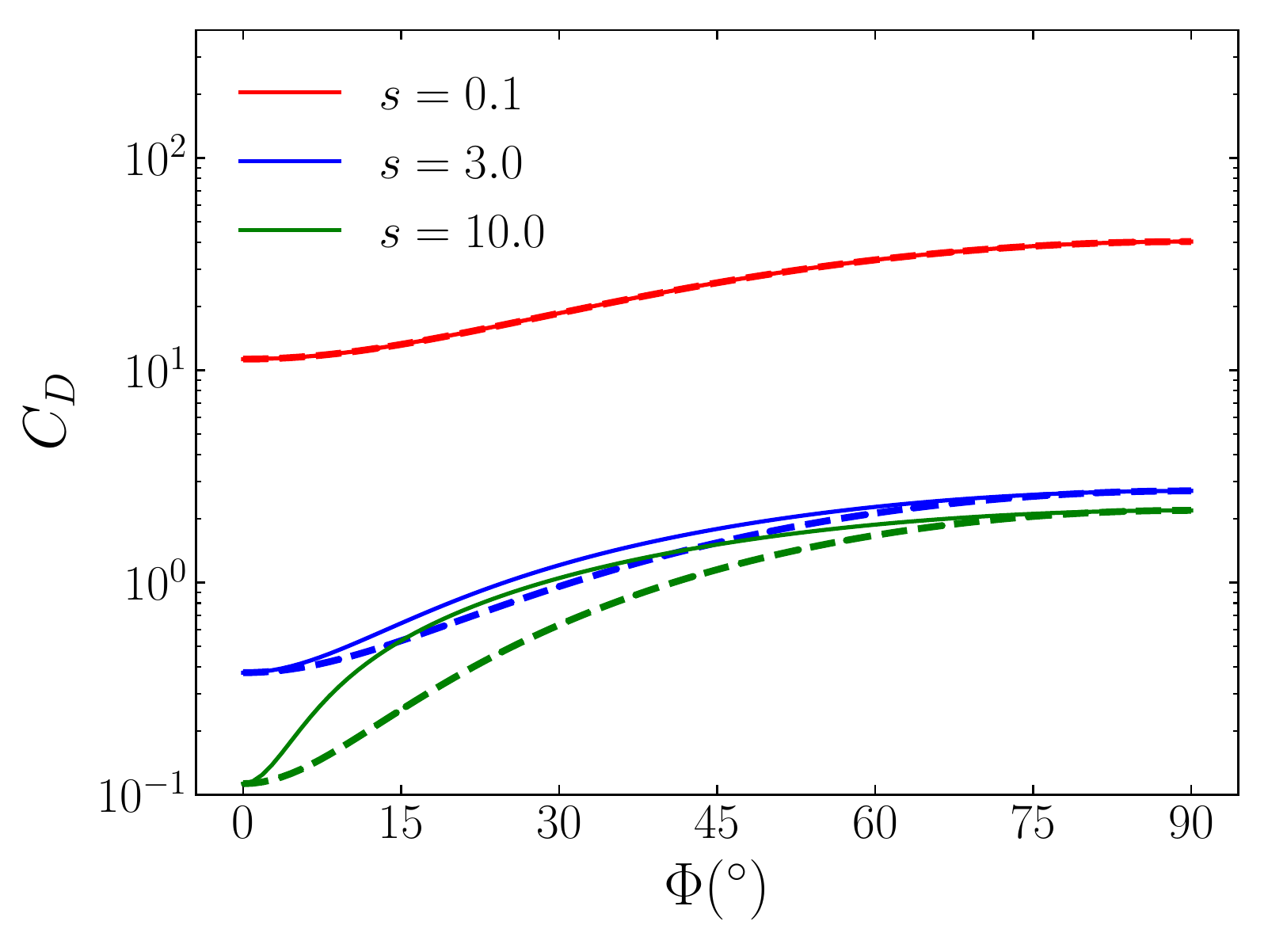}\\
\includegraphics[width=0.40\textwidth]{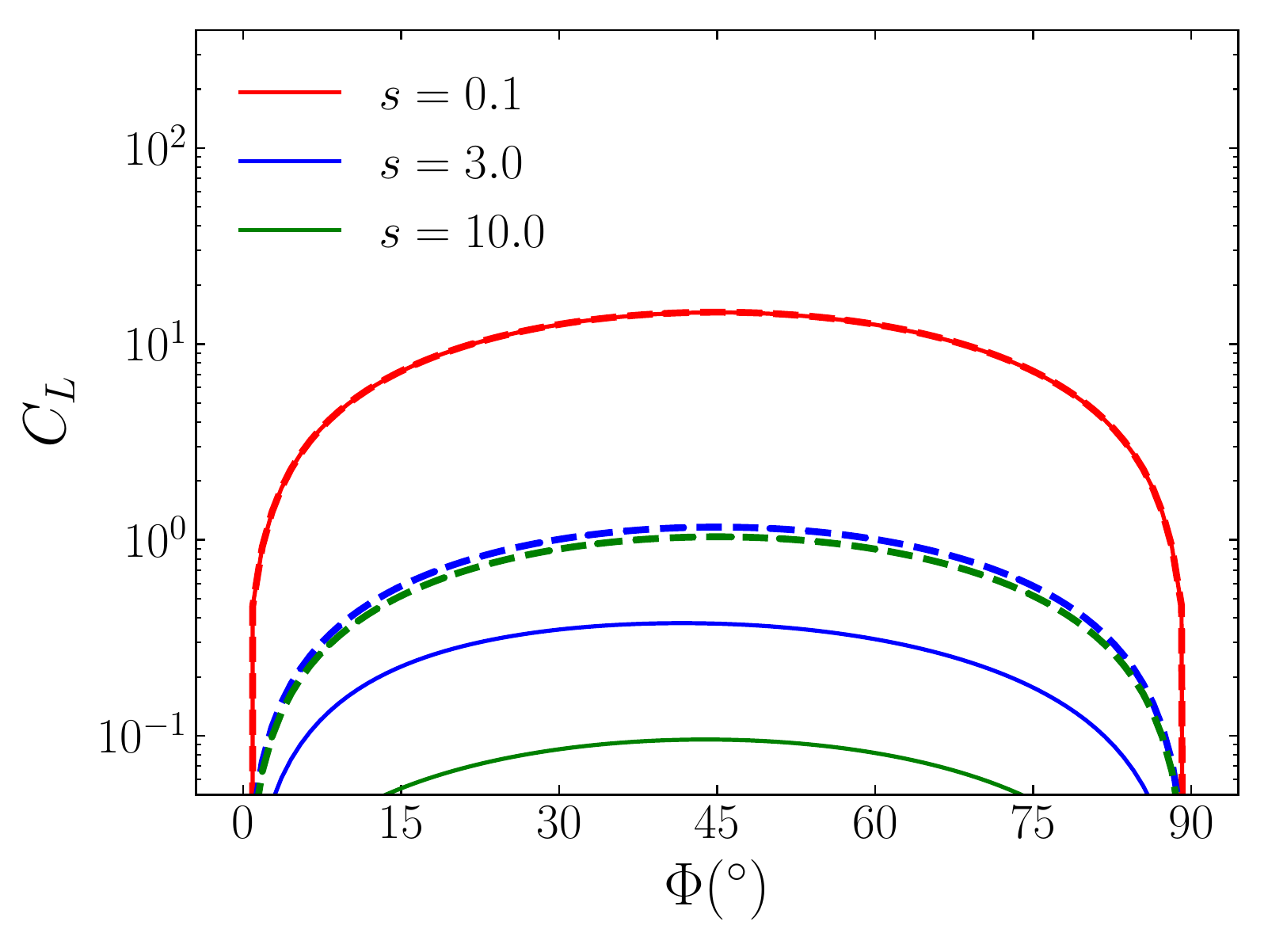}
\caption{\small{Drag (top) and lift (bottom) coefficients (solid lines) as obtained by Bird \cite{bird} for a thin rectangular plate immersed in a collisionless gas flow for different speed ratios $s=U_0 \left(\frac{m}{2k_BT}\right)^{1/2}$ as a function of the angle of attack, $\Phi$. The results are compared with the analytical correlations (dashed lines) proposed by Happel and Brenner \cite{happel} and embodied in  Eqs. (\ref{eq:drag_continuum})-(\ref{eq:lift}). For small velocities the prediction from the continuum regime holds also in the collisionless case, as the solid lines and the dashed lines are perfectly superimposed. This highlights that rarefaction effects do not depend on the relative orientation of the solid body with respect to the incoming flow in the limit of small speed ratios.}}
\label{fig:bird_drag}
\end{figure}

\begin{figure}
\centering
\includegraphics[width=0.25\textwidth]{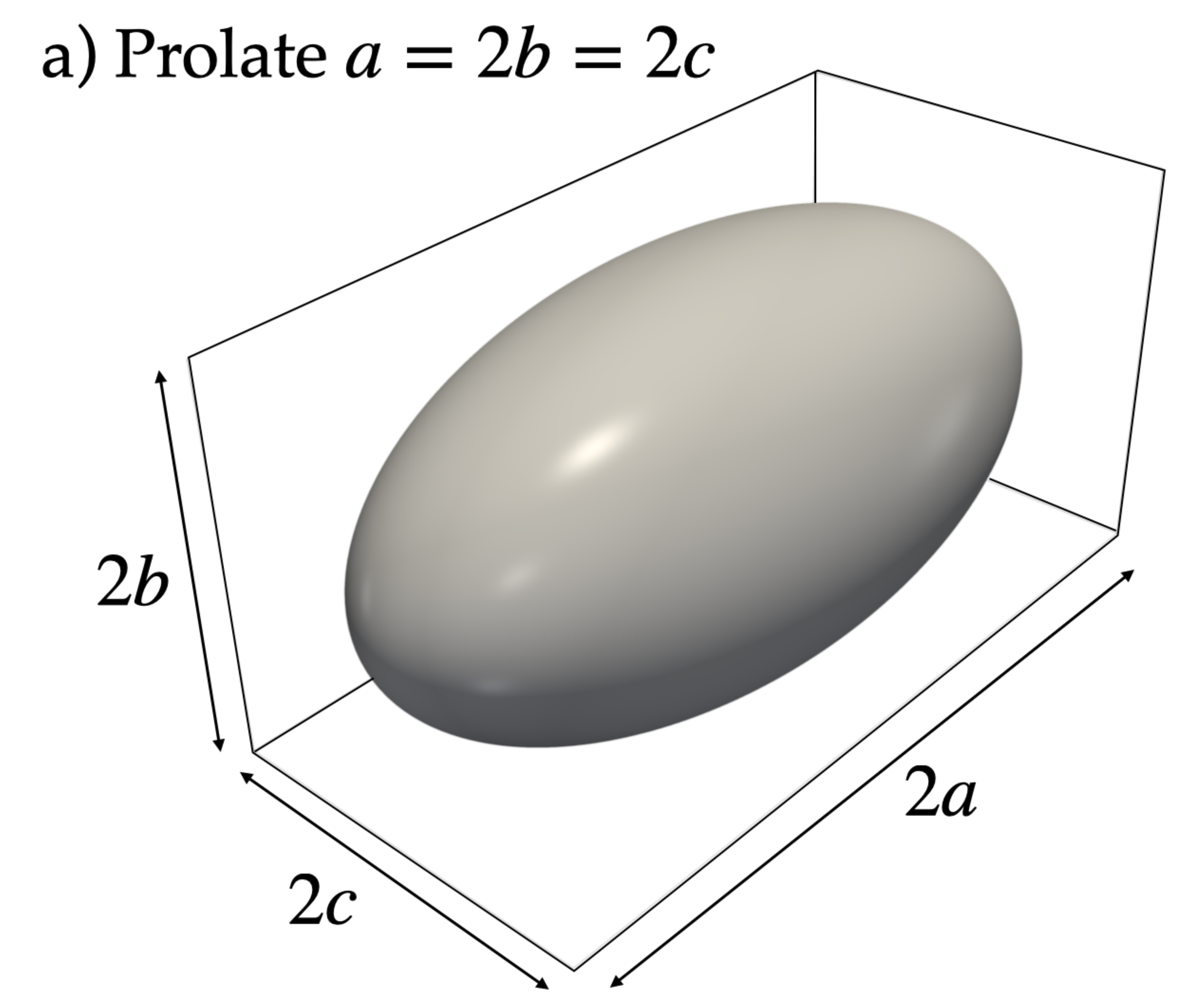}%
\includegraphics[width=0.25\textwidth]{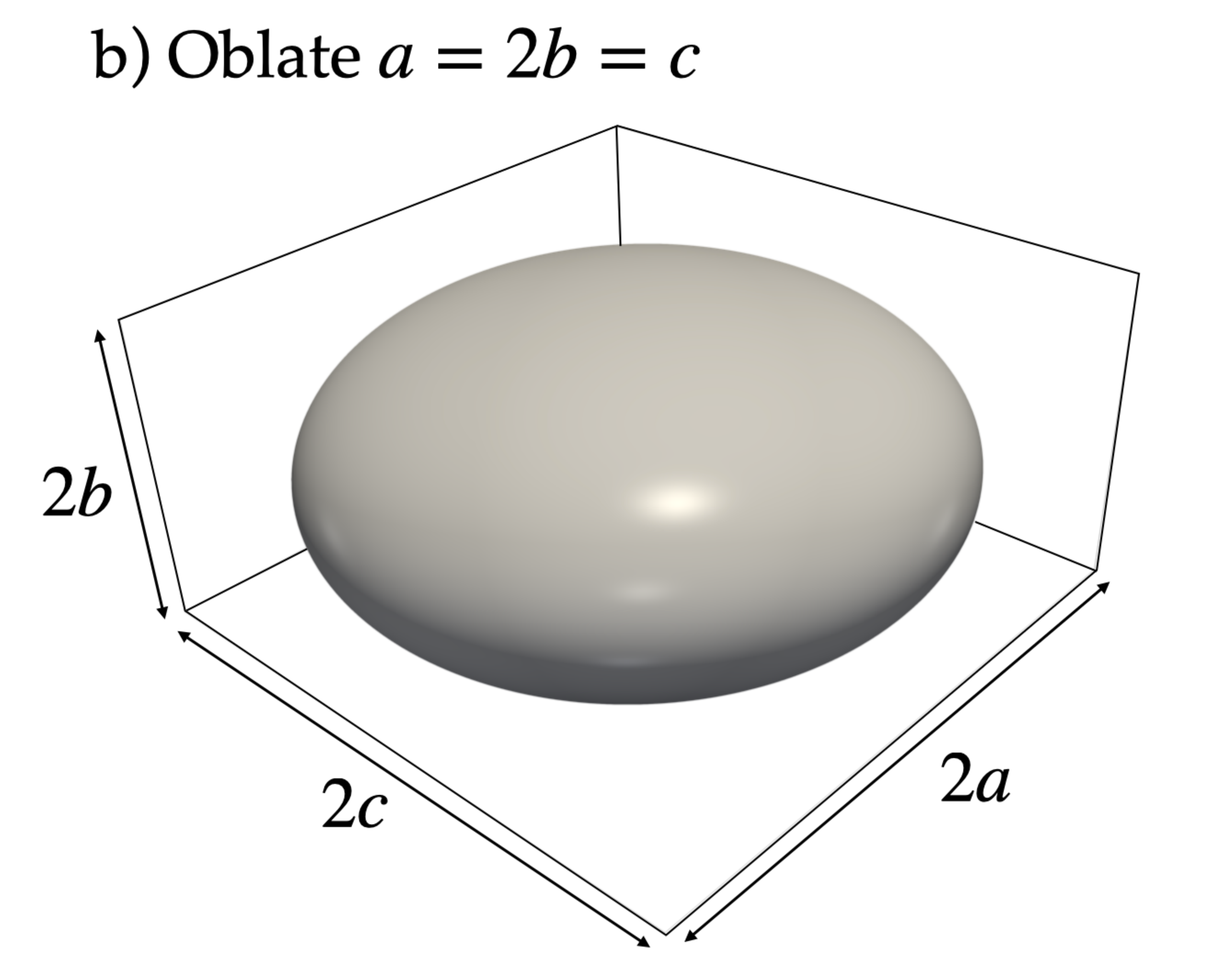}
\caption{\small{Different ellipsoidal particles simulated in this work: a) a prolate ellipsoid with aspect ratio $a/b = a/c = 2$ and b) an oblate ellipsoid with aspect ratio $a/b = 2$ and $ a/c = 1$.}}
\label{fig:part_geom}
\end{figure}

\begin{figure}
\centering
\includegraphics[width=0.45\textwidth]{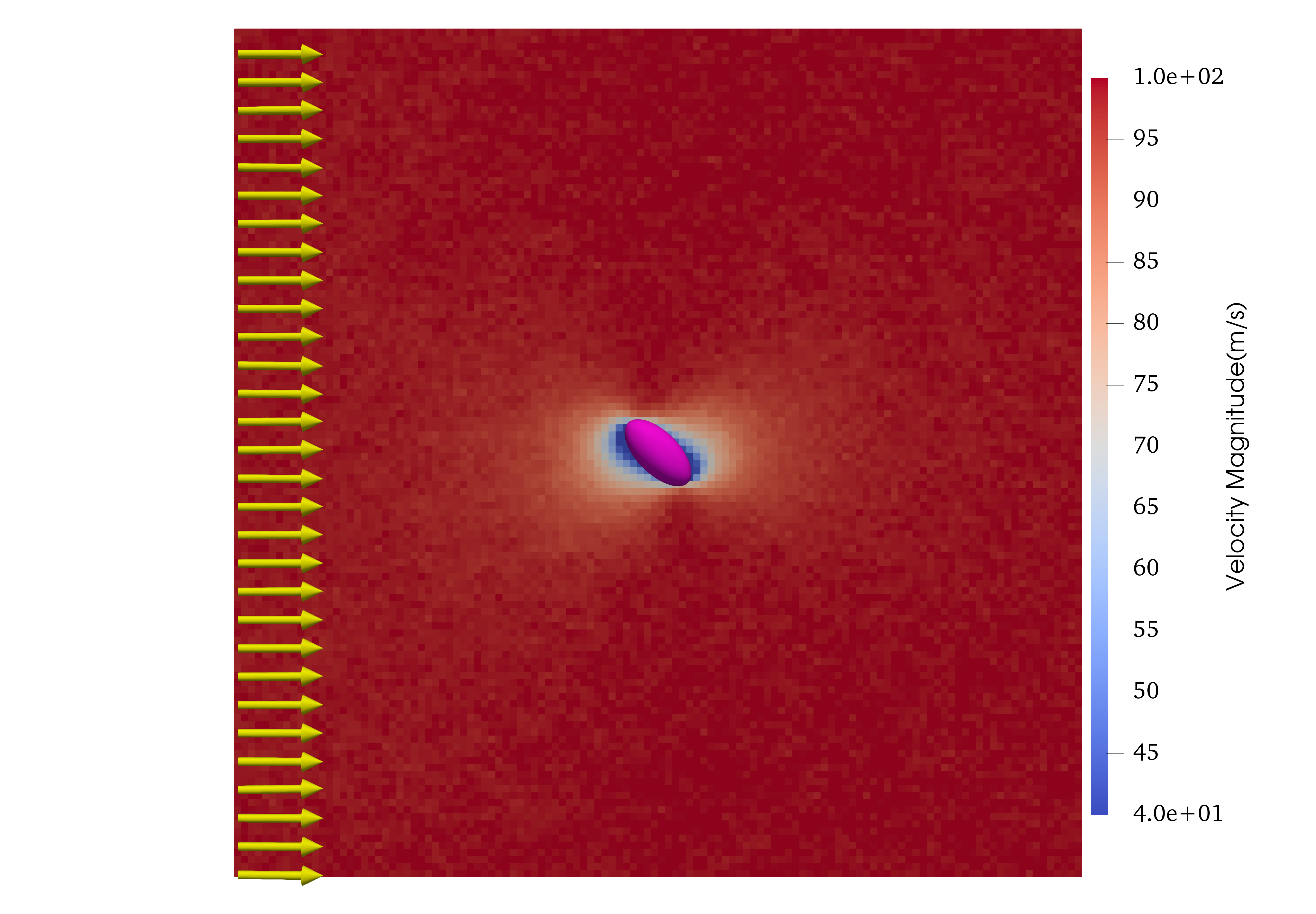}
\caption{\small{Snapshot of the velocity field around a particle from a DSMC simulation. The plot represents a cut on the $xy$ plane, crossing the particle center, of a prolate ellipsoid with aspect ratio $a/b=2$ and major radius $a=0.39\mu \mbox{m}$. The particle is immersed in an argon gas flow with free stream velocity $\mathbf{U} = 99.7\hat{x}\mbox{ m/s}$ (yellow arrows) and is oriented at $\Phi=45^\circ$ with respect to the impinging gas flow. The volume of the simulated particles is fixed to $V=6.5\cdot 10^{-20}\mbox{m}^3$. The simulation domain size is set to $L=8\mu \mbox{m}$, so that $L\geq 20\cdot a$.}}
\label{fig:sim_sketch_ellips}
\end{figure}

\begin{figure}[h!]
\centering
\includegraphics[width=0.45\textwidth]{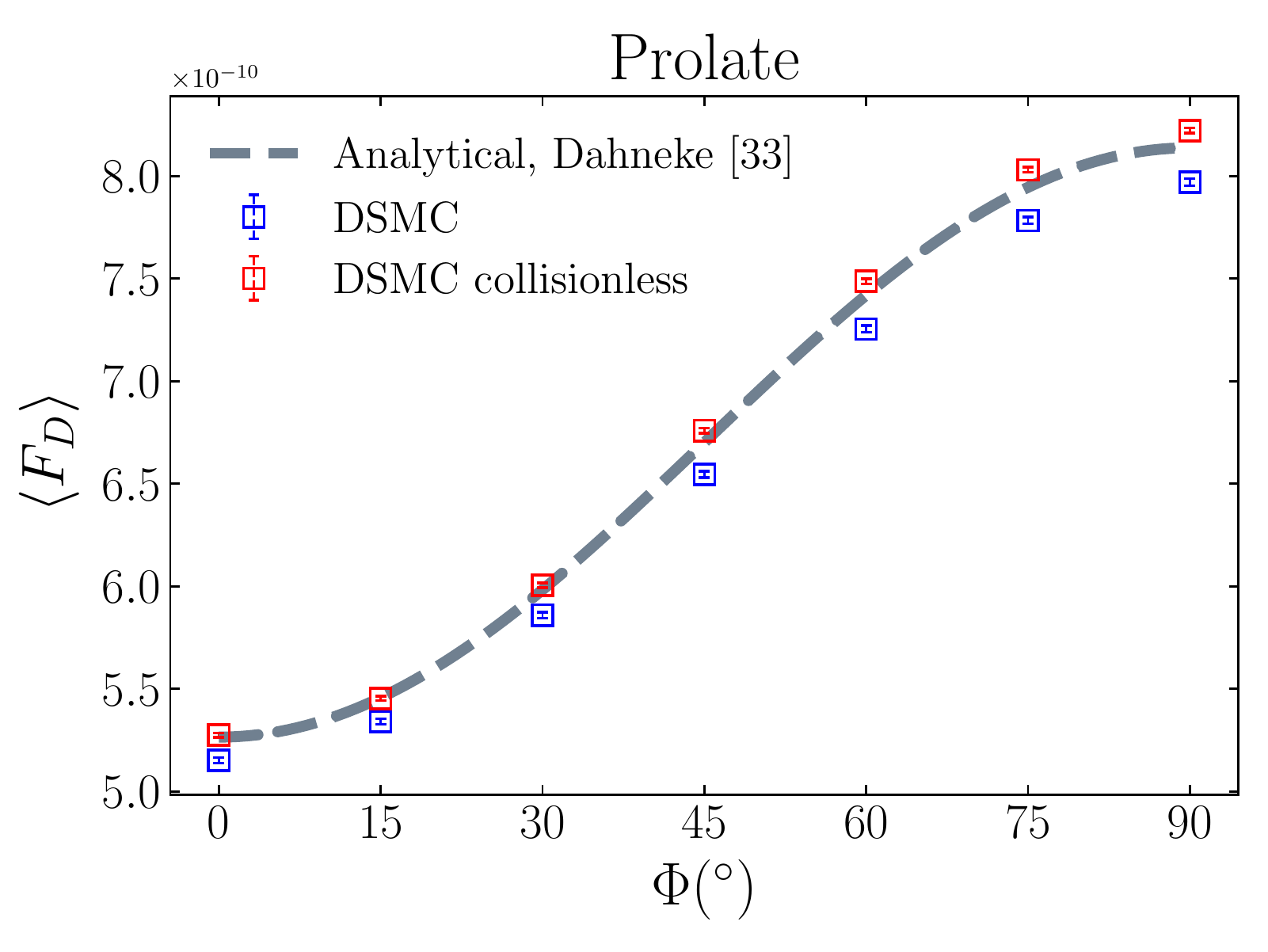}\\
\includegraphics[width=0.45\textwidth]{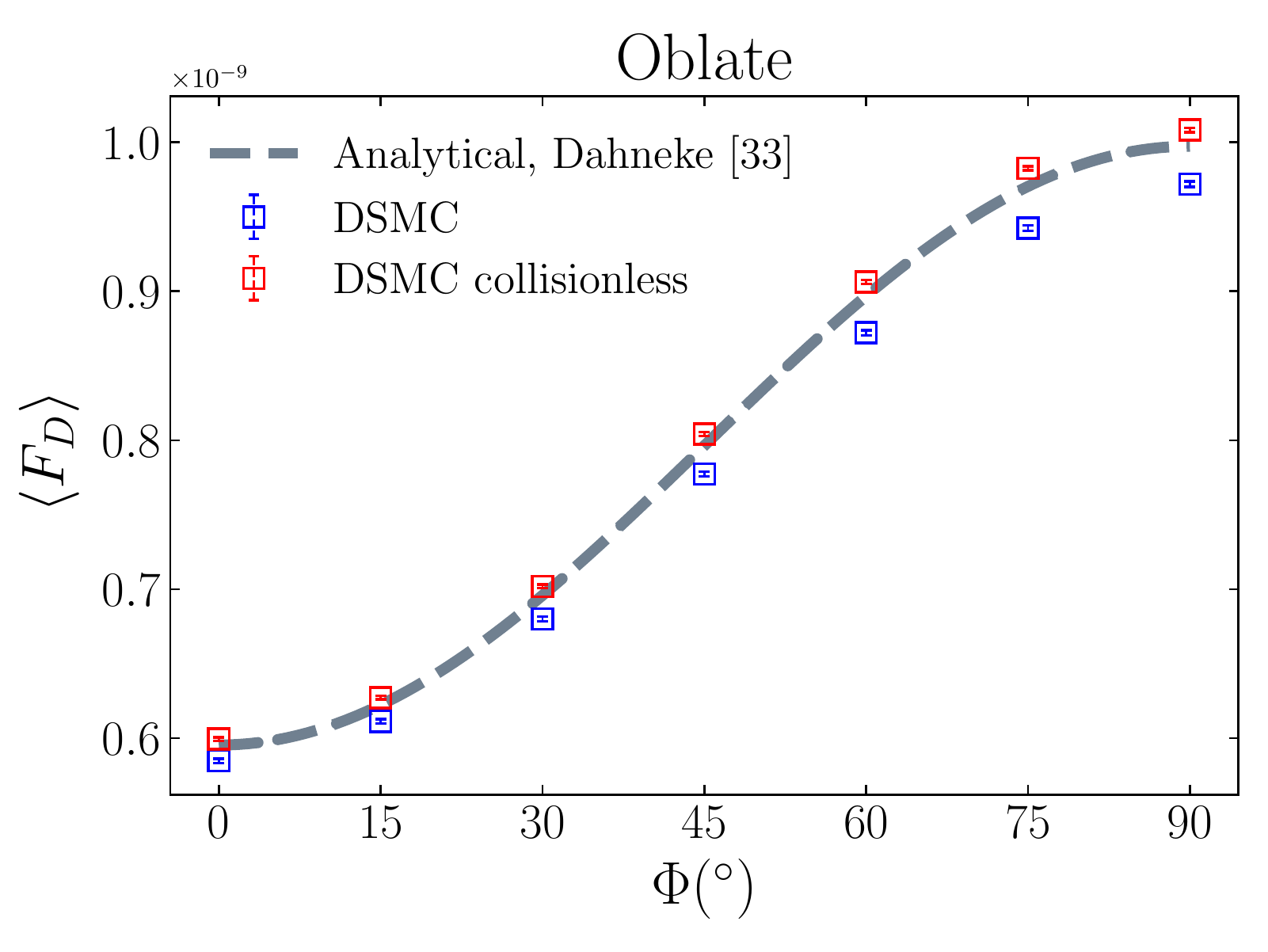}
\caption{\small{Drag force from DSMC simulations for a prolate (top) and oblate (bottom) ellipsoid for different orientations $\Phi$ at fixed $Kn=10$. The aspect ratio of the particles under investigation is $a/b=2$ with a major radius $a=0.39\mu \mbox{m}$ and $a = 0.315 \mu \mbox{m}$ for the prolate and oblate case,  respectively. The results from the DSMC simulations (squares) are compared with the prediction from Dahneke \cite{dahneke} (gray dashed lines) for the collisionless case given by Eqs. (\ref{eq:pr_th})-(\ref{eq:ob_th}). Results from DSMC collisionless simulations (red) show excellent agreement with the theoretical prediction, with a small deviation ($\leq 1\%$) for larger orientations due to velocity effects. The results of Dahneke are, in fact, obtained from an expansion based on the assumption of small streaming velocities, while in our case  $U_0 \sim 100\mbox{m/s}$ for $Kn=10$ to impose $Re=0.1$. For completeness, we also present simulation results when intermolecular collisions are present (blue). The error bars are obtained using Eq. (\ref{eq:errorbars}).}}
\label{fig:dsmc_collisionless}
\end{figure}

\begin{figure}[h!]
\centering
\includegraphics[width=0.45\textwidth]{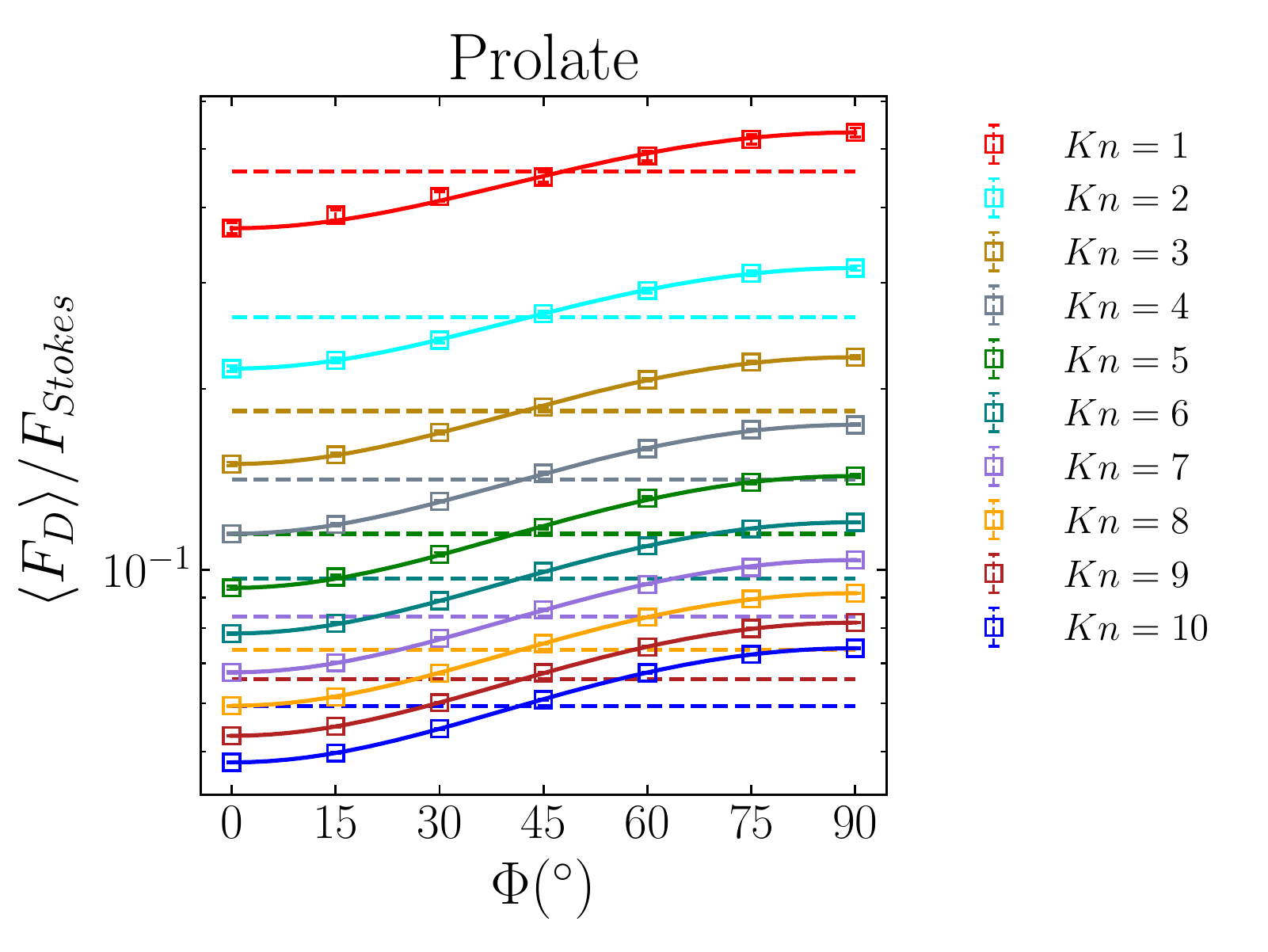}\\
\includegraphics[width=0.45\textwidth]{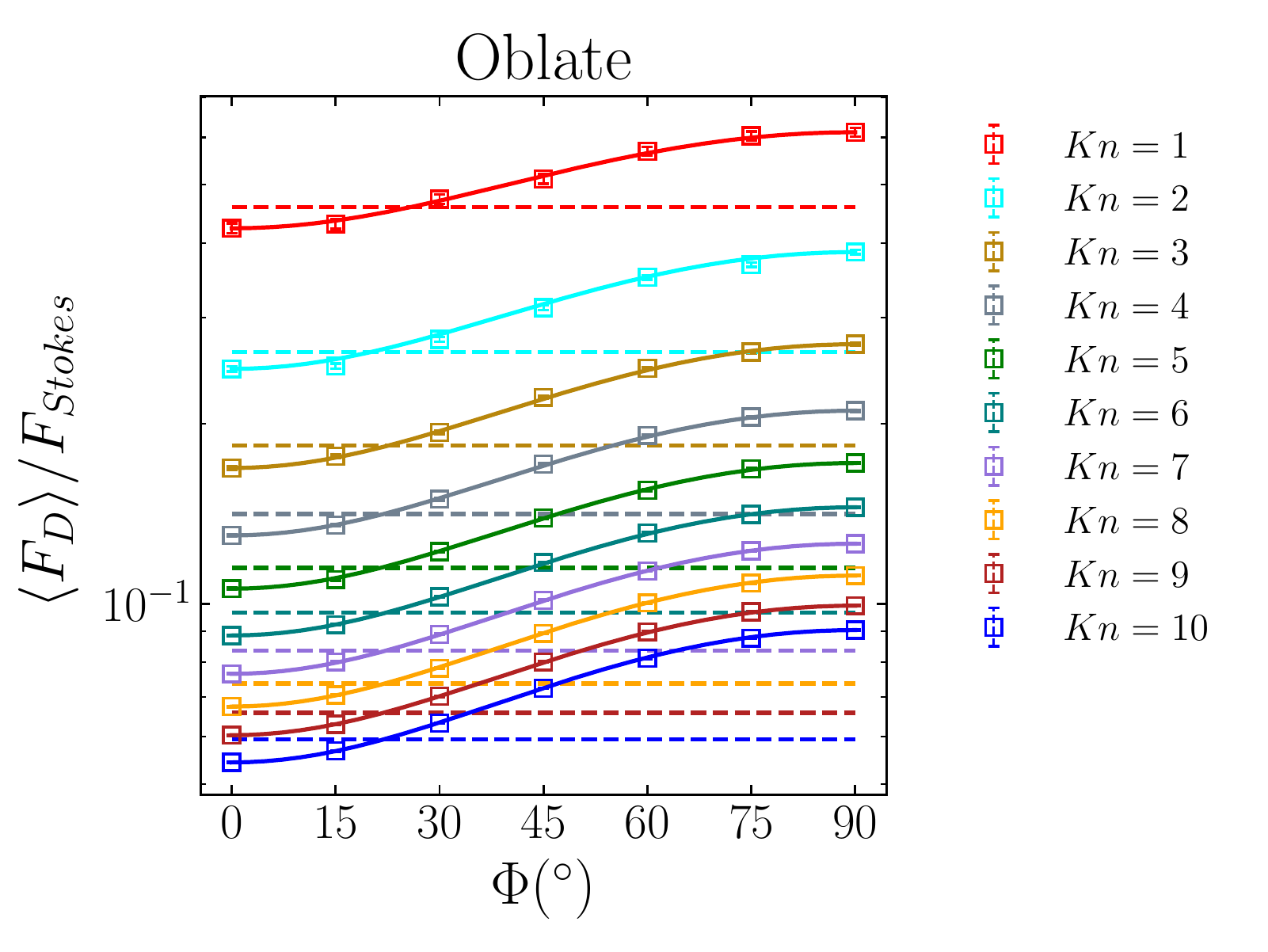}
\caption{\small{Drag force from DSMC simulations for a prolate (top) and oblate (bottom) ellipsoid as a function of the Knudsen number $Kn$ for different orientations $\Phi$, normalized with respect to the Stokes drag of a spherical particle $F_{Stokes} = 6\pi\mu RU_0$. The results from the DSMC simulations (squares) are compared with the prediction for the sphere with equivalent volume of the simulated ellipsoid (dashed lines) given by Eq. (\ref{eq:phil}) and with the theoretical correlation (solid lines) obtained by inserting the values of $F_{D,0^\circ}$ and $F_{D,90^\circ}$ from our DSMC simulations into Eq. (\ref{eq:drag_correlation}). The drag force on the equivalent sphere matches the drag force on a prolate and oblate ellipsoid oriented at $\Phi \sim 41^\circ$ and  $\Phi \sim 22^\circ$, respectively, for all the investigated $Kn$ numbers. For values of $Kn\leq 3$ the statistical fluctuations start to be more evident due to the increased gas density (and thus lower kinetic resolution with respect to larger $Kn$ cases), leading to larger error bars on the simulation data and thus larger fluctuations on the values of the intersection angle (represented by the value of $\Phi$ at which the horizontal lines intersect the drag force on the ellipsoidal particle). The error bars are obtained using Eq. (\ref{eq:errorbars}).}}
\label{fig:force_ellipsoid}
\end{figure}

\begin{figure}[h!]
\centering
\includegraphics[width=0.4\textwidth]{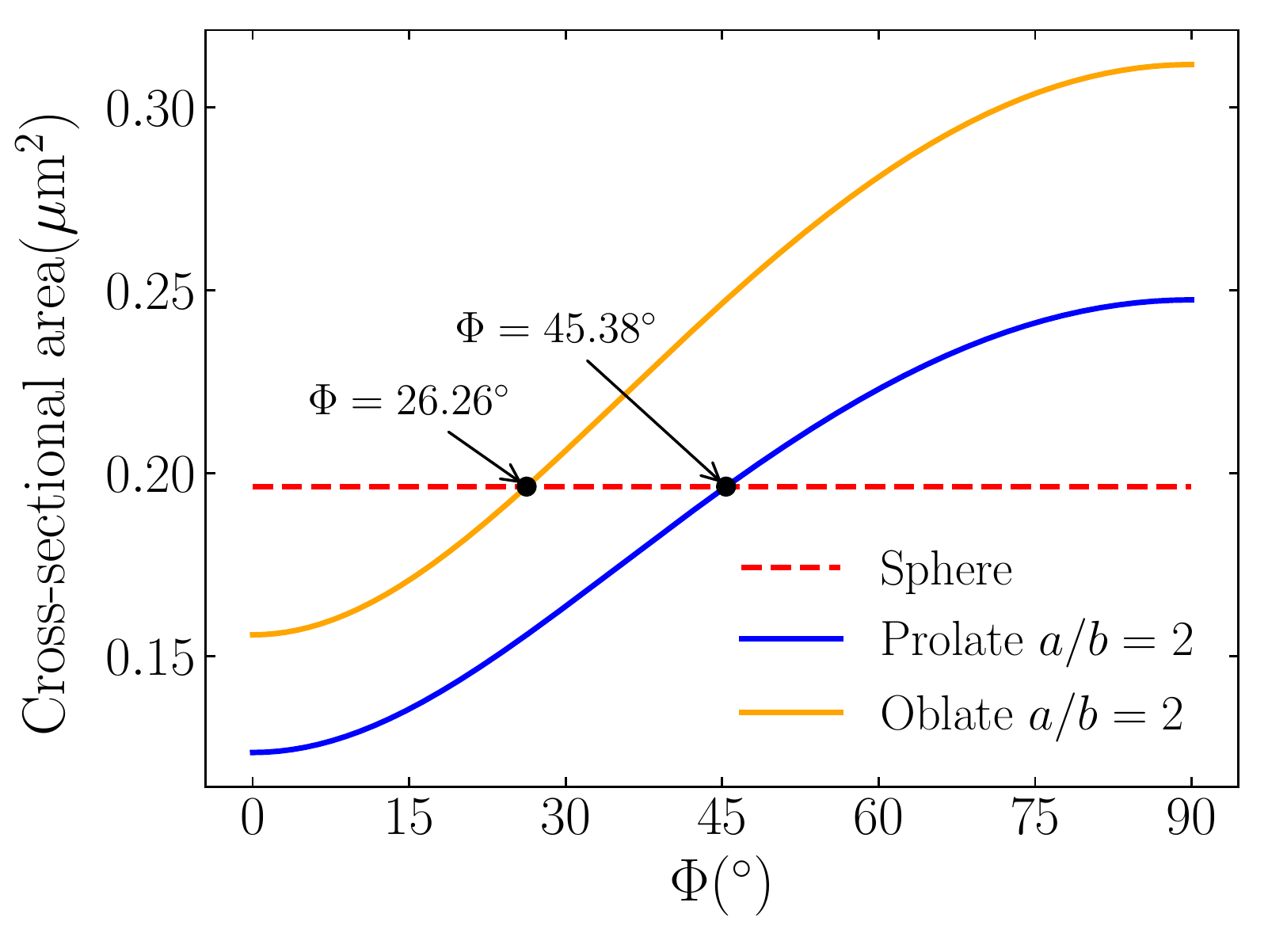}%
\caption{\small{Projected areas on the transverse plane with respect to the uniform ambient gas flow of a spherical (dashed red line), prolate (solid blue line) and oblate (solid orange line) particle. The projected areas are the same when the orientation of the prolate and oblate ellipsoids are $45.38^\circ$ and $26.26^\circ$, respectively. The volume of the particles is kept fixed to  $V=6.5\cdot 10^{-20}m^3$ while the aspect ratio of the ellipsoidal particles is $a/b = 2$.}}
\label{fig:areas}
\end{figure}

When particles with more complex shapes are investigated, a new degree of complexity arises in relation to the choice of the characteristic length needed to define the relevant dimensionless quantities, such as the particle-based Reynolds number, $Re$, and the Knudsen number, $Kn$. For a prolate or oblate ellipsoidal particle, in fact, at least two separate characteristic lengths (embodied by its major and minor axes) are typically available and this can lead to some difficulties in the understanding of which intrinsic dimensions of the particle actually play the dominant role in relation to the characterization of the dynamics of such particles with respect to the flow.\\
To address the same problem in the continuum limit, a number of authors \cite{sanjeevi1,sanjeevi2,ouchene,ouchene2} proposes to use the radius of the sphere with equivalent volume, $R_{eq}$, to define the Reynolds number. In this work we decide to use the same approach, extending this choice also to the Knudsen number, so that the relevant dimensionless numbers read:
\begin{align}
Re = \frac{2U_0R_{eq}}{\nu},
\label{eq:re}
\end{align}
\begin{align}
Kn = \frac{\lambda}{R_{eq}},
\label{eq:kn}
\end{align}
where $\nu = \mu/\rho$ is the kinematic viscosity of the gas. \\
In the following of this Section we show that the proposed definition of $Kn$ is a good approximation to describe rarefaction effects for ellipsoidal particles. Firstly, it successfully reduces the number of characteristic lengths to one (the radius of the equivalent sphere). This aspect not only defines $Kn$ in an unambiguous way, but also makes this definition unrelated to the aspect ratio of the particle and to its orientation. Additionally, the relation between the drag force acting on the equivalent sphere and the one acting on the ellipsoidal particles is preserved independently of the specific value of $Kn$.\\
Before proceeding, we want to provide an explanation, based on a theoretical analysis of collisionless flows, of the necessity for the Knudsen number to be unrelated with a specific aspect ratio or orientation of the particle. Different authors have investigated analytically the interactions between collisionless gas flows and simple geometries using a gas-kinetic approach. Among those authors, Bird \cite{bird} expressed the drag and lift coefficients of a thin plate (i.e. a flat rectangular surface) immersed in a uniform gas flow as functions of the molecular speed ratio, $s=U_0 /(\frac{m}{2k_BT})^{1/2}$ (which represents the ratio between the ambient flow velocity, $U_0$, and the most probable molecular speed $c_{mp} = (2k_BT/m)^{1/2}$), the angle of attack $\Phi$, and the temperature of the solid surface $T_{wall}$.\\
From these relations, valid in the collisionless limit, we can observe that rarefaction effects are independent with respect to $\Phi$, as for small free stream velocities (i.e. $s\ll 1$) the sine-squared drag law typical of the continuum limit is recovered also for the collisionless case.\\
Using the linearity of velocity fields in creeping flows ($Re\ll1$), Happel and Brenner \cite{happel} show that the drag force on an arbitrary shaped particle, in the continuum regime, as a function of its orientation $\Phi$ can be expressed as:
\begin{align}
F_D(\Phi) = F_{D,0^\circ} + (F_{D,90^\circ} - F_{D,0^\circ} )\sin^2\Phi,
\label{eq:drag_correlation}
\end{align}
where $F_{D,0^\circ}$ and $F_{D,90^\circ}$ are the drag force at $\Phi=0^\circ$ and $\Phi=90^\circ$, respectively. For the rest of the paper we will adopt the same short-hand notation using the subscripts $0^\circ$ and $90^\circ$ to refer at the two cases of interest. It is useful to recall (see Fig. \ref{fig:general_geom}) that Eq. (\ref{eq:drag_correlation}) is a general expression valid for any arbitrary orientation of the particle, as the hydrodynamic force acting on it can always be decomposed in a component orthogonal to the principal axis of the particle (for which the orientation is constant), and one component laying on the same plane of the principal axis of the particle, for which the angle of attack $\Phi$ is the only relevant variable to describe orientation effects.\\
From Eq. (\ref{eq:drag_correlation}), Happel and Brenner obtain the correlations for the drag and lift coefficients, which read:
\begin{align}
C_{D}(\Phi) = C_{D,0^\circ} + (C_{D,90^\circ} - C_{D,0^\circ})\sin^2\Phi,
\label{eq:drag_continuum}
\end{align}
\begin{align}
C_L(\Phi) = (C_{D,90^\circ} - C_{D,0^\circ})\sin\Phi\cos\Phi.
\label{eq:lift}
\end{align} 
In Fig. \ref{fig:bird_drag} we present the scaling of $C_D$ and $C_L$ for a thin rectangular plate, as proposed by Bird, as a function of the angle of attack, $\Phi$, for different values of the speed ratio $s$. The results are then compared with respect to the correlations from the continuum regime given by Eqs. (\ref{eq:drag_continuum})-(\ref{eq:lift}). It is shown that for small speed ratios the correlations from the continuum regime hold also in the collisionless limit, highlighting that rarefaction effects do not depend on the orientation of the solid body with respect to the gas flow. This result is related to the fact that for small free stream velocities, the friction due to pressure effects and the one due to tangential effects are equally important in the interactions between the gas molecules and the surface of the solid object.\\
We now proceed in the investigation of the drag force on ellipsoidal particles from DSMC simulations in relation to the definition of the Knudsen number chosen for this work and given by Eq. (\ref{eq:kn}). We do so by comparing the obtained drag force with the one for  the equivalent sphere at the same Knudsen number, taking into consideration different aspect ratios (prolate and oblate ellipsoids) and orientations of the ellipsoidal particles with respect to the impinging gas flow.\\
In our simulations we fix the volume of the equivalent sphere to the same value as the one used for the spherical particle in Subsection \ref{subsec:validation}  ($V=6.5\cdot 10^{-20}\mbox{m}^3$, corresponding to $R_{eq}=5\mu\mbox{m}$). The aspect ratio of the ellipsoidal particles is fixed to $a/b=2$, leading to a major radius $a = 0.39 \mu \mbox{m}$ for the prolate case and $a = 0.315 \mu \mbox{m}$ for the oblate case. A sketch of the simulated ellipsodal particles is presented in Fig. \ref{fig:part_geom}. The physical simulation box size is increased to $L_{phys}=8\mu \mbox{m} $ in order to avoid finite-size effects due to the vicinity of the simulation domain edges with the solid particle. In terms of DSMC cell units, we use a value of $L=140$ for $Kn=1$ and a value of $L=120$ for all other investigated values of $Kn$ ($2,\dots,10$), leading to $0.23 \leq L_c/\lambda \leq 0.026$. The Reynolds number is set to $Re=0.1$ by changing the free stream velocity $U_0$ to ensure creeping flow conditions. A snapshot from the DSMC simulations for $Kn=10$ and $\Phi=45^\circ$ is presented in Fig. \ref{fig:sim_sketch_ellips}.
All the other simulation parameters, such as the gas density $\rho$ and the number of particles-per-cells, as well as the boundary conditions, are the same as in Subsection \ref{subsec:validation}.\\
We firstly perform different simulations by varying the orientation $\Phi$ of the particle with respect to the ambient uniform flow at fixed $Kn=10$ in the collisionless limit and compare our results with the analytical expressions for the drag force on ellipsoidal particles provided by Dahneke \cite{dahneke}. The author extends the theoretical approach from Epstein, valid for small streaming velocities, to particles with different shapes, assuming that the reflected gas molecules do not interact with the incoming ones (collisionless limit). We can achieve this in the DSMC simulations by artificially switching off intermolecular collisions. The analytical expression from Dahneke \cite{dahneke} for the drag force on prolate ellipsoidal particles is:\\
\hspace{-0.4cm}\begin{minipage}{0.51\textwidth}
\begin{align}
\begin{split}
F_{pr}  = &  \frac{\pi \mu ab U_0}{\lambda} \Biggl[ \sin^2(\Phi) \times \\
   \times  &\biggl(A_{pr}\biggl\{4+\biggl(\frac{\pi}{2}-1\biggr)\sigma\biggr\} + \frac{C_{pr}}{B_{pr}^2}\biggl\{2 + \frac{4B_{pr}^2+\pi-6}{4}\sigma\biggr\}\biggr) + \\
            + &\cos^2(\Phi) \times \\
    \times & \biggl(2A_{pr}\sigma + \frac{C_{pr}}{B_{pr}^2}\biggl\{B_{pr}^2(4-2\sigma) - 4 + \biggl(3 - \frac{\pi b^2}{2a^2}\biggr)\sigma\biggr\}\biggr)\Biggr],
\end{split}
\label{eq:pr_th}
\end{align}
\end{minipage}\\
where  $B_{pr} = \sqrt{1-b^2/a^2}$, $A_{pr} = \sin^{-1}(B_{pr})/B_{pr}$ and $C_{pr} = b/a - A_{pr}$. Similarly, the expression for the drag force on oblate ellipsoids reads:\\
\begin{minipage}{0.45\textwidth}
\begin{align}
\begin{split}
F_{ob}  =  &  \frac{\pi \mu ab U_0}{\lambda} \Biggl[ \sin^2(\Phi) \times \\
      \times &\biggl(A_{ob}^2B_{ob}\biggl\{\frac{6-\pi}{4}\sigma -2 \biggr\} + C_{ob} \biggl\{ 4-\frac{4-\pi}{2}\sigma \biggr\} +\frac{a}{b}\sigma\biggr) + \\
              + &\cos^2(\Phi)  \times \\
      \times & \biggl(A_{ob}^2B_{ob}\biggl\{ 4 - \frac{6-\pi}{2}\sigma\biggr\}\frac{a^2}{b^2} + C_{ob}\sigma + \frac{a}{b}\sigma \biggr)\Biggr]   ,
\end{split}
\label{eq:ob_th}
\end{align}
\end{minipage}\\
with $A_{ob} = 1/\sqrt{a^2/b^2 -1}$, $C_{ob} = A_{ob}\log\left( a/b + 1/A_{ob} \right)$ and $B_{ob} =a/b - C_{ob}$. The agreement between DSMC simulations and Eqs. (\ref{eq:pr_th})-(\ref{eq:ob_th}) is excellent, as presented in Fig. \ref{fig:dsmc_collisionless}.\\
We now proceed in investigating rarefaction effects by performing collisional DSMC simulations at varying $Kn$ numbers. The results are shown in Fig. \ref{fig:force_ellipsoid}, where we compare the drag force on the ellipsoidal particles with the one acting on the equivalent sphere at the same $Kn$ number, as obtained by the prediction of Phillips from Eq. (\ref{eq:phil}). \\
As these results show, the $\Phi$-dependence in Eq. (\ref{eq:drag_correlation}) is well captured and rarefaction effects seem thus in good approximation independent on the orientation of the particle, as the correlation  (\ref{eq:drag_correlation}) obtained from the continuum regime is preserved. Moreover, the relation between the drag force on the equivalent sphere and the one acting on the ellipsoidal particles is maintained for all the investigated range of $Kn$, considering the presence of larger fluctuations at lower $Kn$ cases due to lower kinetic resolution. In all the investigated cases, in fact, $F_{Phil}(R_{eq})$ crosses the curves obtained from the simulations for the drag force on the ellipsoidal particle, and the intersection happens at $\Phi \sim 41^\circ$ and $\Phi \sim 22^\circ$ for the prolate and oblate case, respectively.\\
We explain the reason of these values of the intersection angles from a geometrical perspective.
In Fig. \ref{fig:areas} we compare the values of the cross-sectional area, with respect to the gas flow, of the three particles we are investigating (spherical, prolate and oblate). We show that the cross-sectional area of the sphere with equivalent volume coincides with the one of a prolate ellipsoid tilted by $\Phi\sim 45^\circ$ and with the one of an oblate ellipsoid tilted by $\Phi\sim 26^\circ$. These values are very close to the intersection angles that appear in Fig. \ref{fig:force_ellipsoid}. We should mention that the geometrical intersection angles at which the projected areas coincide are not expected to match exactly with the angles at which the drag force on different particles coincide, independently of the rarefaction. Small deviations are expected to occur due to the different surface curvature of the particles, to their elongation and due to the presence of a finite Reynolds number.\\
\section{Predictive model for the Drag and Lift coefficients}
\label{sec:drag}

\begin{figure*}[!]
\centering
\includegraphics[width=0.45\textwidth]{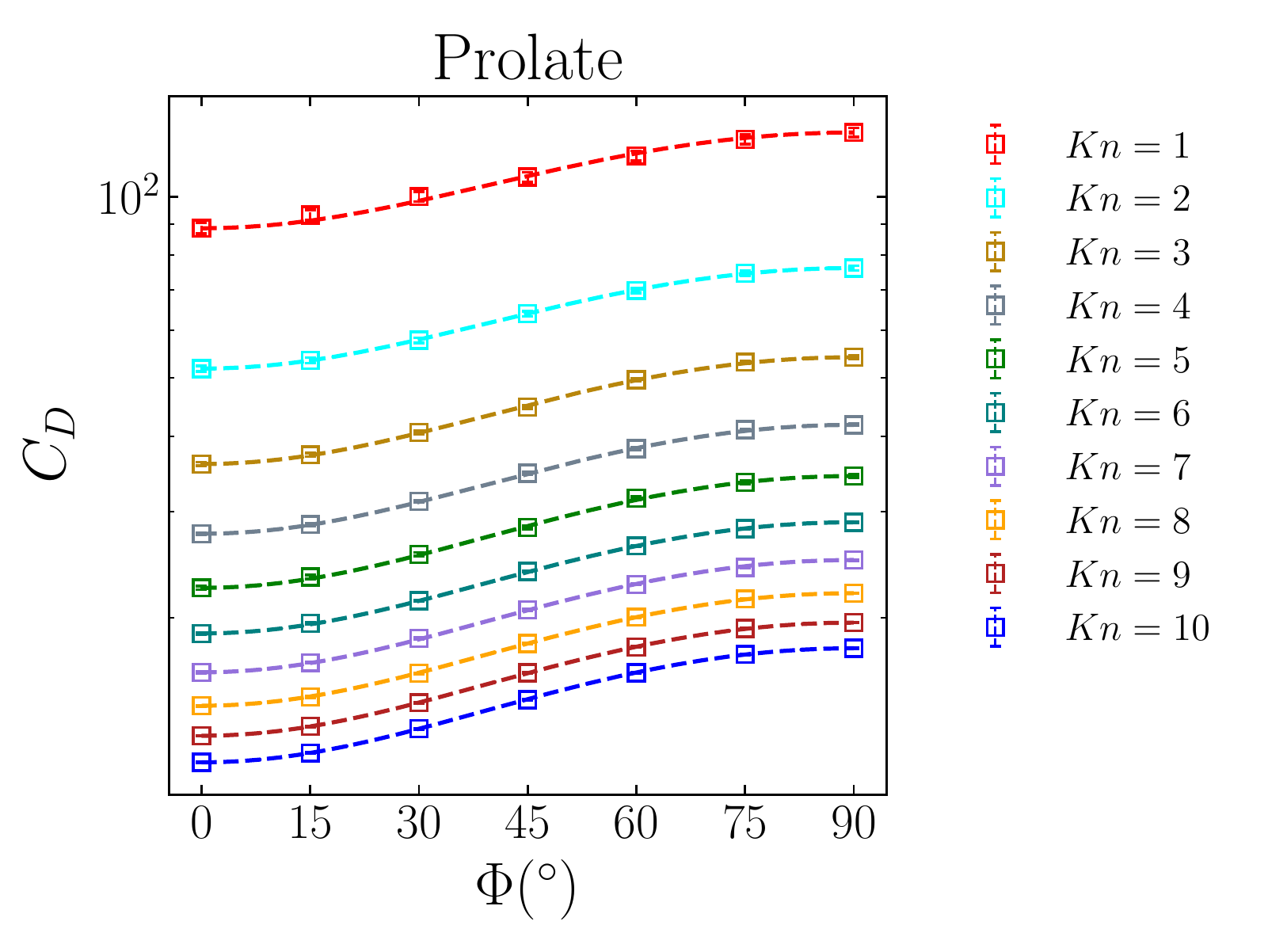}%
\includegraphics[width=0.45\textwidth]{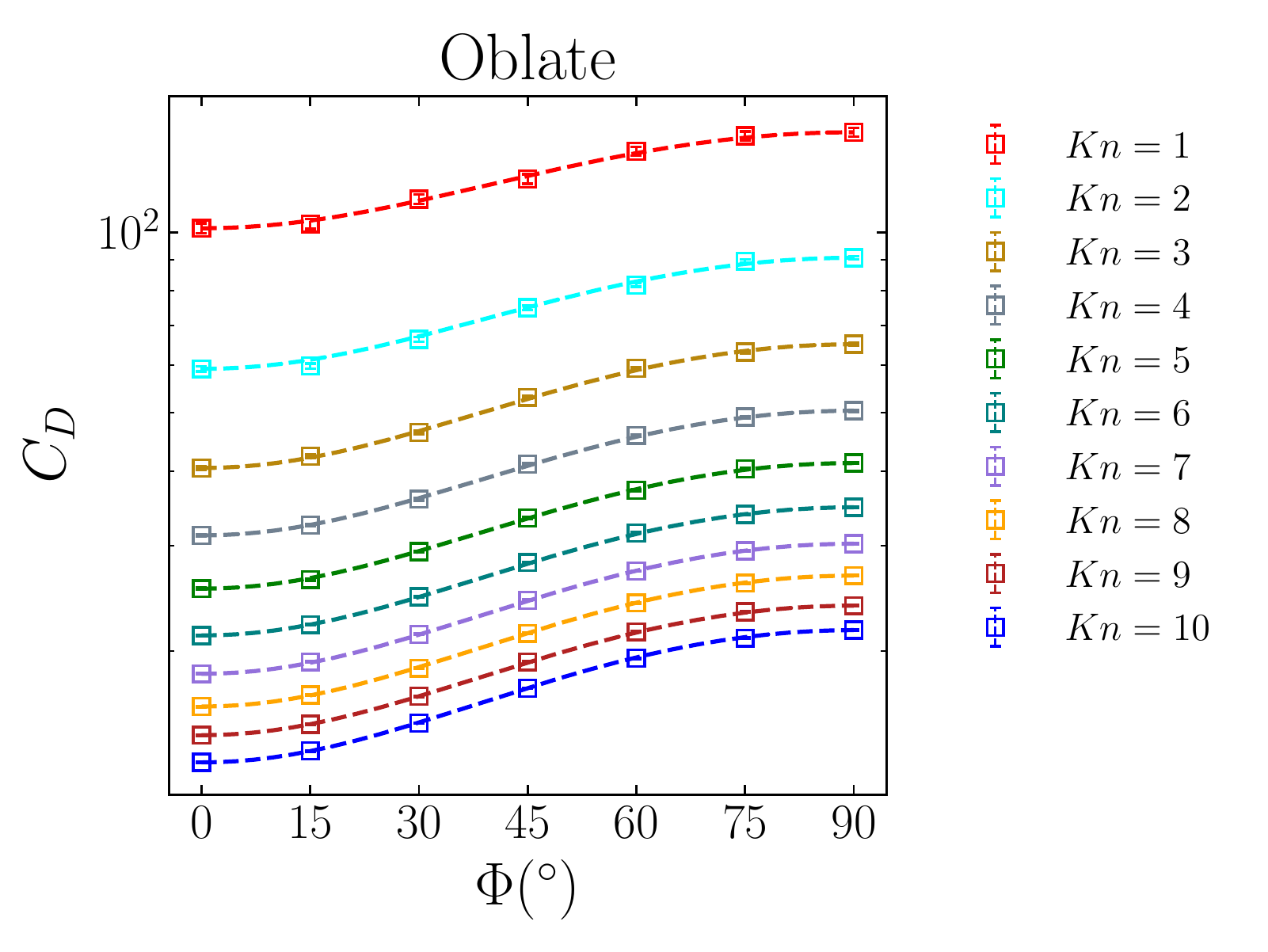}\\
\includegraphics[width=0.45\textwidth]{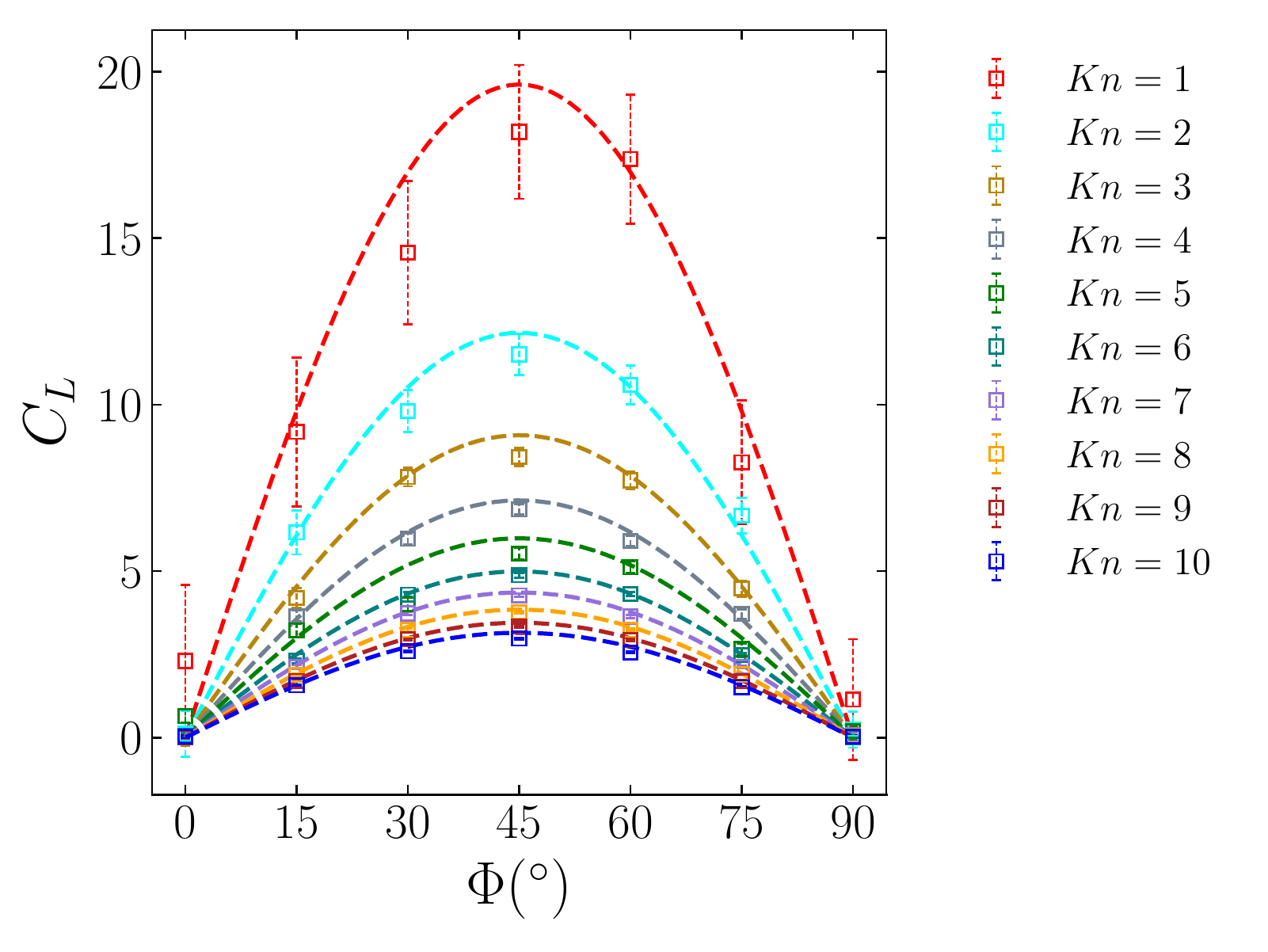}%
\includegraphics[width=0.45\textwidth]{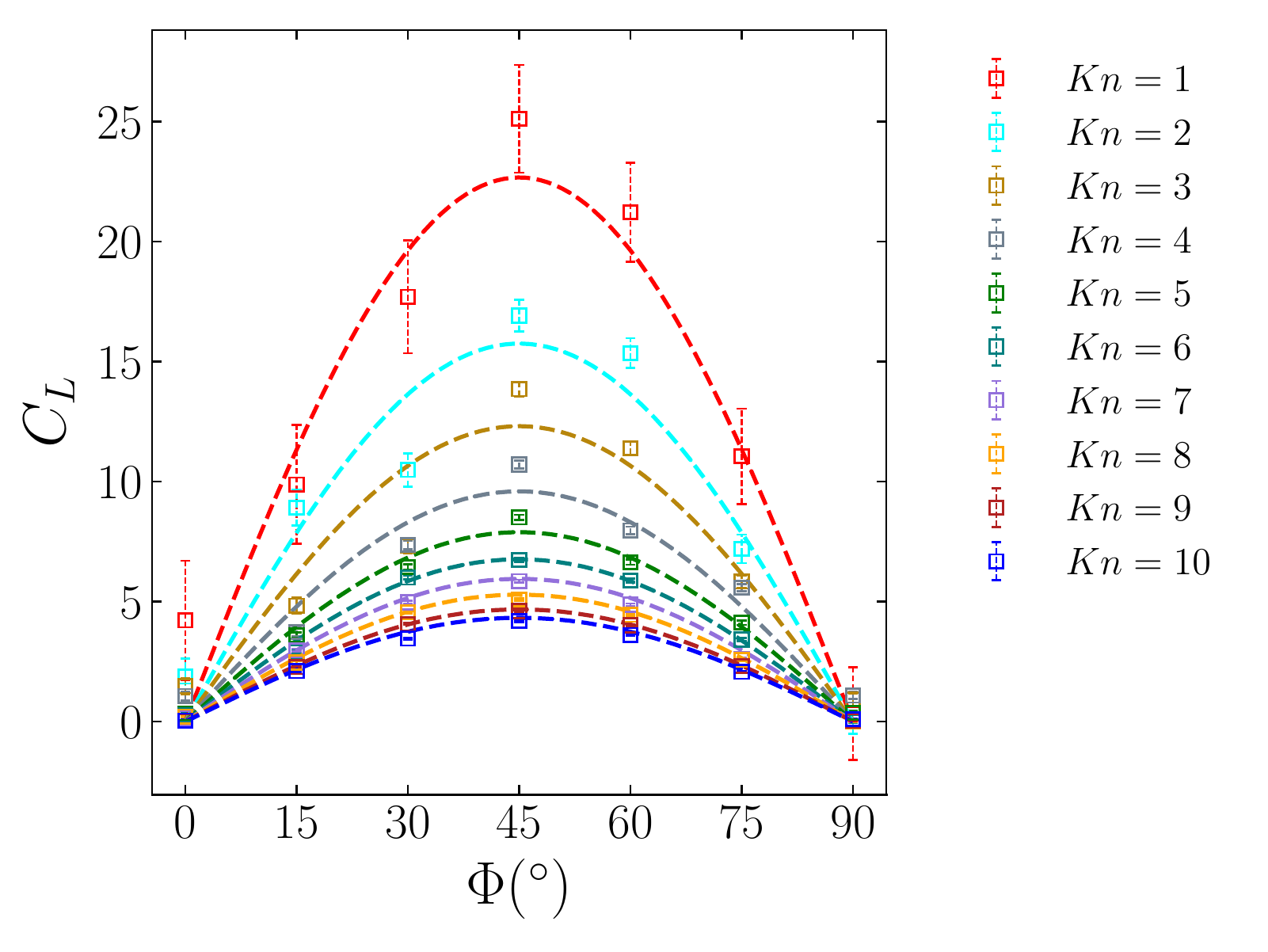}
\caption{\small{Comparison between  Eqs. (\ref{eq:drag_continuum})-(\ref{eq:lift}) (dashed lines) and  results from DSMC simulations (squares) for the drag ($C_D$) and lift ($C_L$) coefficients of a prolate (left column) and an oblate (right column) ellipsoid for different orientations $\Phi$ and $Kn$. The volume of the particles is $V=6.5\cdot 10^{-20}\mbox{m}^3$. Both ellipsoids have aspect ratio $a/b = 2$ with a major radius of $a = 0.39 \mu \mbox{m}$ for the prolate case and $a = 0.315 \mu \mbox{m}$ for the oblate case. In order to use  Eqs. (\ref{eq:drag_continuum})-(\ref{eq:lift}) in this context, we computed $ C_{D,0^\circ}$ and $C_{D,90^\circ}$ directly from our DSMC simulations. The agreement between the theoretical predictions and the DSMC simulations is excellent. The $C_D$ data is plotted in semi-log scale for a better readability and the error bars are obtained using Eq. (\ref{eq:errorbars}).}}
\label{fig:drag_ellipsoid}
\end{figure*}

\begin{figure}[!]
\centering
\includegraphics[width=0.4\textwidth]{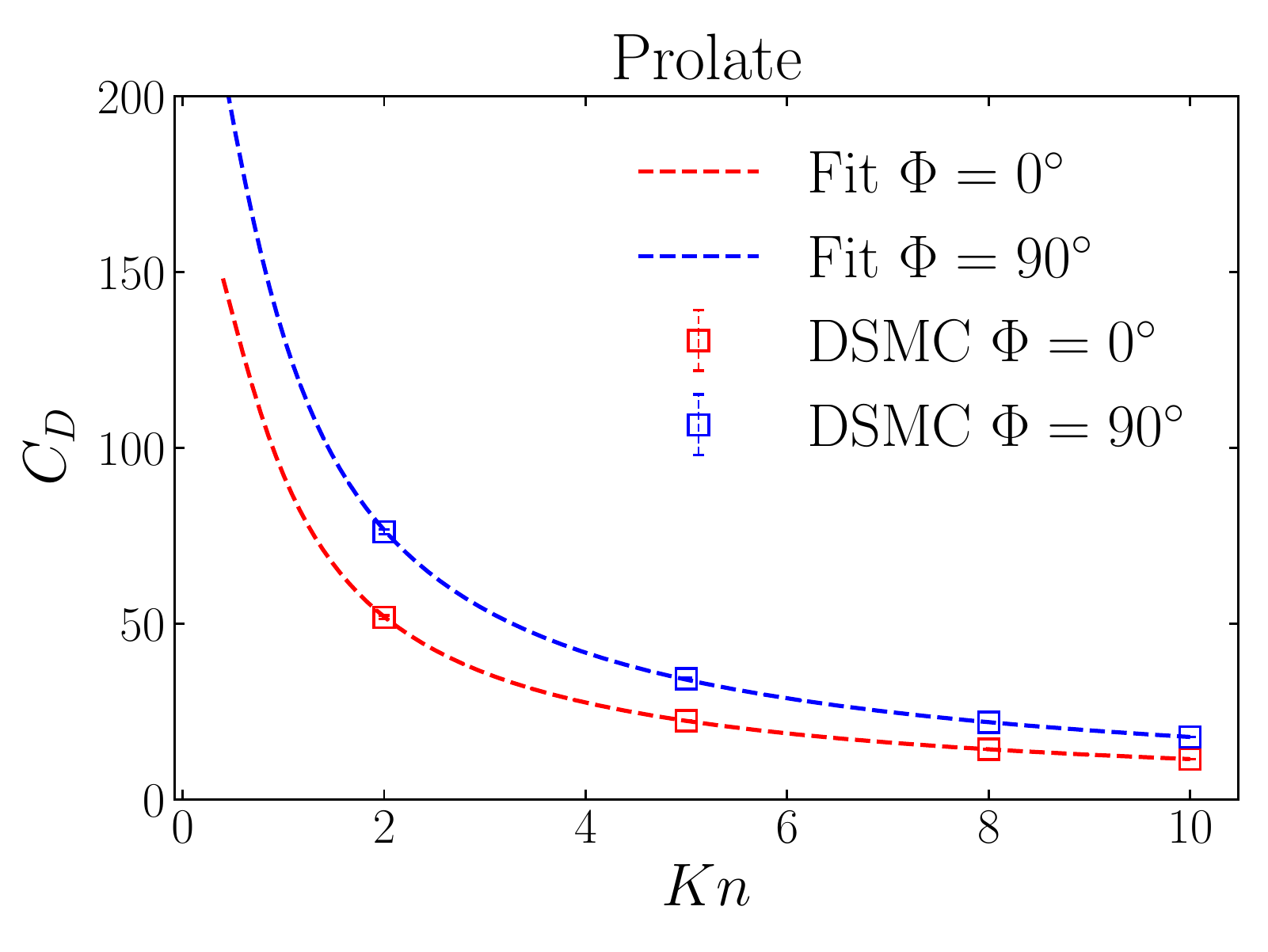}\\
\includegraphics[width=0.4\textwidth]{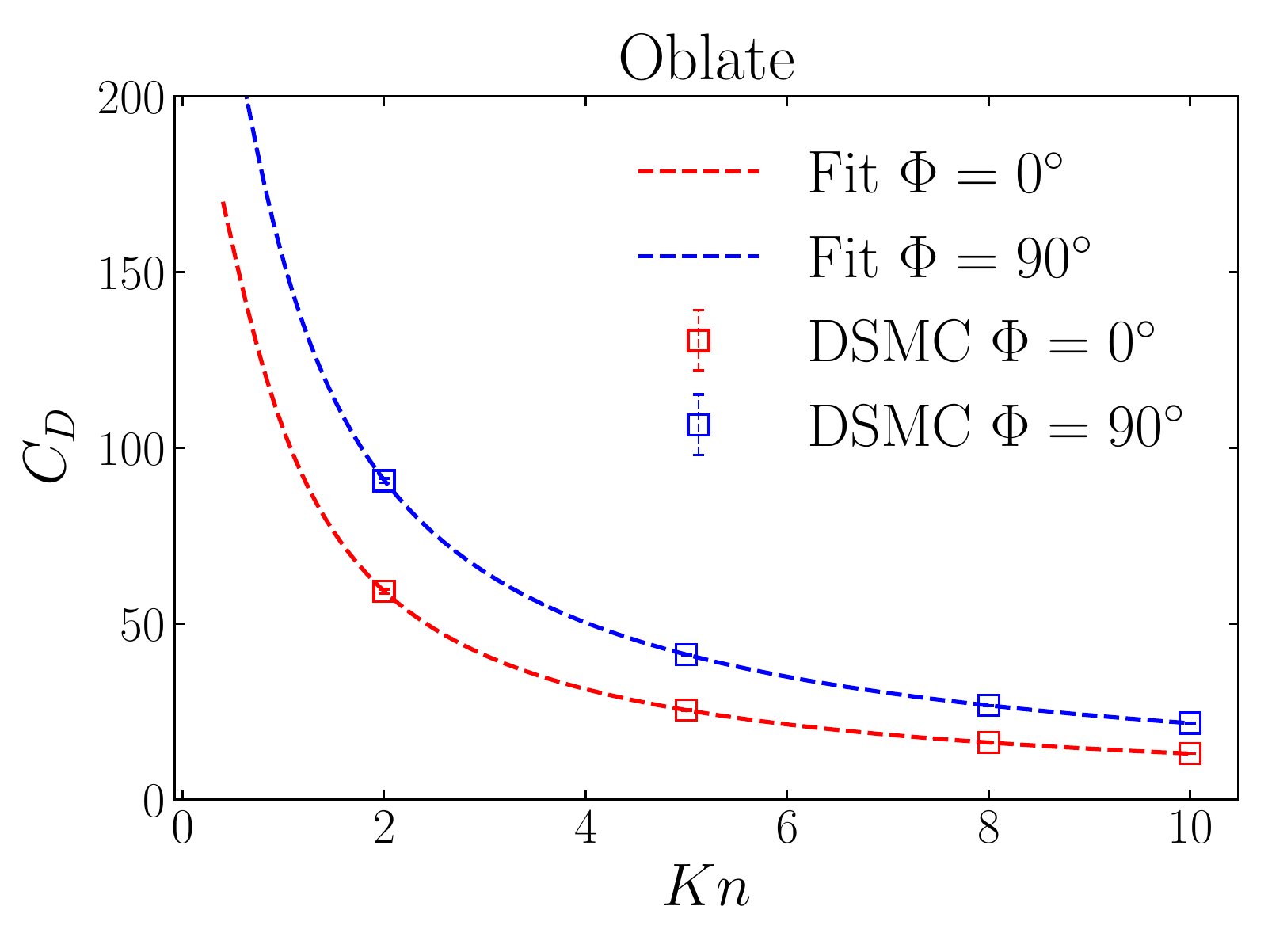}
\caption{\small{Fit of DSMC simulation data of $ C_{D,0^\circ}$ (red) and $C_{D,90^\circ}$ (blue) using the fit functions given by Eqs. (\ref{eq:g_0})-(\ref{eq:g_90}), for a prolate ellipsoid (top) and an oblate ellipsoid (bottom) with volume $V=6.5\cdot 10^{-20}\mbox{m}^3$. Fitted curves (dashed lines) represent the model functions given by Eqs. (\ref{eq:ellipsoid_prediction_0}) and (\ref{eq:ellipsoid_prediction_90}). The error bars used for the fitting are obtained using Eq. (\ref{eq:errorbars}) and the resulting fit parameters are presented in Table \ref{tab:fit}.}}
\label{fig:fit}
\end{figure}

\begin{figure*}[!]
\centering
\includegraphics[width=0.45\textwidth]{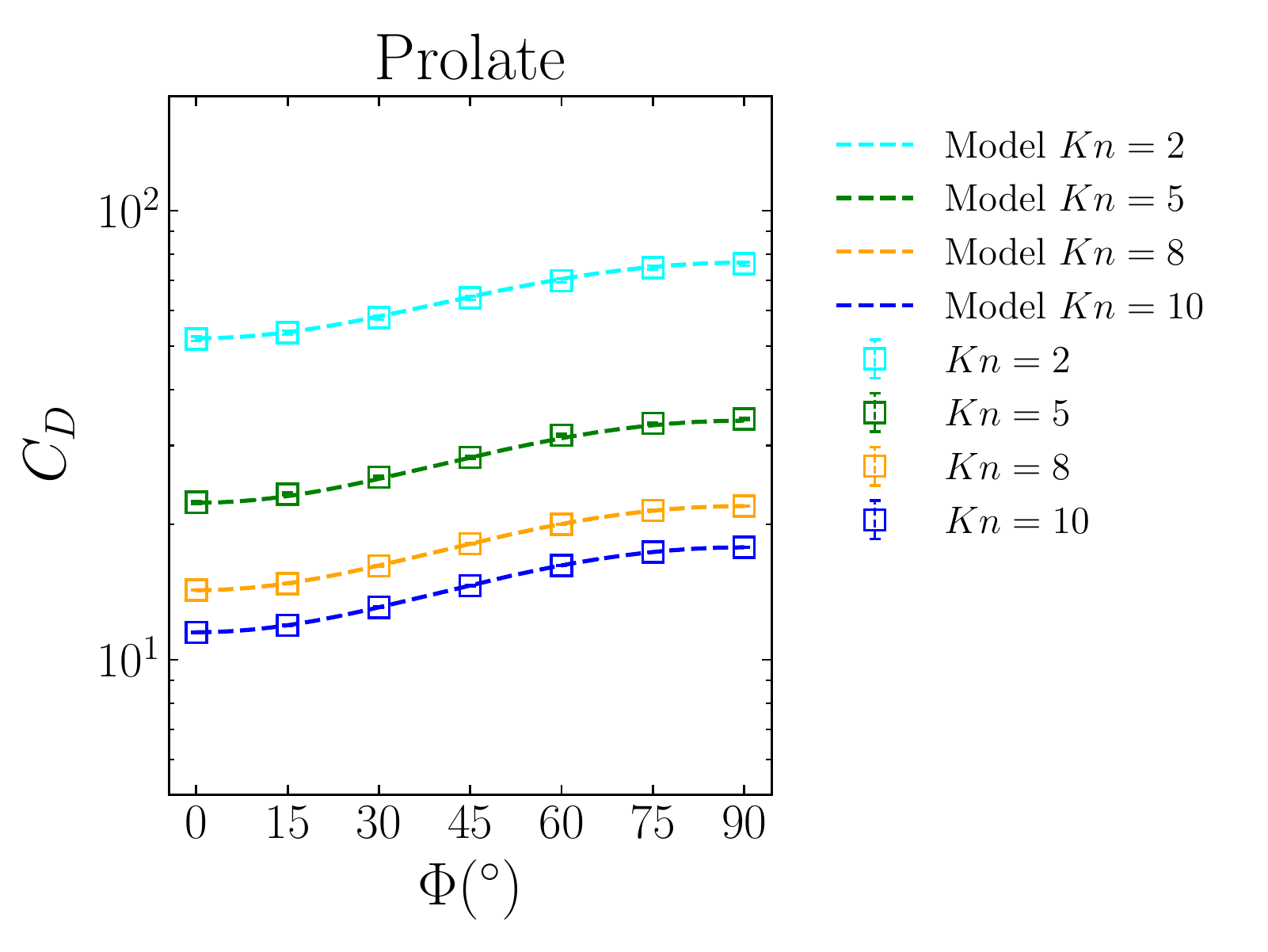}%
\includegraphics[width=0.45\textwidth]{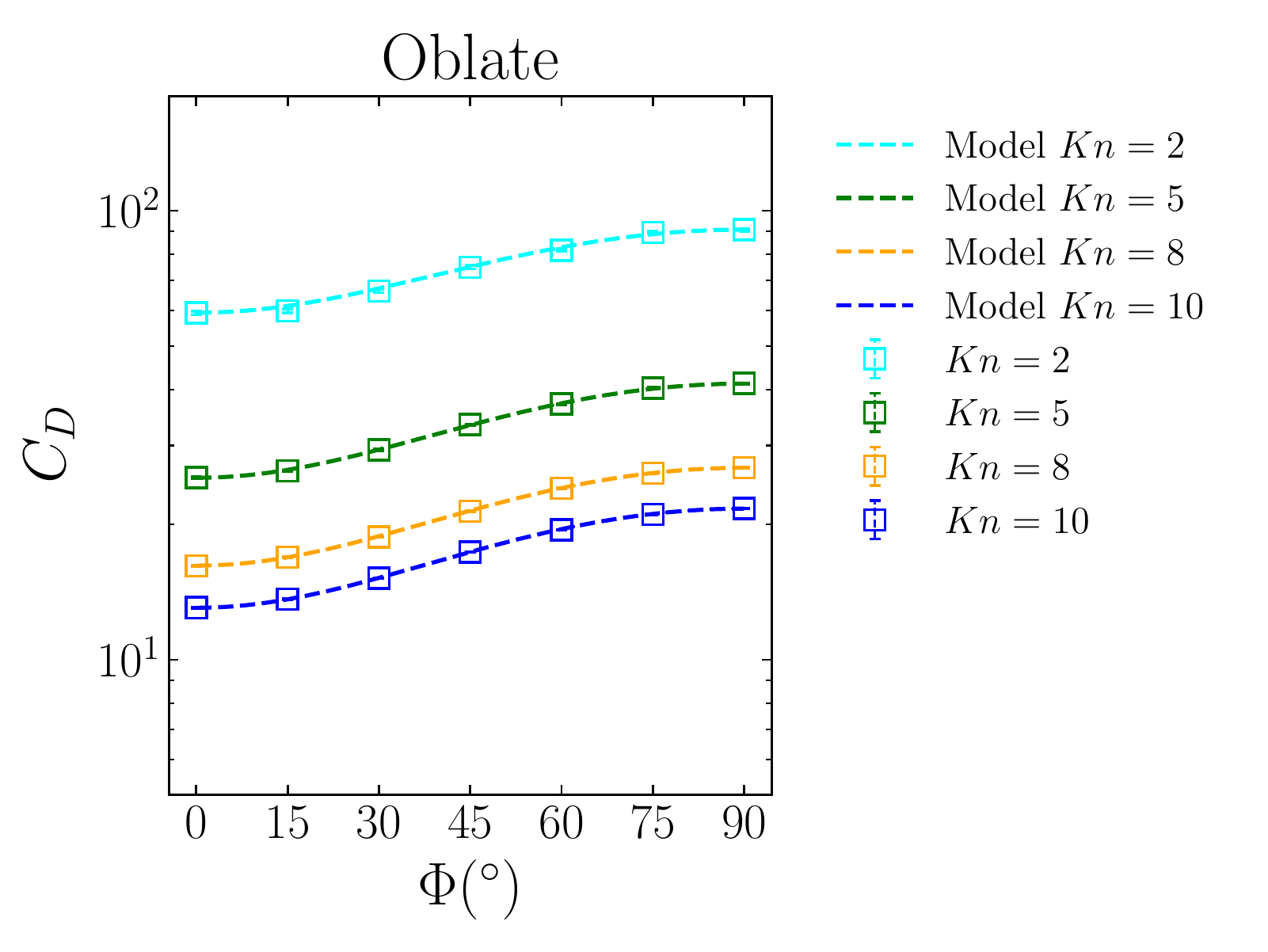}\\
\includegraphics[width=0.45\textwidth]{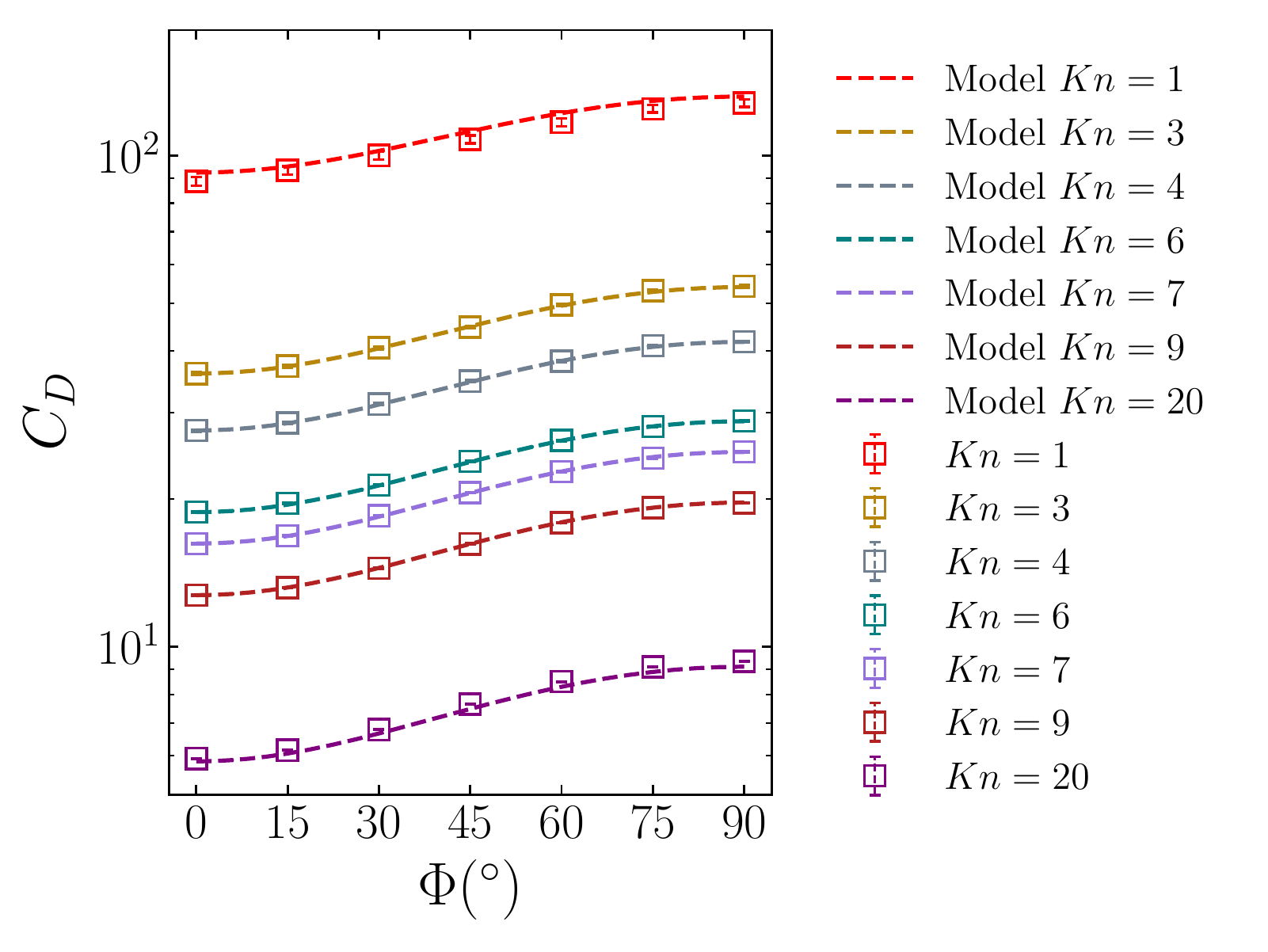}%
\includegraphics[width=0.45\textwidth]{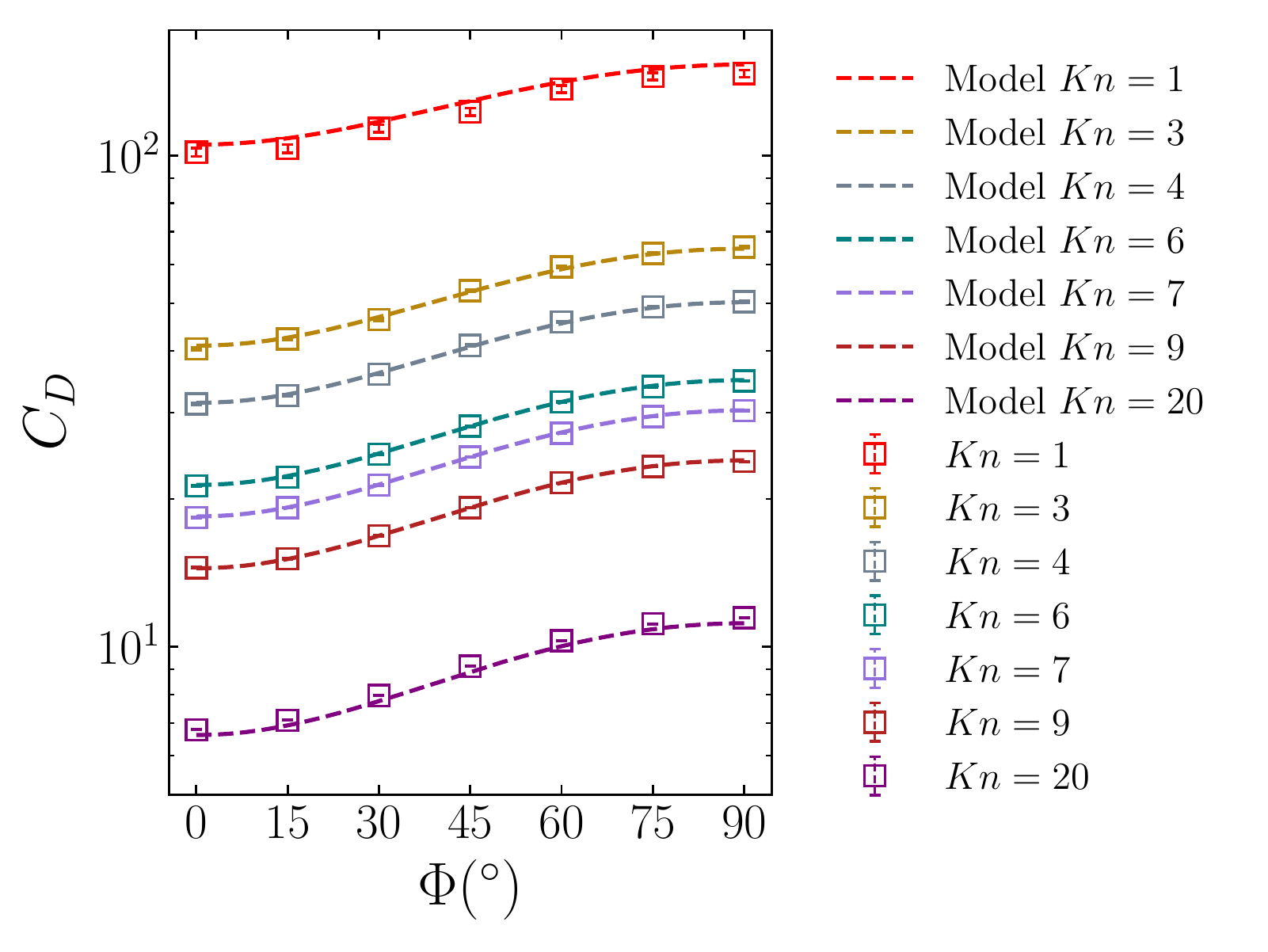}
\caption{\small{Comparison between DSMC simulations (colored squares) and model predictions (colored dashed lines) of the drag coefficient $C_D$ of a prolate (left column) and an oblate (right column) ellipsoid immersed in an uniform gas flow for different $Kn$ as a function of the orientation $\Phi$. The volume of the particles is $V=6.5\cdot 10^{-20}\mbox{m}^3$, the major radius for the prolate ellipsoid with $a=2b=2c$ is $a = 0.39 \mu \mbox{m}$, while for the oblate case with $a=b=2c$ is $a = 0.315 \mu \mbox{m}$.  The data from the DSMC simulations has been split in two sets: $Kn_{fit}$, which is used to obtain the fit for the predictive model, and$Kn_{test} $, which instead is not used during the fit. The model predictions, given by Eq. (\ref{eq:drag_model}), are then compared with the fit set (top row) and the test set (bottom row). In both cases the match is excellent. The $C_D$ data is plotted in semi-log scale for a better readability. The error bars are obtained using Eq. (\ref{eq:errorbars}).}}
\label{fig:cd_model}
\end{figure*}

\begin{figure*}[!]
\centering
\includegraphics[width=0.45\textwidth]{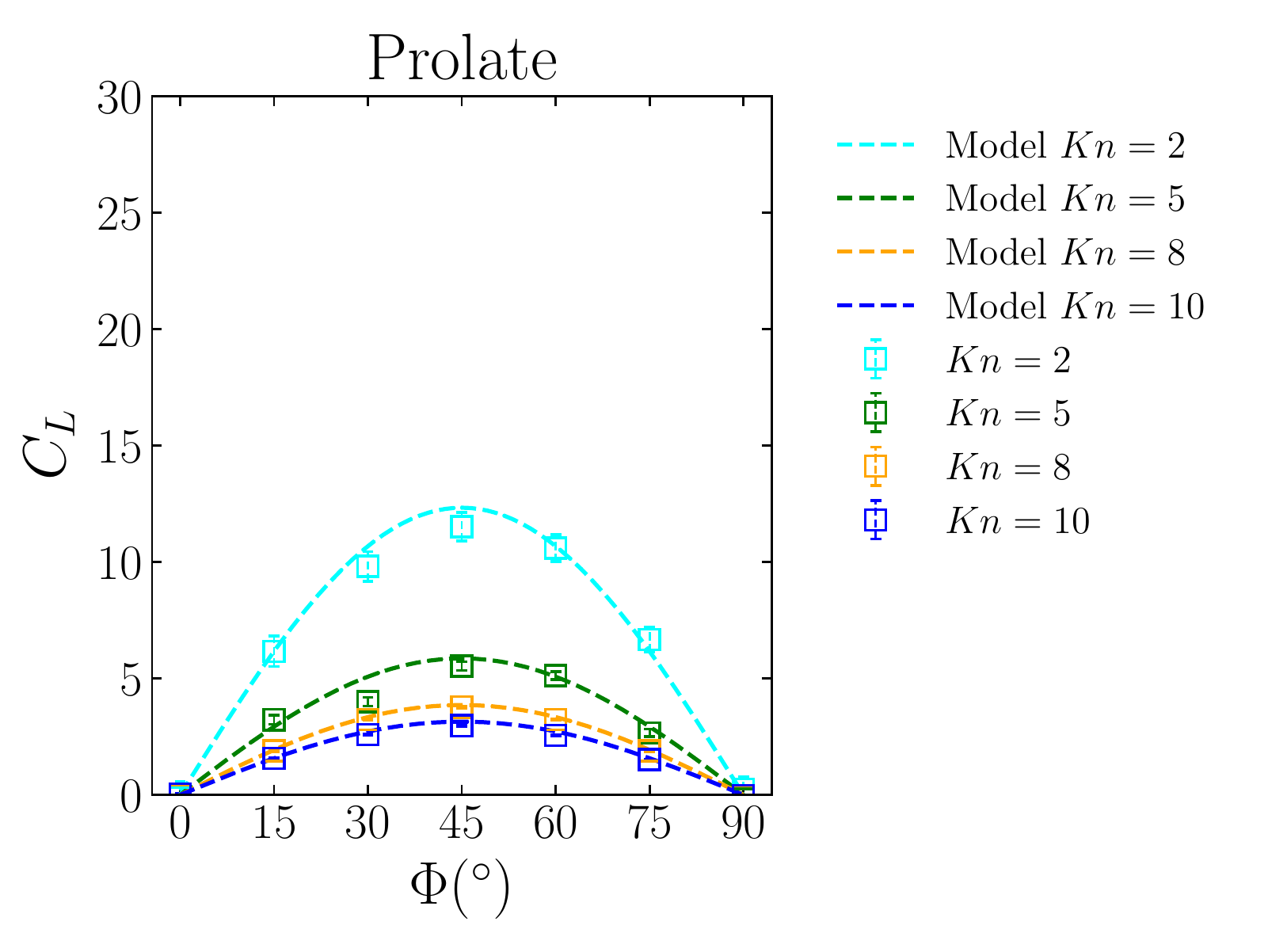}%
\includegraphics[width=0.45\textwidth]{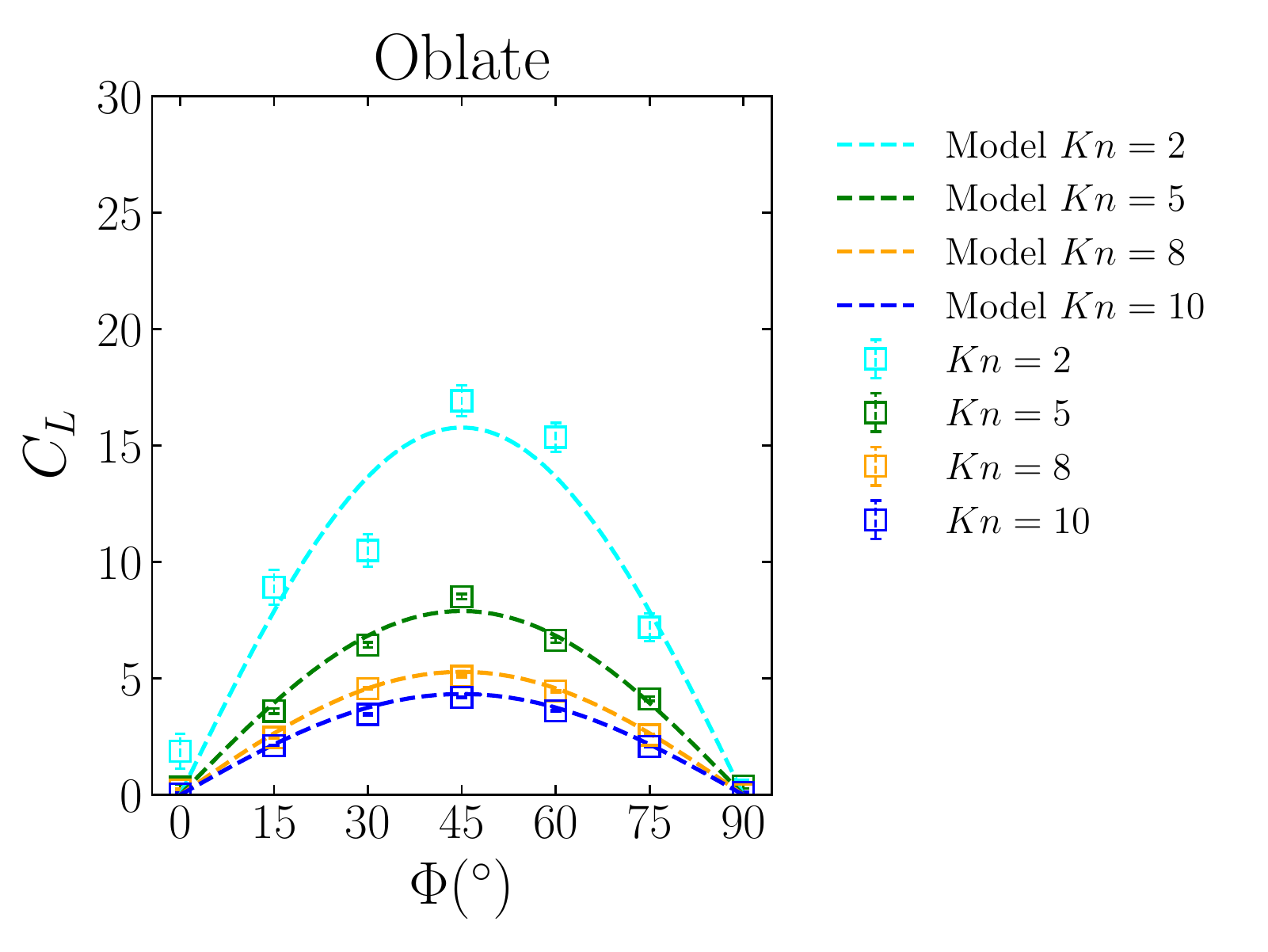}\\
\includegraphics[width=0.45\textwidth]{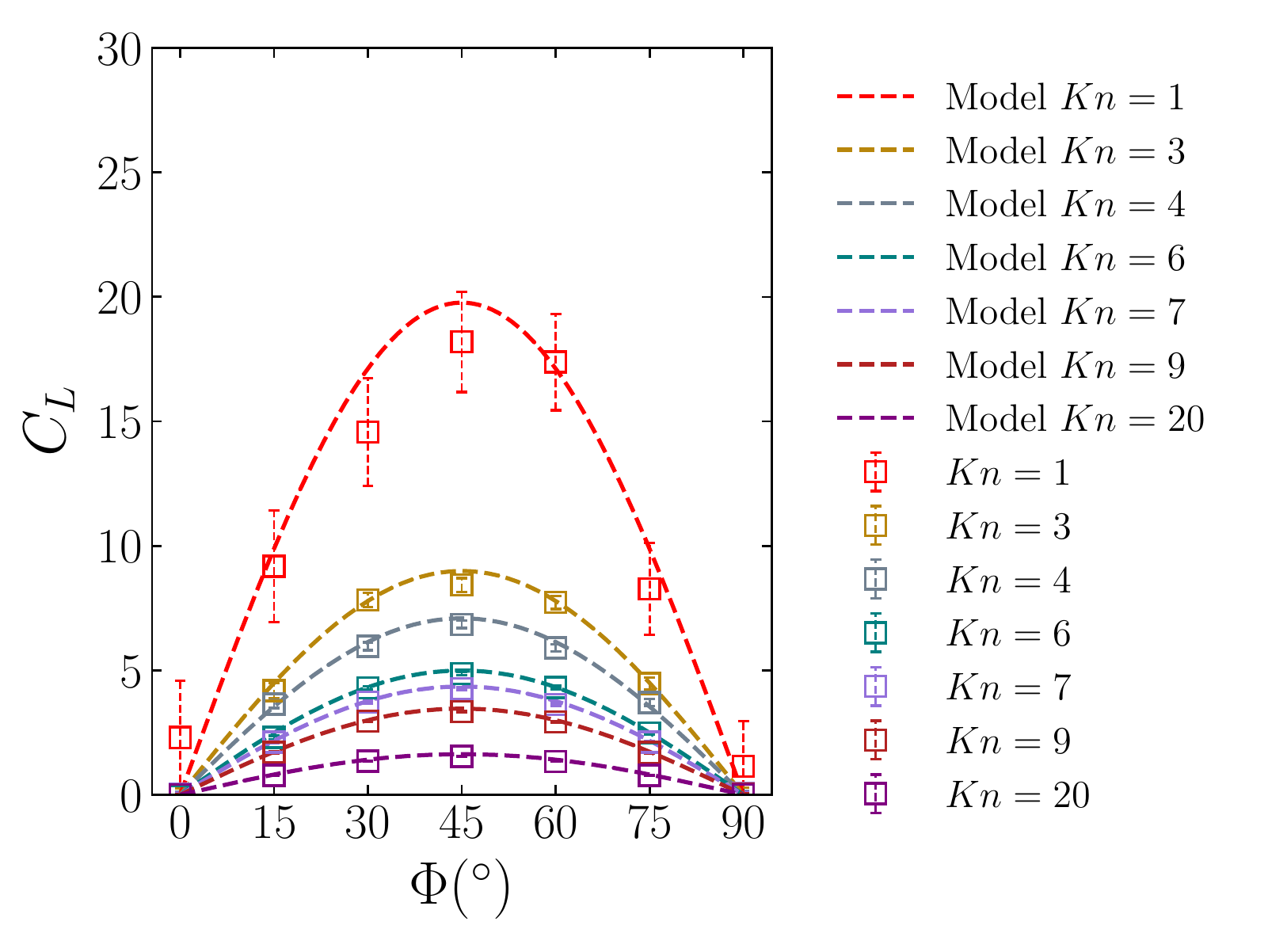}%
\includegraphics[width=0.45\textwidth]{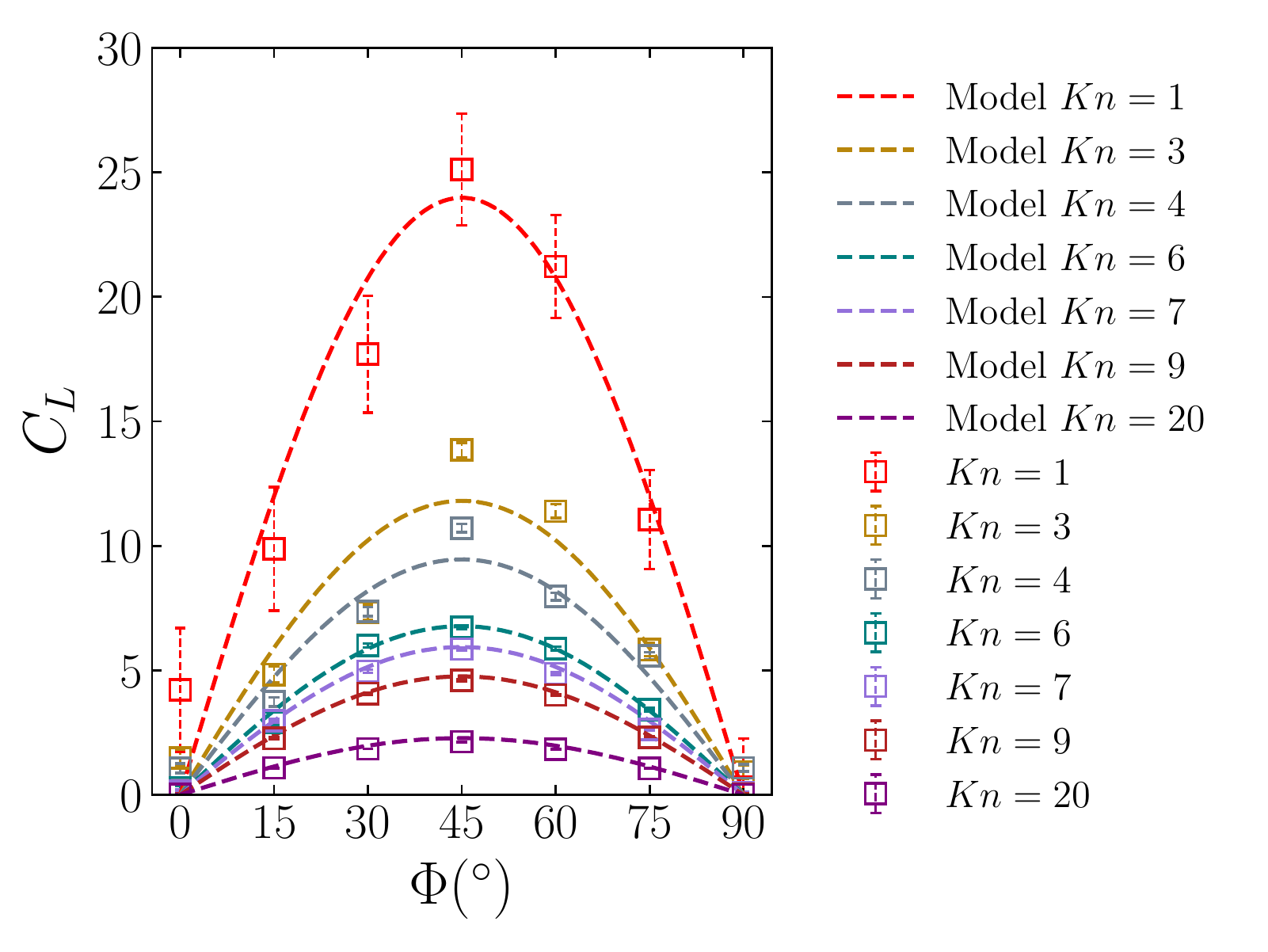}
\caption{\small{Comparison between the lift coefficient $C_L$ from DSMC simulations (colored squares) and the predictive model (dashed lines) for a prolate (left column) and an oblate (right column) ellipsoid.  The data from the DSMC simulations has been split in two sets: $Kn_{fit}$, which is used to obtain the fit for the predictive model, and $Kn_{test} $, which instead is not used during the fit. The model predictions, given by Eq. (\ref{eq:lift_model}) are then compared with the fit set (top row) and the test set (bottom row). Results are in excellent agreement with the simulation data for all investigated $Kn$ numbers. Measurements of $C_L$ are, however, less accurate than the ones of $C_D$ due to the small values of $C_L$ that lead to a smaller signal/noise ratio. This is particularly evident for the oblate case, where the simulation data for $Kn=1$ and $Kn=2$ fail to recover the symmetry with respect to $\Phi$. The error bars are obtained using Eq. (\ref{eq:errorbars}).}}
\label{fig:lift_model}
\end{figure*}

\begin{figure*}[!]
\centering
\includegraphics[width=0.3\textwidth]{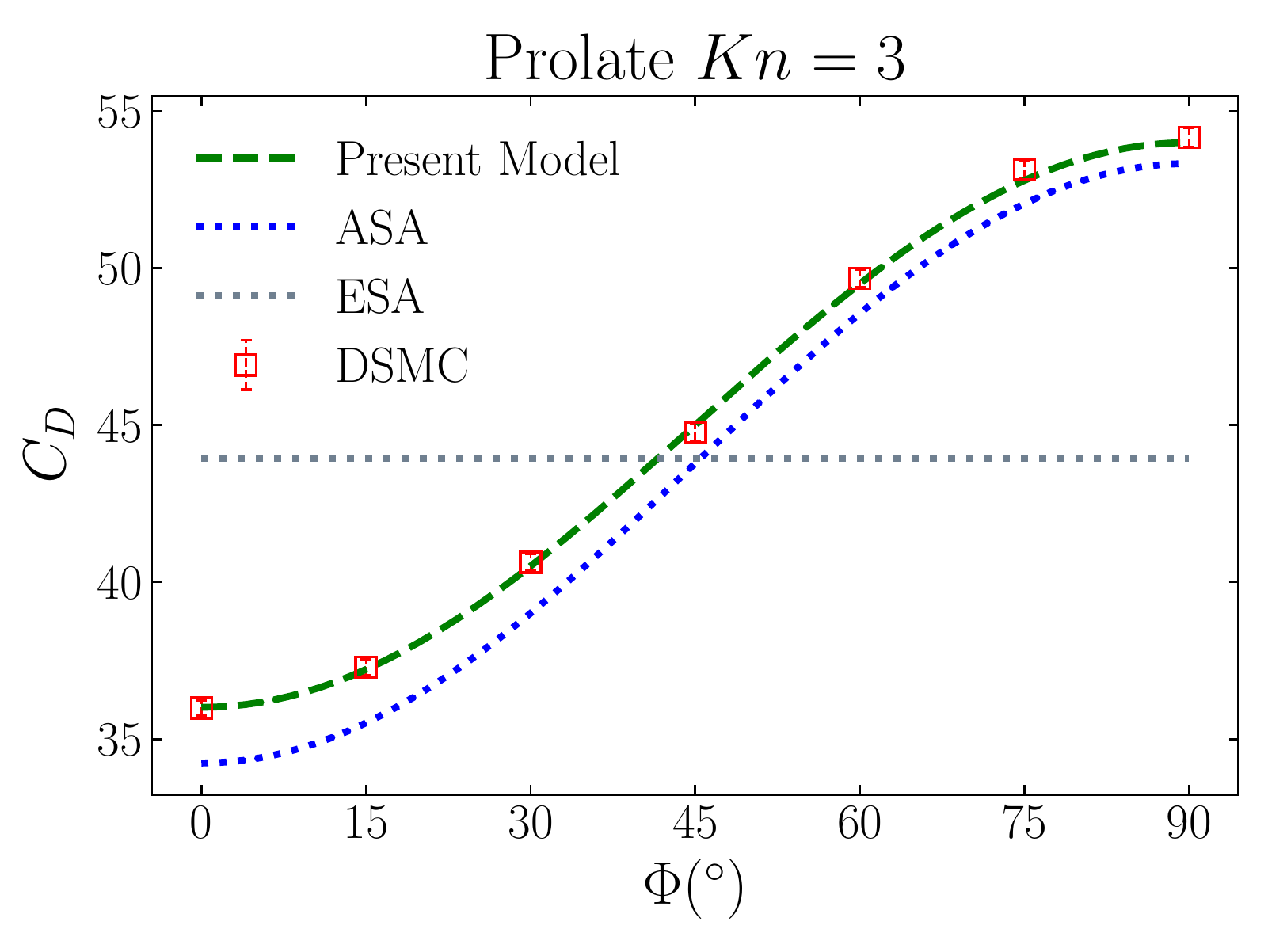}%
\includegraphics[width=0.3\textwidth]{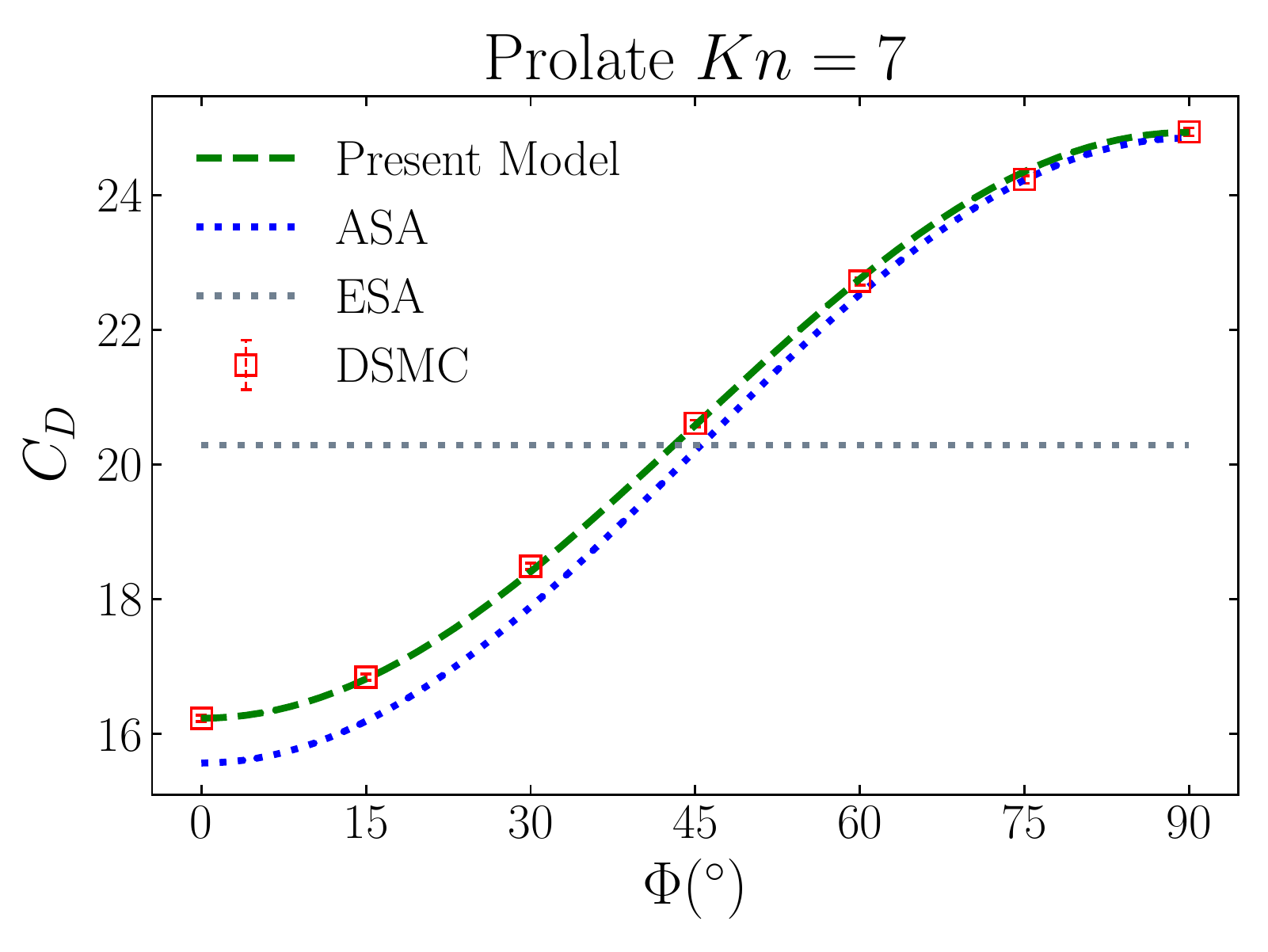}%
\includegraphics[width=0.3\textwidth]{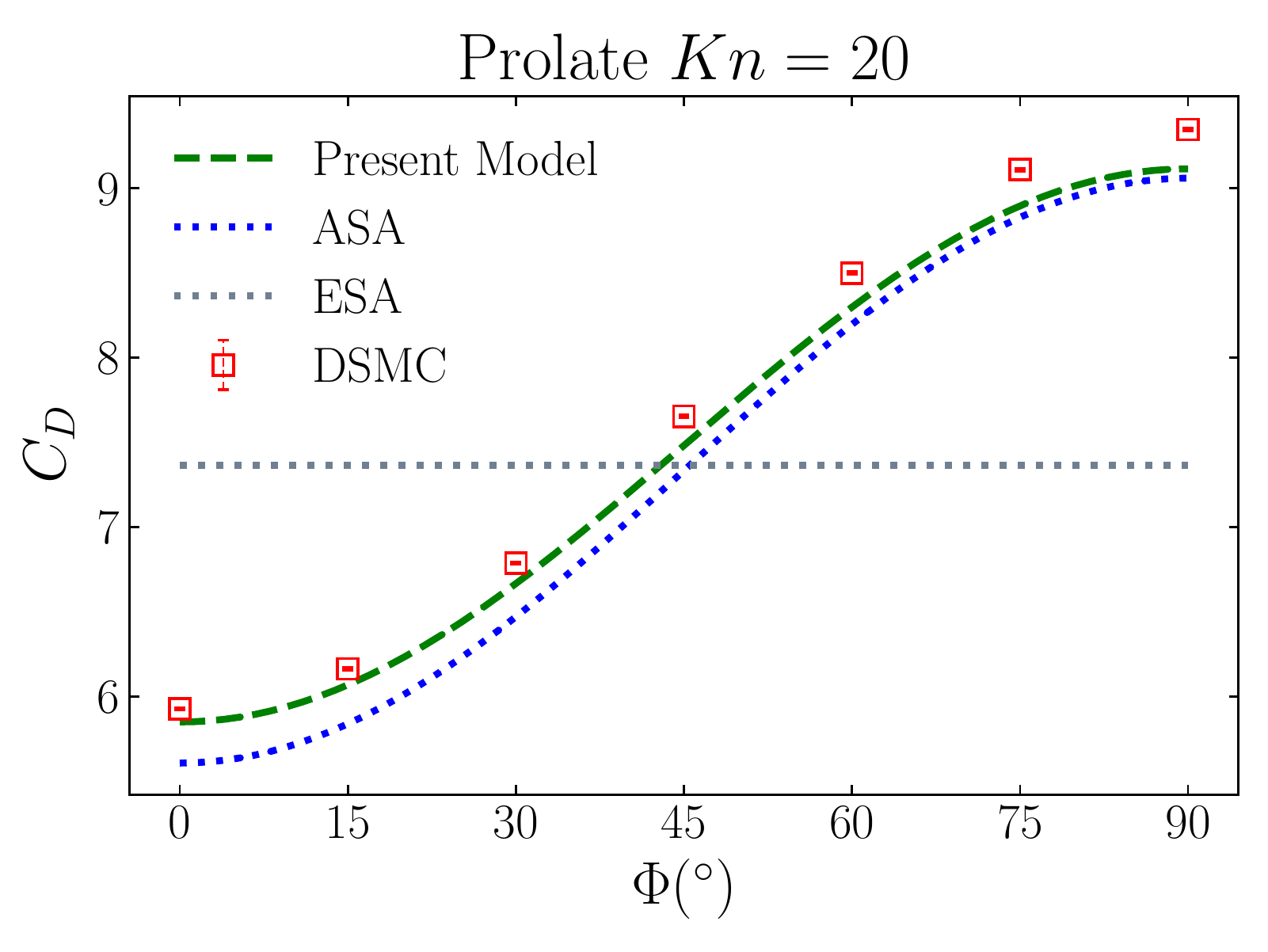}\\
\includegraphics[width=0.3\textwidth]{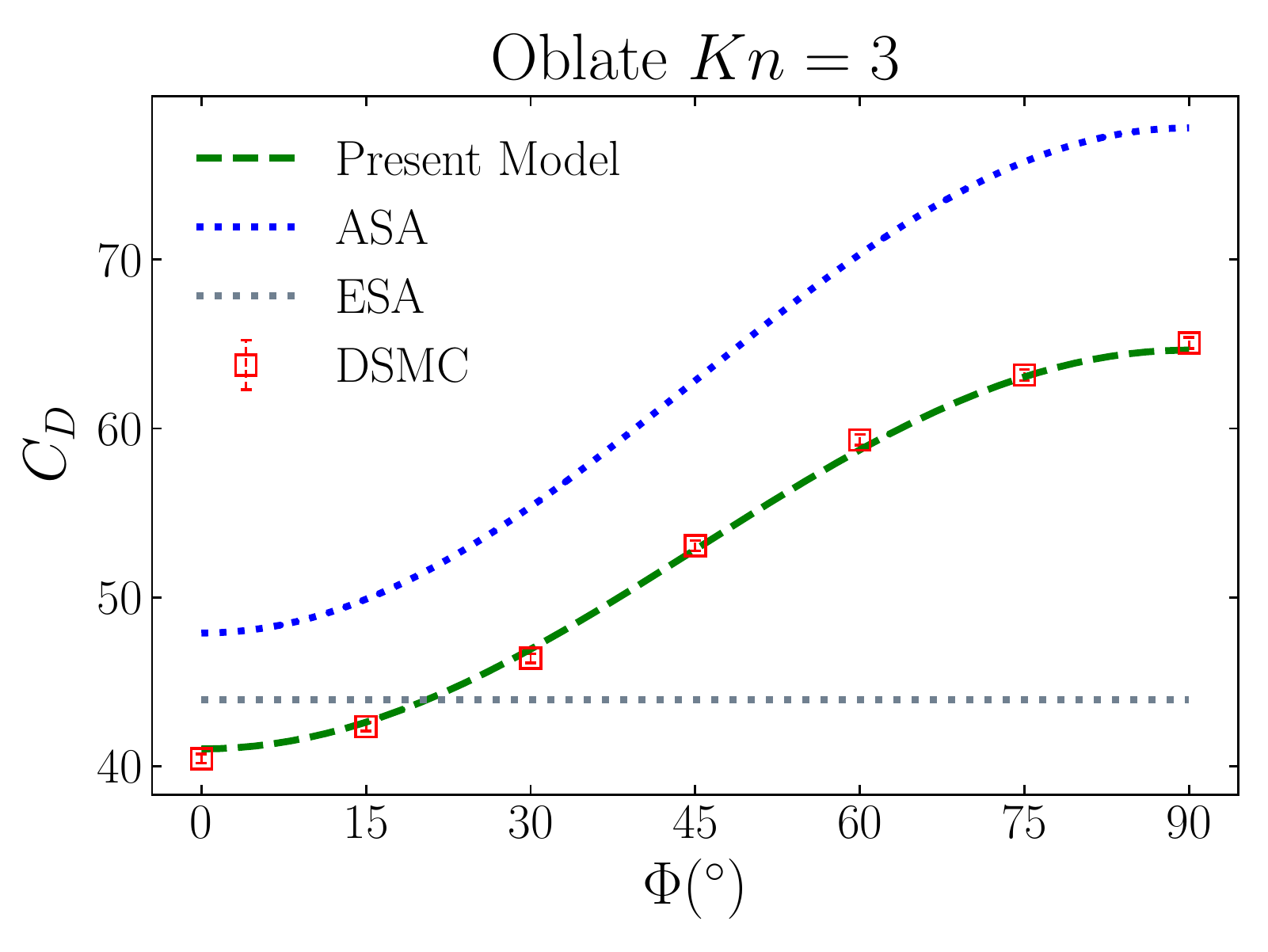}%
\includegraphics[width=0.3\textwidth]{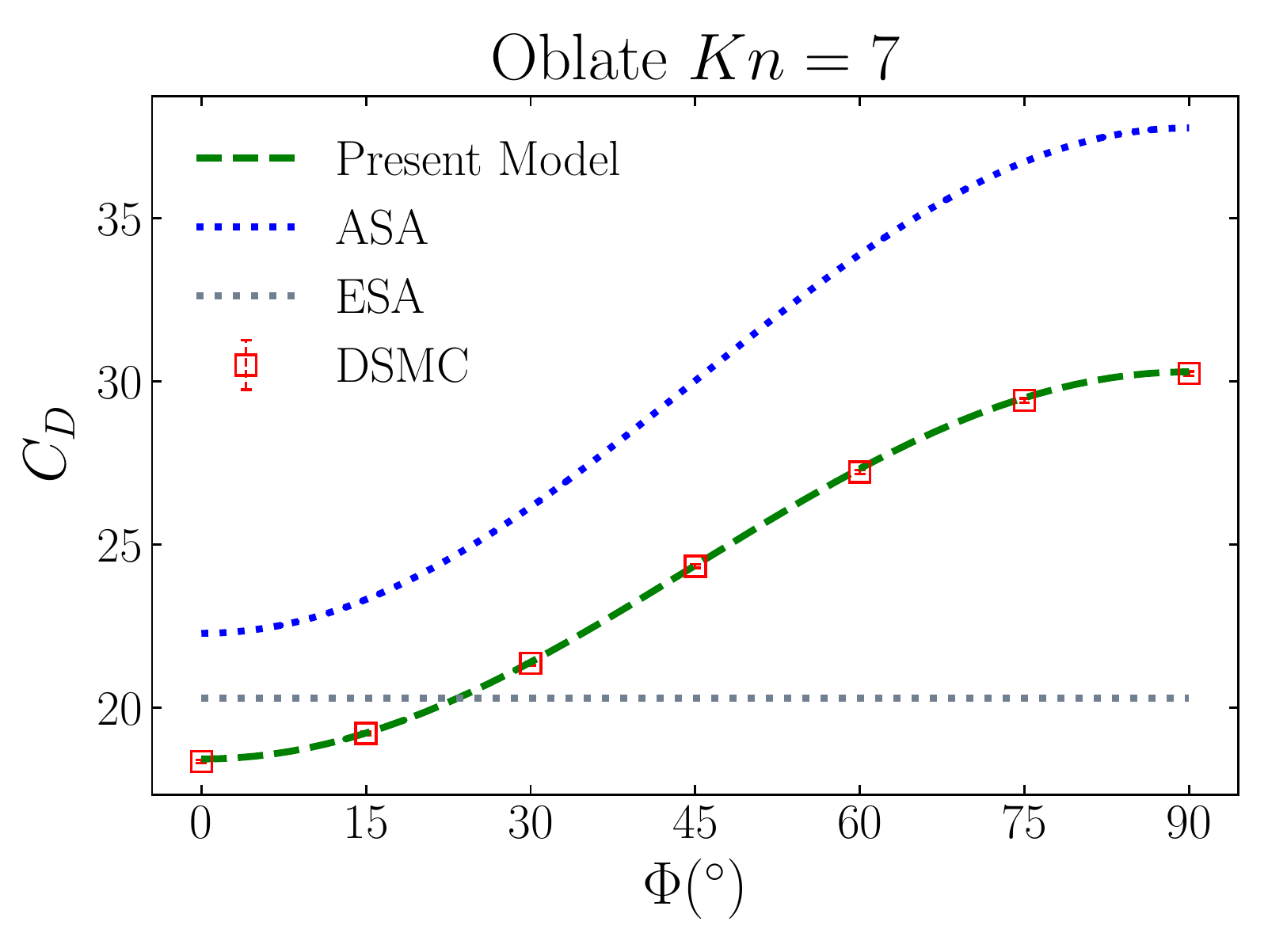}%
\includegraphics[width=0.3\textwidth]{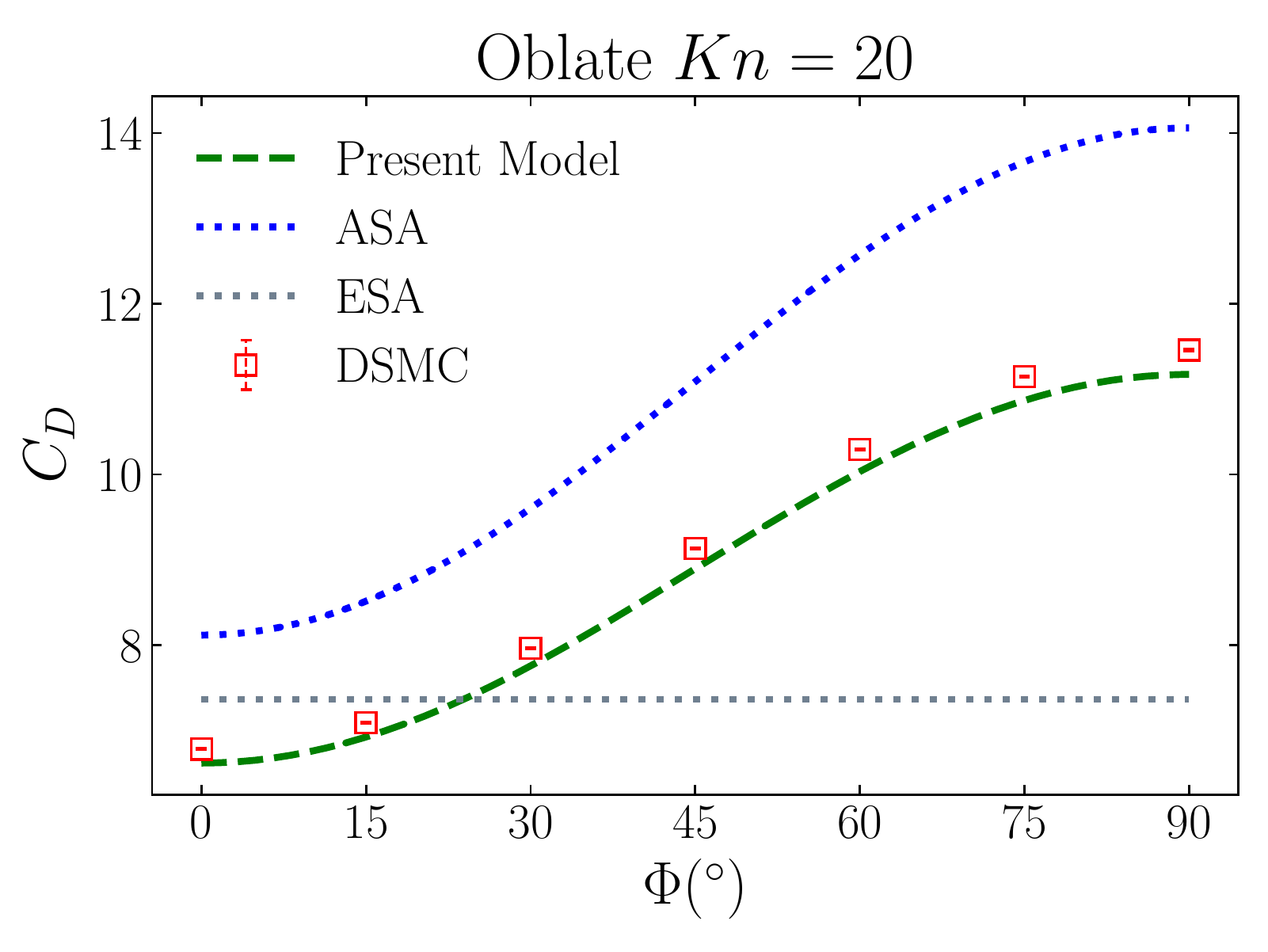}\\
\caption{\small{Comparison of the $C_D$ predictions as obtained with the present model (green dashed line), with the ASA (blue dotted line) and ESA (gray dotted line) models as defined in \cite{dahneke3} for different values of $Kn$ from the $Kn_{test}$ for a prolate (top row) and oblate (bottom row) with internal aspect ratio $a/b=2$. The predictions from the models are also compared with direct results from DSMC simulations (red squares), with related error bars obtained using Eq. (\ref{eq:errorbars}). The performances of the proposed model offer a general improvement with respect to the ASA, especially for the oblate case, where the ASA prediction is less accurate due to the larger deviations from the spherical case when dealing with an oblate ellipsoid.}}
\label{fig:model_comparison}
\end{figure*}

In this Section we investigate the relation between the drag and lift coefficients with respect to the orientation of the ellipsoidal particles (prolate and oblate) and the level of rarefaction, in an attempt to provide a heuristic model to extend the available correlations from the continuum to the rarefied regime. We consider ellipsoidal particles with aspect ratio $a/b = 2$, while all the remaining simulation parameters are the same as the ones used in Section \ref{sec:kn}.\\
Following the work of Sanjeevi \textit{et al.} \cite{sanjeevi1}, we define the drag and lift coefficients for an ellipsoidal particle as:
\begin{align}
C_{D} = \frac{|\mathbf{F_D}|}{\frac{1}{2} \rho U_0^2 \pi R_{eq}^2},
\label{eq:cd}
\end{align}
\begin{align}
C_{L} = \frac{|\mathbf{F_L}|}{\frac{1}{2} \rho U_0^2 \pi R_{eq}^2},
\label{eq:cl}
\end{align}
where $\mathbf{F_D}$ and  $\mathbf{F_L}$ are drag and lift force acting on the particle, $\rho$ is the density of the gas, $U_0$ is the gas free-stream velocity and $R_{eq}$ is the radius of the sphere with equivalent volume. Since we are investigating uniform flows in the Stokes regime, the pitching torque is known to vanish in such conditions \cite{guazzelli} due to the absence of an external rotational field, and thus the analysis of the pitching torque is not relevant in the scope of this work.\\
We can directly apply Eqs. (\ref{eq:drag_continuum})-(\ref{eq:lift}) to obtain the analytical relations for the drag and lift coefficients, with respect to the angle of attack $\Phi$, of the simulated ellipsoidal particles. Here $C_{D,0^\circ}$ and $C_{D,90^\circ}$ can be obtained from our DSMC simulations and the results are shown in Fig. \ref{fig:drag_ellipsoid}. As expected from the analysis presented in the previous Section, simulations results are in excellent agreement with the theoretical predictions.\\
An interesting feature of  Eqs. (\ref{eq:drag_continuum})-(\ref{eq:lift}) is that in order to predict $C_D$ and $C_L$ at any given orientation, it is sufficient to know the value of $C_D$ at $\Phi=0^\circ$ and $\Phi=90^\circ$. In the following of this Section we will provide a heuristic model for $ C_{D,0^\circ}$ and $C_{D,90^\circ}$ that takes into account rarefaction effects, in the attempt to include in   Eqs. (\ref{eq:drag_continuum})-(\ref{eq:lift}) a dependence on the Knudsen number.\\
The starting point is to observe that the drag force for the spherical case given by Eq. (\ref{eq:phil}) consists in the product between the Stokes drag in the continuum regime and a function of the Knudsen number $f(Kn)$ that captures rarefaction effects. Our assumption is that a similar relation holds also for ellipsoidal particles, and that an equation for $C_D$ that includes rarefaction effects can be written as a product between $C_D$ in the continuum limit and a function $g(Kn)$ which represents a small perturbation with respect to the spherical case:
\begin{align}
C_{D,0^\circ}(Kn) = \underbrace{C_{D,0^\circ}^{cont}}_{\mbox{continuum}} \cdot \underbrace{g_{0^\circ}(Kn)}_{\mbox{rarefaction effects}},
\label{eq:ellipsoid_prediction_0}
\end{align} 
\begin{align}
C_{D,90^\circ}(Kn) = \underbrace{C_{D,90^\circ}^{cont}}_{\mbox{continuum}} \cdot \underbrace{g_{90^\circ}(Kn)}_{\mbox{rarefaction effects}},
\label{eq:ellipsoid_prediction_90}
\end{align} 
where $C_{D,0^\circ}^{cont}$ and  $C_{D,90^\circ}^{cont}$ are the drag coefficients in the continuum regime, while $g_{0^\circ}(Kn)$ and $g_{90^\circ}(Kn)$ are model functions to be evaluated. In order to use Eqs. (\ref{eq:ellipsoid_prediction_0}) and (\ref{eq:ellipsoid_prediction_90}), we first need to compute the values of $C_{D,0^\circ}^{cont}$ and  $C_{D,90^\circ}^{cont}$ for the ellipsoidal particles investigated in this work. This can be done using the Schiller-Neumann \cite{schiller} drag expression for the drag coefficient of a spherical particle,
\begin{align}
C_D^{sph} = \frac{24}{Re}\left(1+0.15Re^{0.687}\right),
\label{eq:cd_sphere}
\end{align}
which has proven to be quite accurate up to a moderate Reynolds number. We can then obtain  $C_{D,0^\circ}^{cont}$ and  $C_{D,90^\circ}^{cont}$ for the simulated ellipsoidal particles using the heuristic relations from Happel and Brenner \cite{happel}:
\begin{align}
C_{D,0^\circ}^{cont} = C_D^{sph}K_{0^\circ},
\label{eq:cd_from_sphere_0}
\end{align}
\begin{align}
C_{D,90^\circ}^{cont} = C_D^{sph}K_{90^\circ}.
\label{eq:cd_from_sphere_90}
\end{align}
The correction factors $K_{0^\circ}$ and $K_{90^\circ}$ depend on the shape of the particle and for regular prolate and oblate ellipsoidal particles in creeping flow conditions, the exact analytical expressions for the correction factors were derived by Oberbeck \cite{oberbeck} as a function of their major and minor axes $a$ and $b$, respectively:
\begin{align}
K^{pr}_{0^\circ} &= \frac{(4/3)(a/b)^{-1/3}(1-(a/b)^2)}{a/b - \frac{(2(a/b)^2-1)\ln\left((a/b)\sqrt{(a/b)^2-1}\right)}{\sqrt{(a/b)^2-1}}}, \label{eq:corr_p_0}\\
K^{pr}_{90^\circ} &= \frac{(8/3)(a/b)^{-1/3}((a/b)^2-1)}{a/b + \frac{(2(a/b)^2-3)\ln\left((a/b)+\sqrt{(a/b)^2-1}\right)}{\sqrt{(a/b)^2-1}}}, \label{eq:corr_p_90}\\
K^{ob}_{0^\circ} &= \frac{(8/3)(b/a)^{-1/3}((b/a)^2-1)}{b/a - \frac{(3-2(b/a)^2)\cos^{-1}(b/a)}{\sqrt{1-(b/a)^2}}} ,\label{eq:corr_o_0}\\
K^{ob}_{90^\circ} &= \frac{(4/3)(b/a)^{-1/3}(1-(b/a)^2)}{b/a + \frac{(1-2(b/a)^2)\cos^{-1}(b/a)}{\sqrt{1-(b/a)^2}}}\label{eq:corr_o_90}.
\end{align}
Ouchene \textit{et al.} \cite{ouchene} show that for creeping flows, the set of Eqs. (\ref{eq:corr_p_0})-(\ref{eq:corr_o_90}) predicts the drag coefficients of prolate ellipsoids with different aspect ratios with very high accuracy. We then compute $C_{D,0^\circ}^{cont}$ and  $C_{D,90^\circ}^{cont}$  by substituting Eq. (\ref{eq:cd_sphere}) into Eqs. (\ref{eq:cd_from_sphere_0}) and (\ref{eq:cd_from_sphere_90}), using the corrections given by Eqs. (\ref{eq:corr_p_0})-(\ref{eq:corr_o_90}). The results are shown in Table \ref{tab:cd_cont}.\\
\begin{table}
\begin{center}
\begin{tabular}{ |c|c|c| } 
\hline
  &   Prolate   & Oblate\\[0.1pt]
\toprule
$C_{D,\Phi=0^\circ}$      & $236$   				     &  $247$  			      \\ 
$C_{D,\Phi=90^\circ}$   & $270$ 					& $282$ 					  \\ 
\hline
\end{tabular}
\end{center}
\caption{\small{Values of $C_{D,0^\circ}$ and $C_{D,90^\circ}$ (three-digit accuracy) for a prolate and oblate ellipsoid with aspect ratio $a/b=2$. Results are obtained from the theoretical prediction given by Eqs. (\ref{eq:cd_from_sphere_0}) and (\ref{eq:cd_from_sphere_90})  using the corrections factor from Eqs.  (\ref{eq:corr_p_0})-(\ref{eq:corr_o_90}) and the expression for $C_D^{sph}$ from Eq. (\ref{eq:cd_sphere}).}}
\label{tab:cd_cont}
\end{table}
To address rarefaction effects, we propose the following choice for the general expression of the functions $g_{0^\circ}(Kn)$ and $g_{90^\circ}(Kn)$, where we assume that such effects on ellipsoidal particles can be described as small variations with respect to the function $f(Kn)$ for the spherical case: 
\begin{align}
g_{0^\circ}(Kn) = f(Kn) + \frac{a_{0^\circ}}{b_{0^\circ} + c_{0^\circ}Kn},
\label{eq:g_0}
\end{align}
\begin{align}
g_{90^\circ}(Kn) = f(Kn) + \frac{a_{90^\circ}}{b_{90^\circ} + c_{90^\circ}Kn}
\label{eq:g_90}
\end{align}
where $a,\ b$ and $c$ are free parameters to be determined separately for $C_{D,0^\circ}$ and $C_{D,90^\circ}$. \\
We show that with the proposed choice of the $g(Kn)$ functions, it is sufficient to fit the model on a small set of $Kn$ to obtain a robust predictive model for rarefaction effects on ellipsoidal particles. The simulation data is split in two groups:
\begin{align}
Kn_{fit} = 2,5,8,10.
\label{eq:kn_fit}
\end{align}

\begin{align}
Kn_{test} = 1,3,4,6,7,9,20.
\label{eq:kn_test}
\end{align}
We then perform a fit of $ C_{D,0^\circ}$ and $C_{D,90^\circ}$ from simulation data as a function of $Kn$, using only the $Kn_{fit}$ set from Eq. (\ref{eq:kn_fit}) and the fit functions given by Eqs. (\ref{eq:ellipsoid_prediction_0}) and (\ref{eq:ellipsoid_prediction_90}). The results of the fit are shown in Fig. \ref{fig:fit} and the obtained fit parameters are given in Table \ref{tab:fit}.\\
\begin{table}
\begin{center}
\begin{tabular}{ |c|c|c|c|c| } 
\hline
 &  \multicolumn{2}{|c|}{   Prolate}   & \multicolumn{2}{|c|} {Oblate}\\[0.1pt]
\hline
& $\Phi=0^\circ$ &$\Phi=90^\circ$ &$\Phi=0^\circ$ & $\Phi=90^\circ$  \\[0.1pt]
\toprule
$a$      & $-0.171$   				     &  $0.115$  					  &  $-0.219$ 					 & $0.223$  \\ 
$b$   & $1.023$ 					& $2.492$ 					  & $3.762$ 				      & $1.523$ \\ 
$c$   & $1.482$  			         & $1.603$   & $1.192$ 					  & $1.126$ \\ 

\hline
\end{tabular}
\end{center}
\caption{\small{Fit parameters obtained using Eqs. (\ref{eq:ellipsoid_prediction_0}) and (\ref{eq:ellipsoid_prediction_90}) to fit $C_{D,0^\circ}$ and $C_{D,90^\circ}$ as obtained from DSMC simulations. These parameters are used to define the functions $g_{0^\circ}(Kn)$ and $g_{90^\circ}(Kn)$, which general expression is given in Eqs. (\ref{eq:g_0})-(\ref{eq:g_90}).}}
\label{tab:fit}
\end{table}
Once the functions $g(Kn)$ are determined for the needed orientations, we can verify if the model succeeds in the prediction of  $C_D$ at different values of $Kn$ and $\Phi$. In order to do so we plug Eqs. (\ref{eq:ellipsoid_prediction_0}) and (\ref{eq:ellipsoid_prediction_90}) into Eq. (\ref{eq:drag_continuum}) to obtain the final model equation for $C_D(\Phi, Kn)$:
\begin{equation}
\begin{split}
&C_{D}(\Phi,Kn) = (C_{D,0^\circ}^{cont}g_{0^\circ}(Kn) + \\ &(C_{D,90^\circ}^{cont}g_{90^\circ}(Kn) - C_{D,0^\circ}^{cont}g_{0^\circ}(Kn))\sin^2\Phi.
\end{split}
\label{eq:drag_model}
\end{equation} 
The comparison between Eq. (\ref{eq:drag_model}) and the results from DSMC simulation is performed on both the data sets $Kn_{fit}$ [Eq. (\ref{eq:kn_fit})] and $Kn_{test}$ [Eq. (\ref{eq:kn_test})], where the latter set has not been used during the fit process. The results are shown in Fig. \ref{fig:cd_model}, where we can observe an excellent agreement between the proposed model and the simulations. Particularly relevant is the agreement between the model and the data for $Kn_{test}$, showing that the model correctly predicts rarefaction effects on values of $Kn$ that were not included in the fitting process and it can be extended to the regimes with $Kn\leq 2$ and  $Kn\geq 10$. \\
In an analogous way, we investigate the capability of the predictive model given by Eqs. (\ref{eq:ellipsoid_prediction_0})-(\ref{eq:ellipsoid_prediction_90}) to address the lift coefficient $C_L$ of the different ellipsoids. Following the same approach we used for the evaluation of $C_D$, we plug Eqs. (\ref{eq:ellipsoid_prediction_0}) and (\ref{eq:ellipsoid_prediction_90}) into Eq. (\ref{eq:lift}) to obtain the model equation for $C_L$:
\begin{equation}
\begin{split}
&C_L(\Phi,Kn) = \\ & (C_{D,90^\circ}^{cont}g_{90^\circ}(Kn) -  C_{D,0^\circ}^{cont}g_{0^\circ}(Kn) )\sin\Phi\cos\Phi.
\end{split}
\label{eq:lift_model}
\end{equation} 
We can now compare the prediction from Eq. (\ref{eq:lift_model}) with the $C_L$ obtained from the DSMC simulations,  again using the same approach to separate the data into $Kn_{fit}$ and $Kn_{test}$. The results of the comparison are shown in Fig. \ref{fig:lift_model}.
The model prediction is in reasonable agreement with the simulation data, considering that $C_L \ll C_D$, leading to a lower signal/noise ratio from the DSMC simulations for $C_L$ with respect to $C_D$. \\
In the last part of this paper, we compare the performances of the predictive model proposed in this work with existing phenomenological models available in the literature used to predict the drag on non-spherical particles at different Knudsen numbers, namely the previously mentioned ESA and ASA, as defined by Dahneke \cite{dahneke3}. The results of the comparison, limited to some of the values in $Kn_{test}$ set, are presented in Fig. \ref{fig:model_comparison}, where we use the values provided by \cite{dahneke3} to compute the ASA prediction for the spheroidal particles investigated in this work. The performances of the proposed model in reproducing $C_D$ show a general improvement with respect to the ASA model, and this is particularly evident for the oblate case. This is due to the larger departure from the spherical case of oblate ellipsoids, as the ASA model appears to be less accurate the higher this departure is.\\
While the presented approach is proven to be successful in predicting rarefaction effects on the forces exerted on ellipsoidal particles, it is limited to the investigated aspect ratio ($a/b=2$) and for fully-diffusive surfaces ($\sigma = 1$). In our future works we plan to extend the predictive model to a larger range of aspect ratio, including also effects from a momentum accommodation coefficient $\sigma$ lower than unity to take into account the presence of specular reflections.
\section{Conclusions}
\label{sec:conclusions}

In conclusion, we developed a two-way coupled algorithm to address interactions, under rarefied conditions, between gas flows and spheroidal particles based on momentum exchange for our in-house DSMC numerical code. The surface of the solid particle is defined by its analytical expression and the interactions between the gas and the rigid body are computed from a microscopic approach. The collision points at which the computational molecules impinge on the solid surface are obtained through a ray-tracing technique, allowing an exact computation of the collision points of the gas molecules impinging on the solid surface.\\
The algorithm features the cut-cell method to address the DSMC grid cells that are partially covered by the solid volume. We use a Monte-Carlo approach to evaluate the volume of the boundary cells, showing that for an appropriate number of trials, it is possible to recover the volume of the interested cells with great accuracy. \\
The algorithm is validated by computing the drag force on a spherical particle immersed in a uniform argon gas flow at different rarefaction levels. We show that the measurements from our simulations are in good agreement with different results available in the literature, especially when compared with analogous DSMC methods. The accuracy scaling of the mean value and of the standard deviation of the drag force is investigated with respect to the spatial and kinetic resolutions of the system. The mean value shows a second-order scaling with respect to spatial resolution in cases where the DSMC grid size is not too small when compared to the mean free path of the gas, and a first-order scaling with respect to the kinetic resolution. The standard deviation on the measured drag force is constant in the former case and with an exponent of roughly $0.5$ in the latter case. An interesting feature of the algorithm is that it is possible to achieve an accuracy as high as $97\%$ (with respect to a reference value obtained from simulations at higher resolution) in the evaluation of the mean value of the drag force also when the size of the spherical particle is comparable with the size of a single DSMC grid cell, making this approach valuable to address particles with complex surfaces (providing they can be described by an analytical function) or with a small size, when compared with the size of the simulation grid.\\
We then address the impact of shape, orientation and rarefaction on the drag force for a prolate and an oblate ellipsoid. We firstly propose a suitable definition of the Knudsen number, $Kn$, for ellipsoidal particles based on the radius of the sphere of equivalent volume. Orientation and rarefaction effects are, in fact, not related, and it is possible to address them separately. We come to this conclusion by inspecting analytical results available in the literature for the collisionless case. For small values of the molecular speed ratio, in fact, we observe that the scaling of the drag and lift coefficients, with respect to the orientation, typical of the continuum regime is recovered also in the collisionless case.\\
When dealing with ellipsoidal particles, the definition of $Kn$ proposed in this work recovers the expected scaling of the drag force experienced by different particles with respect to their cross-sectional areas, allowing to meaningfully capture rarefaction effects on more complex shapes using only one reference length.\\
Finally, we develop a heuristic model to predict the drag and lift coefficients for ellipsoidal particles in a range of $Kn$ that includes the transition and the free-molecular regimes. The predictive model is based on the assumption that rarefaction effects on ellipsoidal particles can be represented as small perturbation with respect to the spherical case. These perturbations are obtained through a fit of our simulation data. The model obtained with this procedure shows robust performances in predicting drag and lift coefficients in the investigated range of $Kn$, and we show that the model can be successfully applied outside of the range of $Kn$ used for its derivation. Moreover, the model proposed in this work offers better predictions when compared to similar phenomenological models such as the Equivalent Sphere Approximation and the Adjusted Sphere Approximation, especially for the oblate ellipsoid case, where the shape of the particle largely deviates from the spherical case.\\
This work can improve the available models used in unresolved Euler-Lagrangian simulations of particles in rarefied conditions, as the drag and lift correlations can now be extended to the rarefied regimes. This allows in principle to include shape and orientation effects in point-particles simulations, greatly increasing the capability to simulate suspensions of non-spherical particles in rarefied gas flows, which are expected to follow different trajectories with respect to the spherical case. This approach presents, however, some limitations as the derivation is currently limited to prolate and oblate ellipsoidal particles in Stokes flows with aspect ratio $a/b=2$ and for fully diffusive gas-surfaces interactions.\\
In our future works we plan to extend the proposed technique to ellipsoids with different aspect ratios and to include a study of the impact of different momentum accommodation coefficients, by taking into account the presence of specular reflections at the gas-solid interface. A further development will be related to take into consideration particles that are free to move and rotate due to the interactions with the gas flow.

\section*{Acknowledgments}
This work was supported by the Netherlands Organization for Scientific Research (NWO-TTW), under the Project No. 15376.

\end{document}